\documentclass[article]{aa}
\usepackage{graphicx,latexsym,longtable,lscape,supertabular}
\usepackage[authoryear]{natbib}

\begin{document}
\title{Obscured and unobscured AGN populations in a hard-X-ray selected
sample of the XMDS survey} 
 
\author{M. Tajer \inst{1,2} 
\and M. Polletta \inst{3}
\and L. Chiappetti \inst{4}
\and L. Maraschi \inst{1}
\and G. Trinchieri \inst{1}
\and D. Maccagni \inst{4}
\and S. Andreon \inst{1}
\and O. Garcet \inst{5}
\and J. Surdej \inst{5}
\and M. Pierre \inst{6}
\and O. Le F{\` e}vre \inst{7}
\and A. Franceschini \inst{8}
\and C. J. Lonsdale \inst{3}
\and J. A. Surace \inst{9}
\and D. L. Shupe \inst{9}
\and F. Fang \inst{9}
\and M. Rowan - Robinson \inst{10}
\and H. E. Smith \inst{3}
\and L. Tresse \inst{7}
}

\institute{INAF -- Osservatorio di Brera, via Brera 28, 20121 Milano, Italy
\and
Universit\`a degli Studi di Milano - Bicocca, Dipartimento di Fisica, Piazza della Scienza
3, 20126 Milano, Italy
\and 
Center for Astrophysics \& Space Sciences, University of California, San Diego, La Jolla, CA
92093, USA
\and
INAF -- IASF Milano, via Bassini 15, 20133 Milano, Italy
\and
Institut d'Astrophysique et de G\'eophysique, Universit\'e de Li\`ege, All\'ee du
6 Ao\^ut 17, 4000 Li\`ege 1, Belgium
\and
CEA/DSM/DAPNIA Service d'Astrophysique, Saclay, 91191 Gif sur Yvette, France
\and
Laboratoire d'Astrophysique de Marseille, UMR 6110 CNRS-Universit\'e de Provence, BP 8, 13376 Marseille Cedex 12, France
\and
Dipartimento di Astronomia, Universit\`a di Padova, Vicolo Osservatorio 2, 35122,
Padova, Italy
\and
\textit{Spitzer} Science Center, California Institute of Technology, 100-22,
Pasadena, CA 911125, USA
\and
Astrophysics Group, Blackett Laboratory, Imperial College, Prince Consort Road,
London, SW7 2BW, UK
}

\offprints{M. Tajer} 
\mail{marzia.tajer@brera.inaf.it}

\date{Received <date> / accepted <date>}
\authorrunning{Tajer et al.}
\titlerunning{AGN in the XMDS}

\abstract
{}
{Our goal is to probe the populations of obscured and unobscured AGN
investigating their optical-IR and X-ray properties as a function
of X-ray flux, luminosity and redshift within
a hard X-ray selected sample with wide multiwavelength coverage.
}
{We selected a sample of 136 X-ray sources
detected at a significance of $\geq 3 \sigma$ in the $2-10$ keV band ($F_{2-10} \ga 10^{-14}$
erg cm$^{-2}$ s$^{-1}$) in a $\sim 1$ deg$^2$ area
in the XMM Medium Deep Survey (XMDS). The XMDS area is covered with optical
photometry from the 
VVDS and CFHTLS surveys and infrared Spitzer data from the SWIRE
survey. Based on the X-ray luminosity and X-ray to optical ratio, 132 sources are likely AGN, of
which 122 have unambiguous optical - IR identification.  
The observed optical and IR spectral energy distributions of all identified sources 
are fitted 
with AGN/galaxy templates in order to classify them and 
compute photometric redshifts.
X-ray spectral analysis is performed individually for sources
with a sufficient number of counts and using a stacking
technique for subsamples of sources at different flux levels. Hardness ratios are used to estimate X-ray
absorption in individual weak sources.}
{70\% of the AGN are fitted by a type 2 AGN or a star forming galaxy template.
We group them together in a single class of ``optically obscured'' AGN. These 
have ``red'' optical colors and in about
60\% of cases show significant X-ray absorption (N$_H > 10^{22}$ cm$^{-2}$).
Sources with SEDs typical of type 1
AGN have ``blue'' optical colors and exhibit X-ray absorption in about 30\% of cases.
The stacked X-ray spectrum of obscured AGN is flatter than that of type 1
AGN and has an average spectral slope of $\Gamma = 1.6$.
The subsample of
objects fitted by a star forming galaxy template has an even harder stacked
spectrum, with $\Gamma \sim 1.2 - 1.3$. 
The obscured fraction is larger at lower fluxes, 
lower redshifts and lower luminosities. X-ray absorption is less
common than ``optical'' obscuration and its incidence is nearly constant with
redshift and luminosity. This implies that at high luminosities 
X-ray absorption is not necessarily related to optical obscuration. The
estimated surface densities of obscured, unobscured AGN and type 2 QSOs are
respectively 138, 59 and 35 deg$^{-2}$ at $F > 10^{-14}$
erg cm$^{-2}$ s$^{-1}$.}
{}

\maketitle

\section{Introduction}
One of the goals of deep X-ray surveys is to probe the origin of the X-ray
background (XRB) to the faintest flux levels; they have resolved into discrete
sources more than 90\% of the
XRB in the $0.5 -2$ keV band and up to 80 - 90\% in the $2 - 10$ keV band
\citep{Giacconi etal 02, Alexander etal 03, Moretti etal 03, Deluca&Molendi 04}.
The resolved fraction of the XRB drops however to $\sim 60$\% above $\sim 6$ keV
and $\sim 50$\% above $\sim 8$ keV \citep{Worsley etal 05}.   
Most of the sources detected in deep, pencil-beam X-ray surveys are characterized by poor counting statistics,
preventing a detailed analysis of X-ray spectral properties, and optical
counterparts are often too faint for spectroscopic follow-up.  
Medium deep surveys, covering larger areas, are useful to
bridge the gap between known X-ray source populations at low redshifts
and those required to model the background and to collect a large number of
sources for which X-ray and optical spectral analysis are feasible
\citep[e.g.][]{Piconcelli etal 03, Georgakakis etal 06}. They are also more
effective than deep surveys to find rare objects, such as type 2 QSOs
\citep{Fiore etal 03}.   

The XMM Medium Deep Survey \citep[XMDS, see][ hereafter Paper I]{paper1}
consists of 19 X-ray pointings, of 
nominal exposure of 20 ksec, covering a contiguous area of about 2.6 deg$^2$. It also lies at the heart of the larger, shallower XMM Large Scale
Structure (LSS) Survey \citep{Pierre etal 04, Pacaud etal 06}, which will cover $\sim 10$
deg$^2$ and is principally devoted to clusters study \citep{Pierre etal 06}.
Several surveys at different wavelengths are associated to the XMDS:
the VIRMOS VLT Deep Survey \citep[VVDS,][]{Lefevre etal 04} and the Canada - France - Hawaii Telescope Legacy
Survey (CFHTLS)\footnote{{\tt http://www.cfht.hawaii.edu/Science/CFHLS/}} in the
optical, the UKIRT Infrared Deep Sky Survey \citep[UKIDSS,][]{Dye etal 06,
Lawrence etal 06} in the near-IR and the \textit{Spitzer} Wide-Area
InfraRed Extragalactic Legacy Survey \citep[SWIRE,][]{Lonsdale etal
03}\footnote{\tt http://swire.ipac.caltech.edu/swire/swire.html} in the mid-IR.
Radio observations performed at VLA at 1.4 GHz \citep{Bondi etal 03}
and at 325 and 74 MHz \citep{Cohen etal 03} also cover the XMDS area.

In Paper I the catalogue of XMDS sources detected ($ S/N \geq 4 \sigma$)
in \textit{at least one} of five
energy bands $0.3 - 0.5$, $0.5 - 2$, $2 - 4.5$, $4.5 - 10$ and $2 - 10$ keV within
the VVDS field (area $\sim 1$ deg$^{2}$) was presented together with tentative optical identifications.
The logN-logS distributions
were derived for X-ray sources in the full XMDS area separately in the two 
bands $0.5- 2$ and $2 - 10$ keV.
\citet{Gandhi etal 06} computed the logN-logS in the same
energy bands for X-ray sources in the whole XMM - LSS area, 
finding results in agreement with those of Paper I. 

Here we consider a sample of X-ray sources selected in
the ``hard'', $2 - 10$ keV, band, in order to investigate the populations of obscured 
and unobscured AGN and discuss their multiwavelength properties
in a way unbiased by the intensity of the sources in soft X-rays. 
In order to take advantage of the best multifrequency coverage available,
we consider sources in the VVDS area.
Using the \textit{Spitzer} data we construct the mid-IR/optical to X-ray spectral energy distributions (SEDs) 
of the sources to estimate redshifts and classify AGN into different
categories according to the best fitting template. We use the term ``optically
unobscured'' AGN for objects fitted by a type 1 AGN template, and ``optically
obscured'' AGN for objects fitted by type 2 AGN or a star forming galaxy
template, indicating that the optical-UV emission from the AGN is at least 
partially hidden.
In a companion paper \citep{Polletta etal 07} the templates used and
the observed SEDs are presented in detail and their dependence on luminosity
and absorption is discussed.
A comparison between X-ray and optical properties
of AGN with optical spectroscopy in the whole XMM-LSS is in
progress (Garcet et al., in prep.). 

Independently of the SED classification, the X-ray spectra and/or hardness ratios allow us
to estimate absorption in the X-ray band, associated to intervening gas. In
unified models for AGN \citep[e.g.][]{Antonucci 93}, obscuration by dust and
absorption by gas are thought to occur in a dusty torus surrounding the AGN.  
The increasing complexity of properties shown by individual AGN has
lead to a revision of this simple scheme proposing different regions
around the AGN as sites of absorption at different wavelengths
\citep[e.g.][]{Elvis 00, Krongold etal 07, Elitzur 06}.
We will therefore distinguish
optically obscured and X-ray absorbed AGN  and examine separately
their dependence on X-ray flux, redshift and luminosity.

The paper is organized as follows: the multiwavelength data set is
presented  in Section~\ref{sample}. Optical and IR identifications
are discussed in Section~\ref{id} while the X-ray, optical and IR 
properties of the sample are derived in Section~\ref{xottir}. 
The template SEDs and fitting process leading to the estimate of 
photometric redshifts and to the AGN classification 
are described in Section~\ref{photoz}.
The X-ray spectral analysis is presented in Section~\ref{xray}: 
X-ray spectra are analyzed individually for sufficiently bright sources, while 
for faint sources absorption is estimated from hardness ratios. 
A stacking technique is used to derive average X-ray spectra of subsamples
of AGN with different SED classification. 
The surface density of optically obscured and unobscured AGN and
of type 2 QSOs is derived in Section~\ref{lognlogs}.
Section~\ref{fractions} is devoted to the comparison of
the fractions of optically obscured or X-ray absorbed
AGN as a function of redshift, X-ray flux and luminosity.
Finally, in Section~\ref{conclusions} the results are summarized.

Throughout the paper H$_0 = 70$ Km s$^{-1}$ Mpc$^{-1}$, $\Omega_\Lambda = 0.7$ and $\Omega_M = 0.3$ are assumed. 

\section{The multiwavelength data set} \label{sample}

\begin{figure}
\centering
\includegraphics[width=8.5cm]{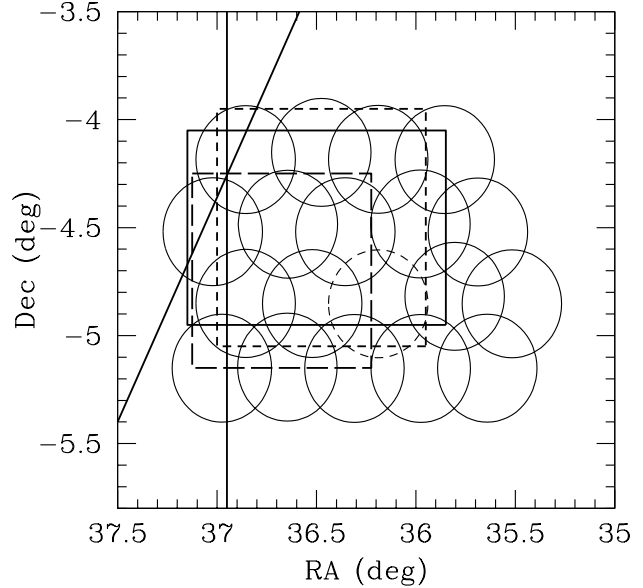}
\caption{Layout of the 19 \textit{XMM - Newton} pointings of the XMDS (circles;
the dashed circle marks field G12, which was not
analyzed because of high background level, see Paper I)
and of the associated surveys: VVDS (solid rectangle), CFHTLS Deep (D1, short
dashed rectangle), and Wide (W1, area on the right side of the vertical solid line),
UKIDSS (long dashed rectangle) and SWIRE (area on the right side of the diagonal
solid line).}
\label{fov}
\end{figure}

The layout of the XMDS observations is shown in Fig.~\ref{fov}; superposed
are the areas covered by the various associated surveys at other 
wavelengths, VVDS and CFHTLS in the optical, UKIDSS in the near-infrared and
SWIRE in the mid-infrared. 
Shallower \textit{XMM - Newton} pointings in the context of the
XMM - LSS lie all around the XMDS area (see Pierre et al., in prep.)     

\subsection{X-ray data} \label{xrayobs}
The sample includes 136 X-ray sources detected at $\geq 3
\sigma$ in the $2 - 10$ keV band (the 3 $\sigma$ hard sample hereafter), which fall
within the sky area covered by the VVDS photometric
survey ($\sim 1$ deg$^2$, solid rectangle in Fig.~\ref{fov}). This area benefits from the richest multiwavelength coverage as
evident from Fig.~\ref{fov}. 

\begin{figure}
\centering
\includegraphics[width=8.5cm]{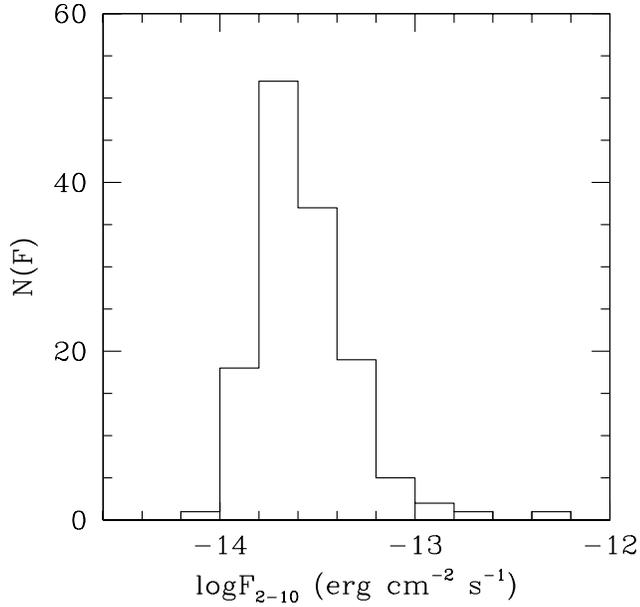}
\caption{2 - 10 keV flux distribution for the 136 X-ray sources in the 3
$\sigma$ hard sample.}
\label{istoflux}
\end{figure}

The sources were extracted from the XMDS catalog described in Paper I, to which
we refer also for details on the \textit{XMM - Newton}
observations and data reduction.
For all the sources count rates and fluxes were
obtained independently in  5 energy bands: 
$0.3 - 0.5$, $0.5 - 2$, $2 - 4.5$, $4.5 - 10$ and $2 - 10$ keV.
Fluxes were computed for a simple power law spectrum with spectral
index $\Gamma = 1.7$ and the average galactic column density in the XMDS
region \citep[N$_H = 2.61 \times 10^{20}$ cm$^{-2}$,][]{Dickey&Lockman 90}
separately in each energy band.
Fig.~\ref{istoflux} shows the $2 - 10$ keV flux distribution; 
the lowest flux that we sample is $\sim 10^{-14}$ erg cm$^{-2}$ s$^{-1}$.

Hardness ratios were computed for all sources (see subsection~\ref{xcolors}). For 55
sources we detect a sufficient number of net counts $(> 50$ in the $2 - 10$ keV
band) to attempt a spectral analysis for each source (see details in subsection~\ref{singlespec}).  

\subsection{Optical and near-infrared data} \label{opticalobs}
Broad band \textit{BVRI} photometric observations 
from the VVDS \citep{Mccracken etal 03} 
are available for XMDS sources over an area of about 1 deg$^{2}$. 
This photometry was obtained at the Canada
France Hawaii Telescope (CFHT) with the CFH12K camera 
at limiting magnitudes of $B_{AB} \sim 26.5, V_{AB} \sim 26.2, R_{AB} \sim 25.9$
and $I_{AB} \sim 25.0$ (50\% completeness for point sources).

The \textit{U} band imaging was performed over an effective area of $\sim 0.71$
deg$^2$ with the Wide Field Imaging (WFI) mosaic camera on the ESO MPI 2.2 m
telescope at La Silla, Chile. Two different \textit{U} filters were used, the ESO
\textit{U}/360 filter and the Loiano Observatory \textit{U} filter. The limiting magnitude is
$U_{AB} \sim 25.4$ \citep[see][]{Radovich etal 04}. 

A small area ($\sim 165$ arcmin$^2$) within the VVDS was also observed in the \textit{J} and \textit{K} bands down to a limiting magnitude of $J_{AB} \sim 24.2$ and
$K_{AB} \sim 23.9$ (50\% completeness for point sources) with the SOFI instrument
mounted on the ESO NTT telescope. A detailed description
of the $K$ band imaging survey is reported in \citet{Iovino etal 05}.

Optical spectroscopy with the VIsible Multi Object Spectrograph (VIMOS) on the
ESO -- VLT UT3 in the VVDS area is still in progress;  the project aims at
observing a representative large subsample of objects down 
to a limiting mag of  $I_{AB} \leq 24$
\citep{Lefevre etal 05}. 9 sources in the present sample have been observed and
for 8 of them a redshift has been derived.
Additional spectroscopic redshifts for 22 sources have recently been obtained from 2dF and
VLT FORS2 observations performed in the context of XMM-LSS follow up campaigns
(see Garcet et al., in prep.). Three other redshifts are from the literature.

The XMDS area lies within  the sky region covered by the CFHTLS, a large collaborative project between the Canadian and French communities.
Observations use the wide field imager MegaPrime equipped with MegaCam,
in the  $u^* g' r' i' z'$ filters. 
Both the ``Wide''
survey field W1 (8 deg $\times$ 9 deg)
and the ``Deep'' survey field D1 (1
deg $\times$ 1 deg)
cover the 
XMDS region at a depth of $i' = 24.5$
\citep{Hoekstra etal 06} and $i' = 26.1$ \citep[50\% completeness
limit,][]{Semboloni etal 06}, respectively. In the following we
will use the D1 notation for data from the CFHTLS Deep and the W1 notation for
data from the CFHTLS Wide. W1 observations have been recently completed; the coverage shown in
Fig~\ref{fov} is deduced from data available to us at the time of
analysis. 

Near-infrared observations of the XMDS area are also in progress in the
context of the UKIDSS
\citep{Dye etal 06, Lawrence etal 06}. The survey uses the Wide Field Camera
(WFCAM) of the 3.8 m United Kingdom Infrared Telescope (UKIRT). The XMM-LSS
region is one of the four target fields of the Deep Extragalactic Survey (DXS).
Observations in the $J$ and $K$ filters down to $J = 22.3$ and $K = 20.8$ (Vega
system) are in progress. About 0.8 deg$^2$ of sky have been observed up to now
and part of the data are available in the UKIDSS Early Data Release
\citep{Dye etal 06}. The photometric system used in the UKIDSS is described in
\citet{Hewett etal 06}.

\subsection{Mid-infrared data} \label{irobs}
The XMDS and XMM-LSS fields are covered by the  
SWIRE survey \citep{Lonsdale etal 03}.
Observations
were performed with the Infrared Array Camera (IRAC) at 3.6, 4.5, 5.8 and 8.0
$\mu$m and with the Multiband Imaging Photometer (MIPS) at 24, 70 and 160 $\mu$m
to a $5 \sigma$ depth of 4.3, 8.3, 58.5, 65.7 $\mu$Jy and 0.24, 15, and 90 mJy, respectively.
The area covered in the XMDS region is about 2.5 deg$^2$, 
corresponding to about 95\% of the field. 
12 sources in our selected sample are outside the region covered by SWIRE.
Details on the SWIRE data and source catalogs are given in 
\citet{Surace etal 05}.

\subsection{Radio data}
There are also radio observations associated with the XMDS: the VLA VIRMOS Survey,
which covers the VVDS area at a depth of 80 $\mu$Jy ($5 \sigma$ limit) and a resolution of
6\arcsec{} at 1.4 GHz \citep{Bondi etal 03, Ciliegi etal 05} and a low frequency
radio survey performed for the XMM-LSS also at VLA, which covers 5.6 deg$^2$ at a depth of 4 mJy
at 74 MHz and 110 deg$^2$ at a depth of 275 mJy at 325 MHz \citep{Cohen etal
03}.

\section{Optical and Infrared identifications} \label{id}
Most (80\%) of the 136 X-ray sources in the present sample were already included in the $4
\sigma$ catalogue presented in Paper I since they are also detected
at $\geq 4 \sigma$ in the softer bands. 24 X-ray sources 
are considered here for the first time. Although most sources had already been
assigned optical counterparts, we have repeated the identification procedure on
the whole sample in a semi automatic way, that takes into account the experience
accumulated in Paper I and the CFHTLS and SWIRE data now available.

Access to the whole VVDS and CFHTLS catalogues and
images is restricted: photometric data and positions are provided
only for optical sources within a fixed radius from the X-ray positions.
The same is true for the SWIRE data.
We then associated to each X-ray source all
combinations of optical and IR objects in the considered catalogues  within a
search radius of 6\arcsec.
Objects in the VVDS, CFHTLS and SWIRE catalogues were matched only 
a posteriori. 

We computed the probability of chance coincidence between an X-ray source
and all optical VVDS, optical CFHTLS and infrared SWIRE candidates within the search
radius using the following equation 
\citep[ see also Paper I]{Downes etal 86} 
\begin{displaymath}
  p = 1 - exp(-\pi~ n(<m)~  r^2 )
\end{displaymath}
where $r$ is the distance between the X-ray source and
the proposed counterpart (with a lower value fixed at 2\arcsec, which 
roughly corresponds to the \textit{XMM - Newton} astrometric
uncertainty), and $n(<m)$ is the density of objects brighter
than the magnitude $m$ of the candidate counterpart.
We used $I$
magnitudes for VVDS sources and $i'$ magnitudes for CFHTLS objects 
(D1 data where available).
We used the density $n(>F_{3.6})$ for infrared candidate counterparts. 
For each candidate counterpart there are therefore from 1 to 3 values of $p$, 
depending if the object is detected in the VVDS, CFHTLS and SWIRE. We classified 
the probabilities as ``good'' ($p < 0.01$),
``fair'' ($0.01 < p < 0.03$) or otherwise ``bad'' and took as identification 
the object with the best probability combination.
All tentative identifications were
then checked by visual inspection using the VVDS finding charts.
As reported in Paper I, astrometrical corrections were already applied to the
X-ray fields, so we find again that 
98\% of the counterparts are within 4\arcsec{} from the X-ray corrected
position, justifying  our conservative choice of 
6\arcsec{} radius. 

The probability criterion allows us to prefer one candidate counterpart in
the majority of cases, giving us 126 secure identifications (out of 136
sources). Of these, 3 have {\it only} IR counterparts (i.e. no optical counterparts are
detected down to $R_{AB} = 25.3$) and will be referred to as optically blank
fields. We notice that all sources covered by SWIRE
(i.e. all but 12) are also detected in the IR.

For the remaining 10 sources, the identification process is ambiguous
leaving  two or more possible counterparts, with similar
probabilities.  However, in 6 cases, the counterparts have similar
magnitudes and colors allowing us to include these sources
in parts of the 
discussion not involving the redshift determination or SED classification.
In the other 4 cases, the sources are completely dismissed.

The X-ray, optical and infrared properties of sources of the 3 $\sigma$ hard sample are
reported in Table~\ref{gentab}. For brevity, not all data used in this work are
reported in the Table. The SWIRE catalogue is available through IRSA/Gator
({\tt http://irsa.ipac.caltech.edu/applications/Gator}). We plan to publish the
Catalogue of all XMDS X-ray sources with optical and IR identifications in a future paper.

We also searched for UKIDSS counterparts of X-ray sources using a
radius of 4\arcsec, finding a near infrared counterpart for 72 X-ray sources.
Generally UKIDSS sources are coincident with optical counterparts. There are
however two exceptions: source 
XMDS 449\footnote{For brevity, in the text we label single sources with their
XMDS identification number. The names of the sources, complying the IAU
standard, along with the associated identifiers, are reported in
Table~\ref{gentab}.} 
for which the UKIDSS source lies
between the two possible counterparts, at a distance of about 3\arcsec{} from
both, and 
XMDS 760, 
for which there are two possible UKIDSS counterparts, both
within $\sim 1$\arcsec{} from the optical counterpart. The first case is one 
of the 4 X-ray sources that we could not identify, and the UKIDSS detection did 
not allow us to resolve the ambiguity. In the second case we
associated to the optical counterpart the brightest UKIDSS source.

33 X-ray sources in the $3 \sigma$ hard sample (24\%)
have a radio counterpart at 1.4 GHz, one of them is also detected at 325 MHz. One
is however associated with the spectroscopically confirmed cluster XLSSC 025.
We will not
use the radio information in this work, but we checked the consistency of the radio fluxes with the templates used to fit the optical
and infrared spectral energy distributions of our objects (see Section~\ref{photoz}).
The correlation between the X-ray and radio luminosities is explored in
\citet{Polletta etal 07}.
 
\section{The AGN sample} \label{xottir}

\begin{figure}
\centering
\includegraphics[width=8.5cm]{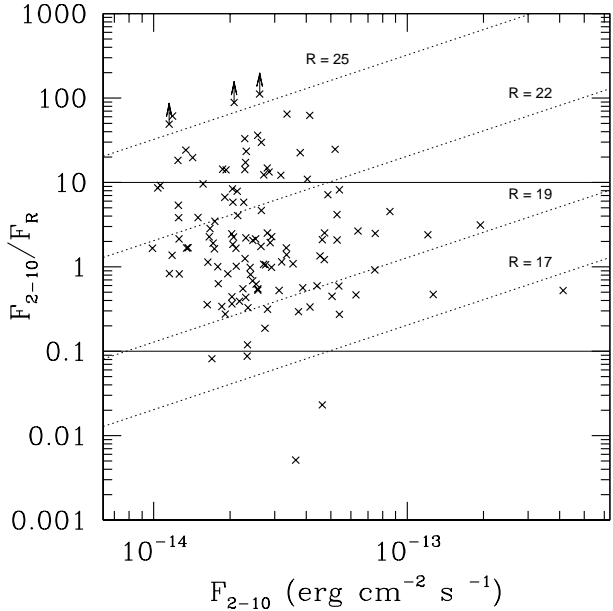}
\caption{X-ray to optical ratio ($2 -10$ keV band vs $R$ band) as a function of
X-ray flux for sources in the 3 $\sigma$ hard sample. Diagonal dotted lines indicate
loci of constant $R$ magnitude while horizontal solid lines mark the region of
canonical AGN ($0.1 < F_X/F_R < 10$). 
Lower limits mark the optically
blank fields, whose $R$ magnitude was fixed to 25.3 \citep[see Fig. 13
in][]{Mccracken etal 03}.}
\label{xott}
\end{figure}

Ignoring two  sources corresponding to spectroscopically
confirmed galaxy clusters \citep[XLSSC 025 and 041, see][]{Pierre etal 06},
we computed the X-ray to optical ratio for 124 sources with secure
identifications, using the equation given in section 6.2 of
Paper I. For about 20 objects for which VVDS
magnitudes were  unreliable because of saturation or unfavorable position in
the field of view, the CFHTLS $r'$ band magnitudes were used,
with the appropriate conversion factor taken from \citet{Silverman etal 05}.
The X-ray to optical ratio is shown as a function of X-ray flux in  Fig.~\ref{xott}. 

About 80\% of the sources fall in the typical range of X-ray to optical ratio
corresponding to the locus of AGN \citep[$0.1 < F_X/F_R < 10$, see
e.g.][]{Akiyama etal 00, Hornschemeier etal 01}, while about 20\% of the sources
have $F_X/F_R > 10$, which corresponds to heavy absorption in the optical
and/or high redshift \citep{Hornschemeier etal 01}. This issue will be further
developed in subsection~\ref{qso2}. 

Only two sources fall significantly below the AGN borderline 
(XMDS 1248 and 842): 
both appear extended in the optical as well as in the 
infrared images. The first 
(XMDS 1248) 
has a low hardness ratio, consistent with no 
intrinsic absorption in the X-ray spectrum, while the second 
(XMDS 842) 
has a higher hardness ratio, possibly indicating
X-ray absorption (N$_H \sim 10^{22}$ cm$^{-2}$). Using
photometric redshifts (see Section~\ref{photoz}), we obtained X-ray luminosities of $\sim
10^{41}$ erg s$^{-1}$ for both of them, even after correcting for absorption (see Table~\ref{gentab}). 
We classify both provisionally as normal 
galaxies, though we can not exclude the presence of a low luminosity AGN or 
even a Compton thick AGN in 
XMDS 842 
\citep[see e.g. FSC 1021+4724 in][]{Alexander etal 05}.
Another source in the sample has $L_{0.5-10} <10^{42}$ erg s$^{-1}$,
XMDS 178, however its X-ray to optical ratio
of 0.27 is in the typical AGN range. On the basis of the X-ray to optical
ratio, we retain this source in the AGN class.
In \citet{Polletta etal 07} slightly different criteria are adopted for these
borderline objects.

To summarize, on the basis of the X-ray to optical flux ratios 122 X-ray sources
with unambiguous identification can be classified as AGN 
and 2 as normal galaxies. 
In the following subsections we will discuss the optical and IR properties of 
this sample. 

\subsection{Optical magnitude and colors} \label{optcol}

\begin{figure}
\centering
\includegraphics[width=8.5cm]{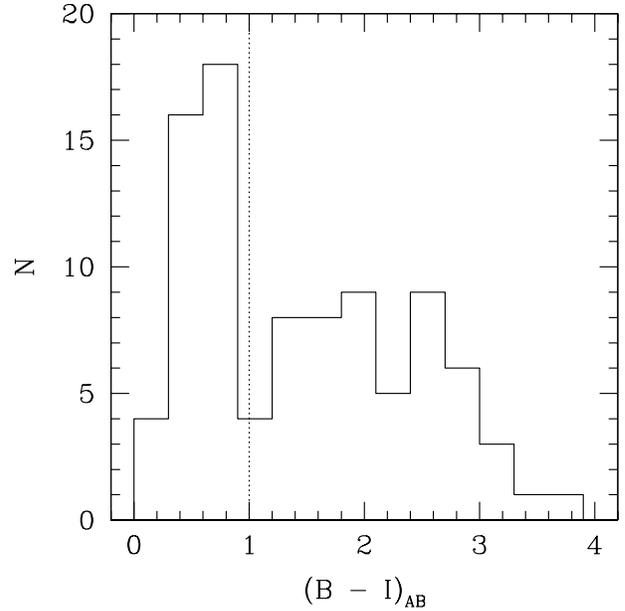}
\caption{$B - I$ distribution for optical counterparts of X-ray sources in the
3 $\sigma$ hard sample. The
dotted line marks the division between the adopted definition of ``blue'' and
``red'' objects (see text).}
\label{istobi}
\end{figure}

\begin{figure}
\centering
\includegraphics[width=8.5cm]{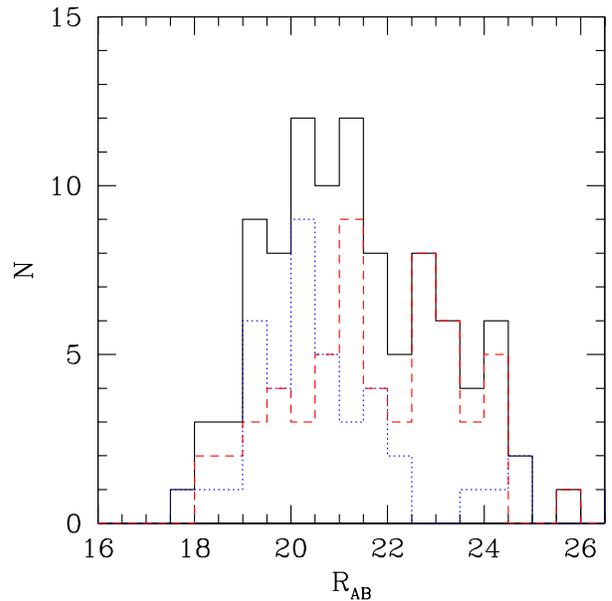}
\caption{$R$ magnitude distribution for the 3 $\sigma$ hard sample. Solid
histogram: total sample; dotted histogram: blue sources (see text);
dashed histogram: red sources (see text).}
\label{istomagr}
\end{figure}

The $B - I$ color and the $R_{AB}$ magnitude 
distributions for the identified sources are shown in
Fig.~ \ref{istobi} and \ref{istomagr}, respectively.

The $B - I$ color distribution shows a high peak at $B - I \leq
1.0$, and a tail extending
up to $B - I \sim 4$. Based on the observed 
color distribution we adopt a somewhat 
arbitrary threshold of $B - I =
1.0$ to divide the sample into two roughly equal size samples of
``blue'' objects, with $B - I < 1.0$ (43\% of all sources), and ``red''
objects, with $B - I > 1.0$ (57\%). As will be  shown later, this
criterion, although crude, proved to be a good one for a rough separation between type 1
(i.e. broad line) AGN and type 2 (narrow line) or star forming galaxy-like AGN based on observed quantities alone,
and is substantially confirmed by the more detailed (but model-dependent) classification based on
the spectral energy distributions. 

The magnitude distributions of these two broad classes are plotted separately in
Fig.~\ref{istomagr}: on average, blue sources ($B - I \leq 1.0$) are
brighter, with a peak at $R \sim 20$ and 90\% of objects at $R < 22$, while red
sources ($B - I > 1.0$) have a broader distribution, extending from $R \sim 18$ to $R \sim 26$.
 
\begin{figure}
\centering
\includegraphics[width=8.5cm]{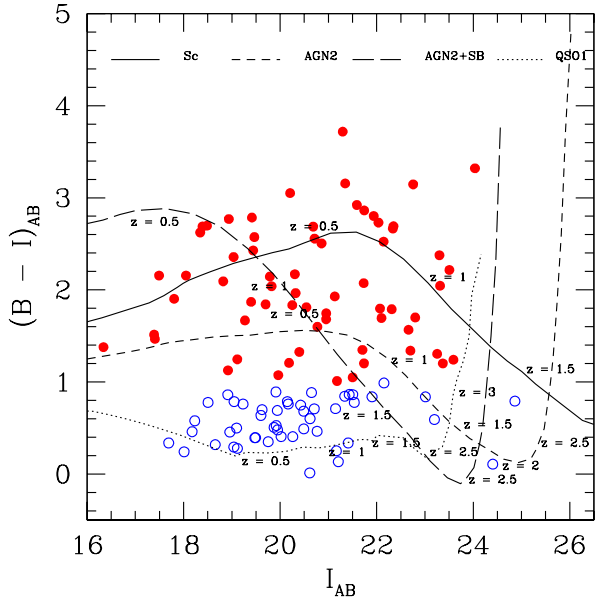}
\caption{$B - I$ color vs $I$ magnitude for sources in the 3 $\sigma$ hard sample.
Empty circles are blue objects, filled circles are red objects.
Overimposed to points are the evolutionary tracks for an Sc galaxy (solid line),
a type 1 QSO (dotted line), a
type 2 AGN (short-dashed line) and a type 2 AGN plus a starburst component
(long-dashed line). The B band absolute magnitudes assumed are $-15.7$ for the
Sc galaxy template, $-22.3$ for the QSO1, $-16.9$ for the type 2 AGN and $-23.9$ for
the type AGN plus starburst.}
\label{ibi}
\end{figure}

In Fig.~\ref{ibi} we show the $B - I$ color
as a function of the $I$ magnitude for our sources, along with the evolutionary
tracks for various templates: a late spiral galaxy (Sc, solid line), a type 1
QSO (dotted line), a
type 2 AGN (short-dashed line) and a type 2 AGN plus a starburst component
\citep[long-dashed line; see below and][]{Polletta etal 07}.
The effects of absorption due to the Intergalactic Medium (IGM) have
been taken into account at high redshift ($z \geq 2.5$) as prescribed in
\citet{Madau 95}. 
Blue sources are near the QSO1 track, while red objects are
generally consistent with star forming galaxies and AGN2 tracks. However for
magnitudes fainter than $I_{AB} = 23$ the different track cross, and type 1 AGN,
type 2 AGN and star forming galaxies have similar colors.

\subsection{X-ray to infrared ratios} \label{ratios}

\begin{figure*}
\centering
\includegraphics[width=6.5cm]{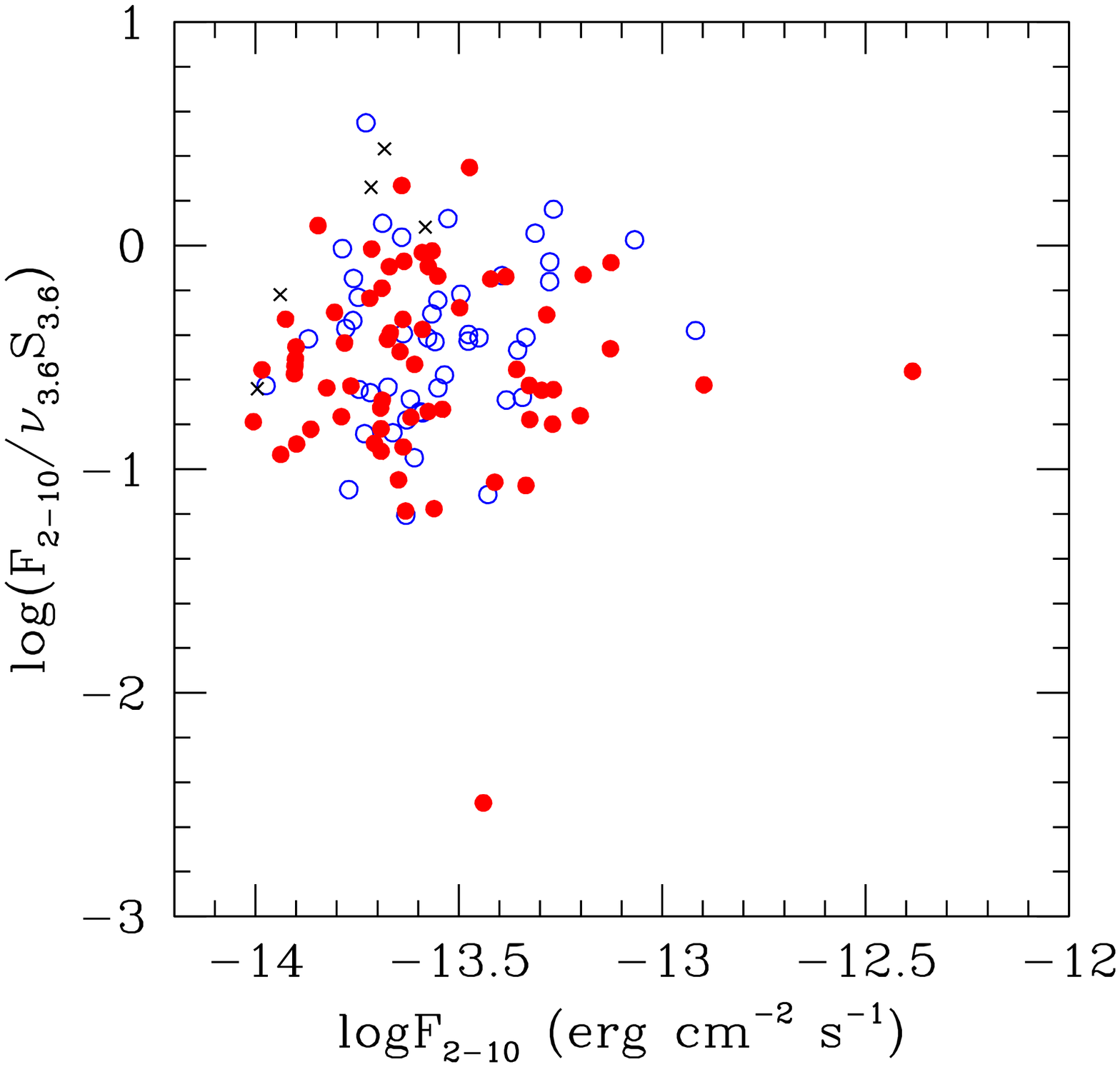}
\includegraphics[width=6.5cm]{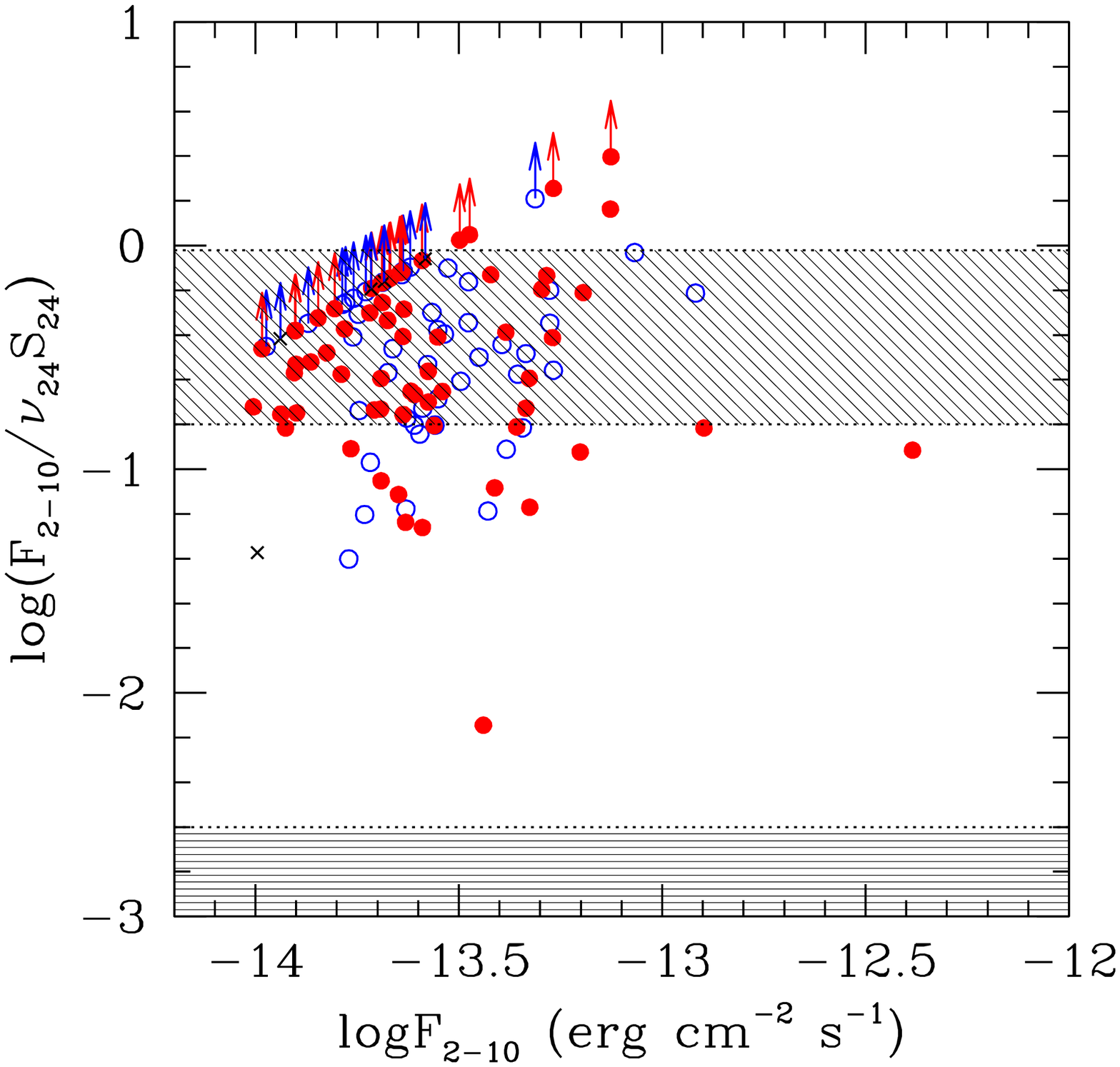}
\caption{X-ray to IR ratio as a function of
X-ray flux for sources in the $3 \sigma$ hard sample. Empty circles are blue sources,
filled circles are red sources, crosses are the optically blank fields or
sources with undefined color classification (because $B$ or $I$ magnitudes are
not available). In the left panel IRAC 3.6
$\mu$m flux is used, while in the right panel MIPS 24 $\mu$m is used. Lower
limits are sources detected in one or more IRAC bands and undetected in the MIPS
24 $\mu$m band, where 5 $\sigma$ upper limit is used. The diagonally shaded area
is the region occupied by hard X-ray selected AGN with IR emission and $z <
0.12$ from \citet{Piccinotti etal 82}; the horizontally shaded area is the region
occupied by local starburst galaxies from \citet{Ranalli etal 03}.}
\label{xir}
\end{figure*}

\begin{figure}
\centering
\includegraphics[width=8.5cm]{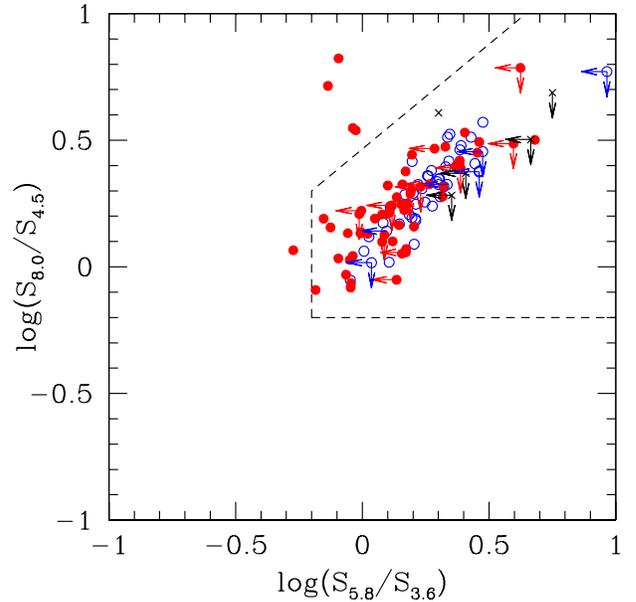}
\caption{IRAC color - color plot for sources in the 3 $\sigma$ hard sample. 
Empty circles are blue sources, filled circles are red sources, crosses are the
optically blank fields or
sources with undefined color classification. Left pointing arrows are
sources undetected in the 5.8 $\mu$m band, down pointing arrows are sources
undetected in the 8.0 $\mu$m band. 5 $\sigma$ upper limits are used. The dashed lines mark the 
region expected for AGN according to \citet{Lacy etal 04}}
\label{lacy}
\end{figure}
 
In Fig.~\ref{xir} we plot the ratios of  X-ray to infrared fluxes at
3.6$\mu$m (left panel) and 24$\mu$m respectively (right panel) as a function of
the X-ray flux.
Sources are all clustered in the same region with no clear separation 
between blue and red sources. 
 
Two typical loci of local sources are shown in the right panel of Fig.~\ref{xir}: 
the area at $-0.8 < \mbox{log}(F_X/F_{IR}) < 0$
is the region occupied by hard X-ray selected AGN \citep[from][]{Piccinotti etal
82} with IR emission 
and $z < 0.12$; the area close to log$(F_X/F_{IR}) = -3$ is the region
occupied by local starburst galaxies from \citet{Ranalli etal 03}, 
adapted from \citet{Alonso-Herrero etal 04}. No objects with X-ray to
infrared ratios typical of local starburst galaxies are found in our sample.
80\% of the objects have X-ray/infrared ratios 
 $-1 < \mbox{log}(F_X/\nu_{3.6}S_{3.6}) < 0$ (i.e. within a factor of
10), and 98\% of them have $-1.2 < \mbox{log}(F_X/\nu_{3.6}S_{3.6}) < 0.6$
(i.e. a factor of $\sim 70$). The most discrepant object
is one of the two normal galaxies with low
X-ray to optical ratio (see above). 
The X-ray to optical ratios for the same sources ranges
 from $\sim 0.1$ to $\sim 60$ (i.e. a factor of 600 excluding lower
limits, see Fig.~\ref{xott}).
This implies that the IR flux is a better diagnostics of the X-ray flux
compared to the optical, a behaviour likely due to the smaller extinction in the
IR and to the fact that nuclear light absorbed by dust is likely
re-radiated in the IR.

The observed range in the $F_X/\nu_{24}S_{24}$ plot is fully consistent with other X-ray and 24 $\mu$m samples, 
\citep[e.g.][]{Alonso-Herrero
etal 04, Franceschini etal 05, Polletta etal 06}, but broader than that of
local hard X-ray selected AGN of \citet{Piccinotti etal 82}.  
This broader dispersion is not surprising given the better sensitivity of X-ray
observations with respect to the \citet{Piccinotti etal 82} data.

A broad range in the X-ray to infrared ratio could
be caused by different amounts of absorption in different sources
that depresses the observed X-ray
flux, but not the infrared emission. Alternatively, it could be an intrinsic
dispersion in the AGN SEDs that is not sampled properly in local objects. If
this dispersion were due only to absorption in
the X-rays, it would imply a broad range of column densities, up to
$1.5 \times 10^{23}$ cm$^{-2}$, consistent with the distribution of measured 
column densities (see subsection~\ref{xcolors}). However, the similarity in the 
distribution of flux ratios of blue and red sources is not observed in the 
column density distribution, the majority of blue sources being unabsorbed and 
the majority of red sources being absorbed. These arguments suggest that
the variety of the intrinsic SED shapes that characterize the AGN
population is a more likely explanation and that
such a variety is also observed for optically blue AGN. In fact a recent study of X-ray and 24$\mu$m-selected
AGN by ~\citet{Rigby etal 05} shows that there is no correlation between the
ratio $F(X)/\nu_{24}F(24\mu\mbox{m})$ and the amount of absorption in the
X-rays, or their optical properties. \citet{Elvis etal 94} measure a
dispersion of a factor of 10 at 24$\mu$m for a large sample of
optically-selected quasars after normalizing their SEDs at 1 $\mu$m,
consistent with the observed dispersion in the X-ray/infrared flux ratios of
our sample. An analysis of the SEDs of the AGN in the sample is presented
in the next Section and in more detail in \citet{Polletta etal 07}.

IRAC infrared colors proved to be a useful diagnostics to identify AGN among IR
sources; in particular, \citet{Lacy etal 04} found that the 8.0/4.5 $\mu$m ratio vs
the 5.8/3.6 $\mu$m ratio plot is effective in isolating  AGN in IR
selected samples, which have red
colors (i.e. high values of the ratios) in
both axis. In Fig.~\ref{lacy} we reproduce the plot of \citet{Lacy etal 04} for
sources in our sample, and we find that the vast majority of them (both optically blue and red) lies in the region expected for
AGN. At the boundaries of the AGN region there could be contamination by
low redshift galaxies \citep{Lacy etal 04}; in fact, all
the objects near the
borders of the AGN region in Fig.~\ref{lacy} have a red optical color. The AGN 
with the reddest IR colors are predominantly blue in the optical, while optically 
red AGN show a broad range of IR colors.

\section{Photometric redshifts and SED classification} \label{photoz}
Taking advantage of the excellent multiwavelength coverage 
from the optical (VVDS, CFHTLS), to near- and mid-infrared  (UKIDSS and SWIRE)
we constructed broad band SEDs for all the 124 identified sources.
We then fitted the observed SEDs   
(taking into account also upper limits) with various templates in order to determine
photometric redshifts. We used 20 templates that represent normal galaxies
(11: 1 elliptical, 7 spirals and 3 starbursts), composite galaxy + AGN (3: starburst +
AGN) and AGN (6: 3 type 1 AGN, 3 type 2 AGN) and cover the wavelength
range from 1000 \AA\ to 500 $\mu$m. These were derived from the observed SEDs
of objects representing the different classes. The effects of dust extinction 
were taken into account by reddening the reference templates according to the 
extinction curve derived in high redshift starbursts by \citet{Calzetti etal 00}. 
In order to limit degeneracies in the best fit solutions we limited the 
extinction A$_{\mathrm{V}}$ to be less than 0.55 mag and included templates 
of highly extincted objects to fit more heavily obscured sources.
The HYPERZ code
\citep{Bolzonella etal 00} was used to fit the SEDs and find the best-fit solution. 
A full description of the templates and a detailed
discussion of the SED fitting procedure and photometric redshift estimates
are presented in \citet{Polletta etal 07}. 

\begin{figure}
\centering
\includegraphics[width=8.5cm]{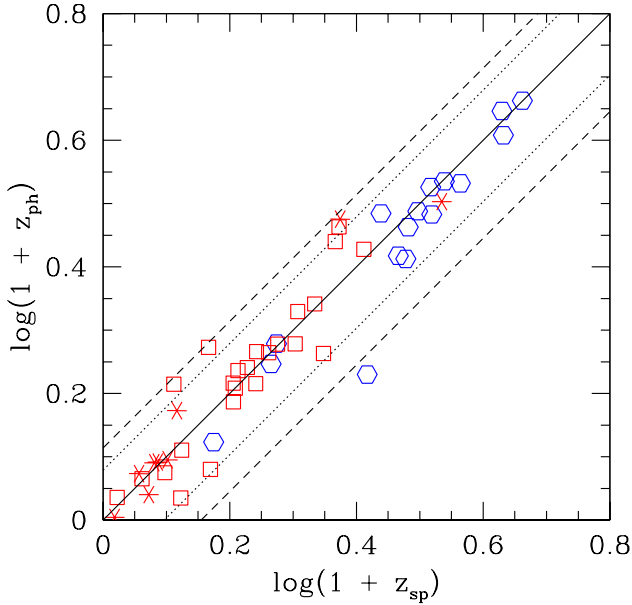}
\caption{Photometric vs spectroscopic redshifts. Solid line is the $z_{ph} = z_{sp}$ relationship, dotted
lines mark the 20\% ``error'' in $1 + z_{ph}$ and dashed lines mark the 30\%
``error'' in $1 + z_{ph}$. Hexagons are type 1 AGN, empty squares are type 2 AGN and
asterisks are SGFs, as characterized by their SEDs discussed in
subsection~\ref{photoclass}.}
\label{zpseczphot}
\end{figure}

A number of spectroscopic redshifts are available to assess the quality of 
our photometric redshift determination. 
For 22 objects redshifts were obtained in the context of
the XMM - LSS follow up programs and made available to us (Garcet et al., in
prep.). Redshifts for two sources were taken from \citet{Lacy etal 06}, who present optical spectroscopy 
of luminous AGN selected in the mid-IR  from \textit{Spitzer} observations.
For XMDS 842  
a redshift is available from
NED\footnote{\tt{http://nedwww.ipac.caltech.edu/}}. 

The VVDS spectroscopic sample (see Gavignaud et al. 2006 for type 1
AGN) yields redshifts for 8 sources in the present sample.             
To obtain a larger redshift comparison set, 
we added 16 additional sources from the larger X-ray sample
discussed in Paper I having a redshift from the VVDS spectroscopic
survey. For the latter similar photometric data are
available so that photometric redshifts could be 
estimated with the same procedure described above.
In total, the spectroscopic comparison sample consists of 49 sources.  
For 3 of them, falling outside the area covered by SWIRE, 
only optical data were available for the SED. 

Photometric and spectroscopic redshifts are compared in Fig.~\ref{zpseczphot}. 
The reliability and accuracy of the photometric redshifts are usually measured via
the fractional error $\Delta z = \left(\frac{z_{phot}-z_{spec}}{1+z_{spec}}\right)$
and the rate of
catastrophic outliers, defined as the fraction of sources with
$| \Delta z| >$ 0.2. 
For our 49 objects, the mean $\Delta z$ is consistent with 0.00, with a 1 $\sigma$
dispersion of 0.12, and the outlier fraction is
10\%. 
These results are significantly better than previously
obtained for AGN samples, where the fraction of outliers is usually
higher than 25\%
\citep{Kitsionas etal 05, Babbedge etal 04}. 
The achieved accuracy still does not allow us to consider photometric redshifts
as fully reliable for individual sources, however it is adequate for a 
statistical analysis of the population. For a more detailed discussion, see
\citet{Polletta etal 07}.

\begin{figure*}
\centering
\includegraphics[width=6.5cm]{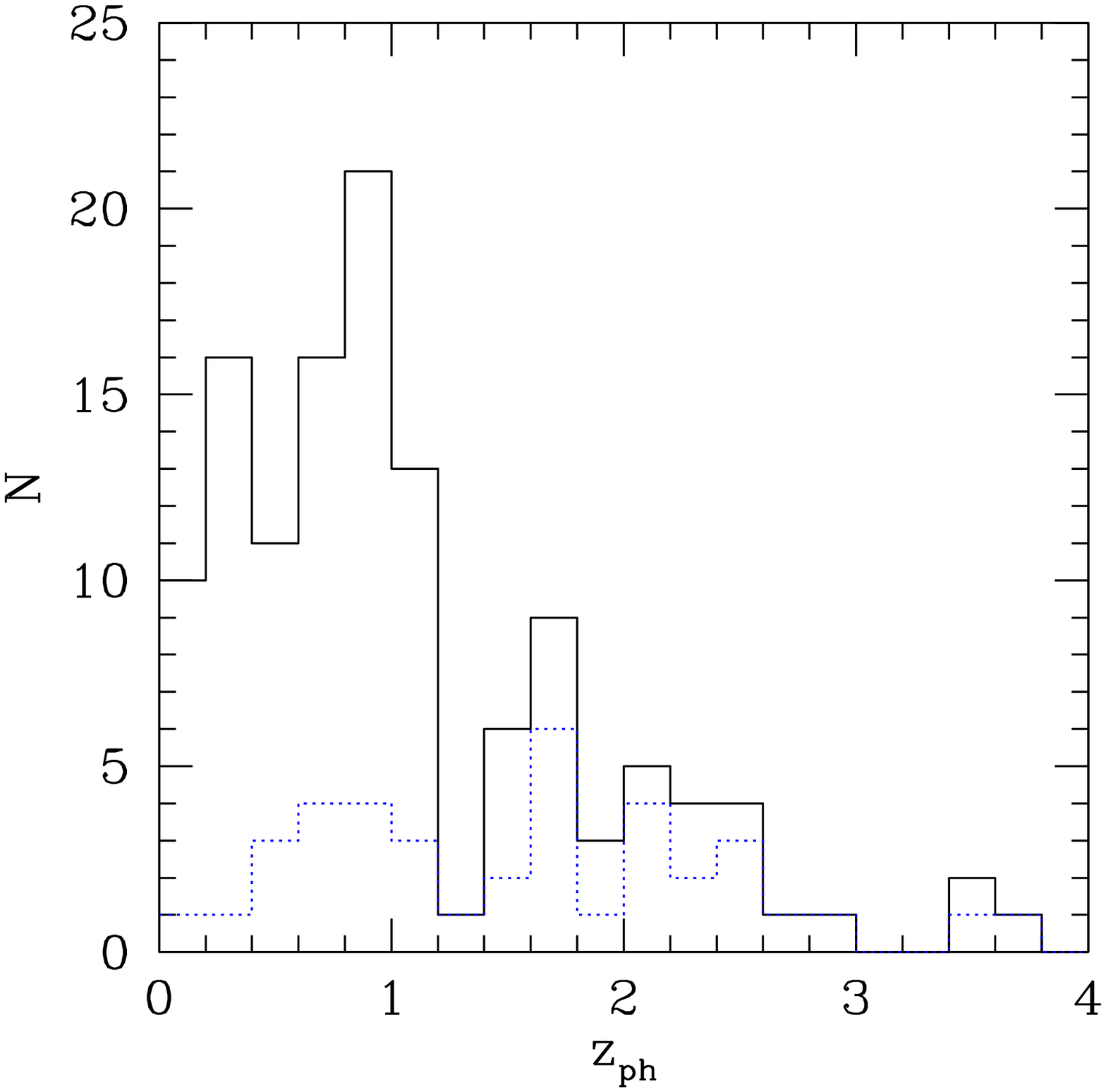}
\includegraphics[width=6.5cm]{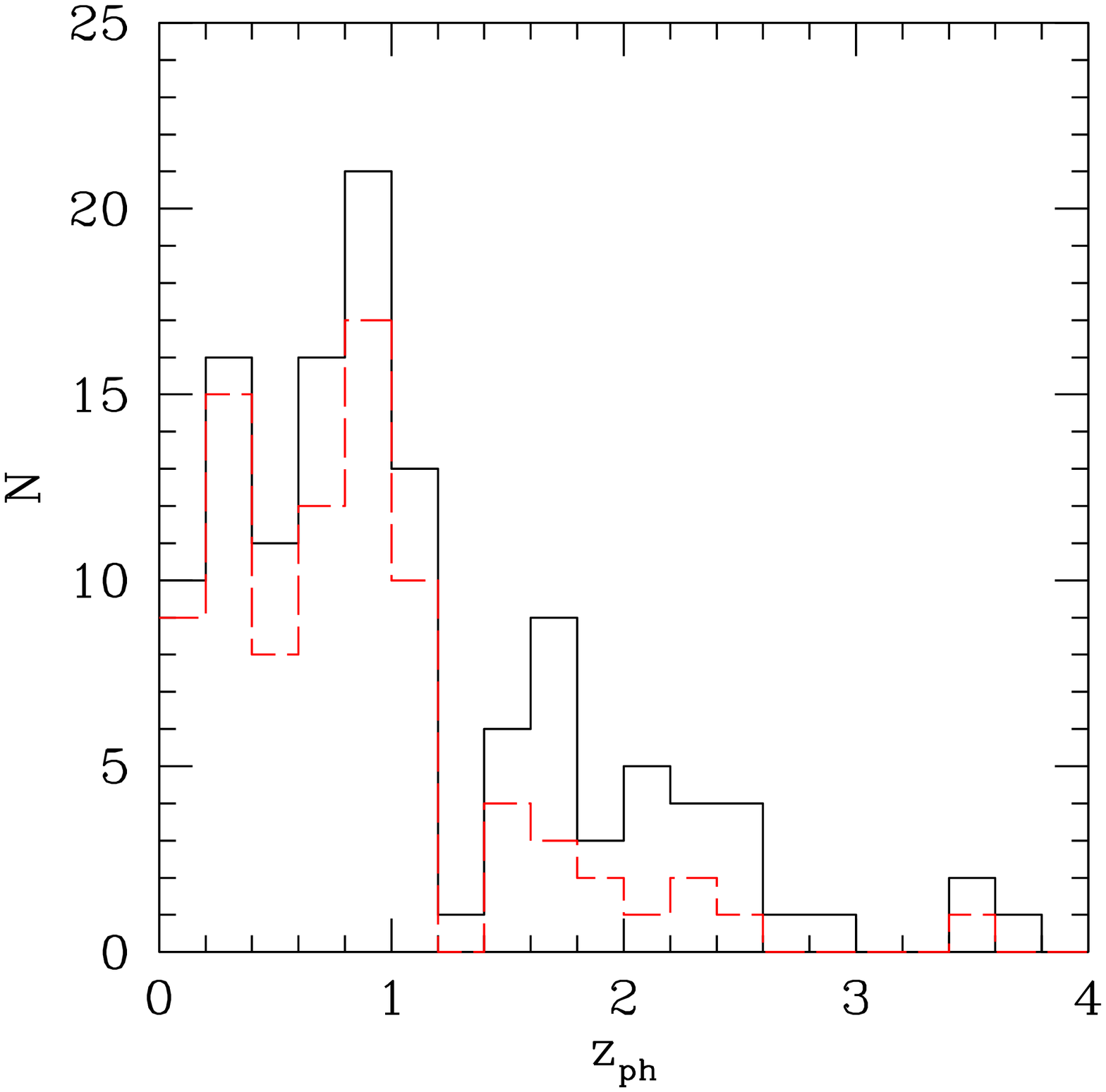}
\caption{The distribution of photometric redshifts is shown as solid histogram
in both panels. The dotted and dashed histograms (left and right panels,
respectively) refer to the subsamples of unobscured and obscured AGN,
respectively, discussed in subsection~\ref{photoclass}.}
\label{istozphot}
\end{figure*}

The distribution of the 124  photometric redshifts 
(including the optically blank fields, for which only IR fluxes were used)
is shown in Fig.~\ref{istozphot}.
The majority (60\%) of sources has $z < 1$, with a tail extending up to $z \sim 4$.
These results are consistent with the redshift distribution of other X-ray
selected samples with similar or deeper flux limit
\citep[e.g.][]{Barger etal 03, Hasinger 03, Barger etal 05}. 

\subsection{Spectral energy distributions and classification}
\label{photoclass}

According to the template which gives the best-fit solution, we assigned sources 
to one of the following broad classes: type 1 AGN, type 2 AGN, or
SFG. The type 1 AGN class corresponds to sources best-fitted with a QSO1
template. The type 2 AGN class includes sources best-fitted with either the
Seyfert 2 templates, or the composite AGN + starburst templates, or
the QSO2 template. The SFG class includes sources fitted by a spiral or a
starburst template. Elliptical templates never yielded best fit solutions. 

\begin{figure*}
\centering
\includegraphics[width=13cm]{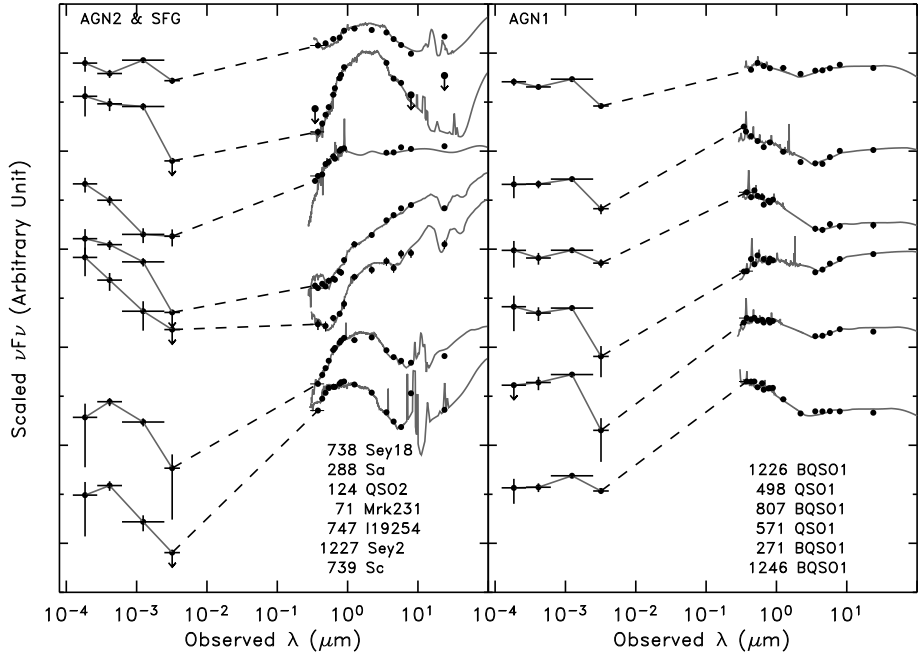}
\caption{Observed SED (filled circles) and redshifted best-fit templates (grey
solid curves) of 13 sources: 2 with star-forming like 
SEDs (left panel),
5 with type 2 AGN SEDs (left panel), and 6 with type 1 AGN SEDs (right
panel). Downward pointing arrows correspond to 5$\sigma$ upper limits.
The source sequence number and best-fit template names are listed in the
same order as the SEDs are plotted.}
\label{sedes}
\end{figure*}

Examples of observed SEDs with their best fit templates are presented in
Fig.~\ref{sedes}.
For sources with both optical, near and mid-IR data, the photometric
classification should be
reliable since the SED shape of the different classes has specific signatures
that can be easily identified. 
Interestingly, while photometric redshifts for
type 1 AGN might be the most uncertain, their classification is instead
rather easy. Note however that the Seyfert 1.8 template appears intermediate
between type 1 and type 2 AGN (see Fig.~\ref{sedes}). 
There is a large variety 
of SED shapes among the templates used for type 2 AGN, composite and star forming 
galaxies. 
In case the fit is not optimal or when only few IR data points are 
available, the
separation between the various classes is uncertain as can be guessed
comparing the SEDs in the left panel in Fig.~\ref{sedes}.

The SED fitting procedure yields 39 type 1 AGN (32\%),
61 type 2 AGN (49\%) and 24 SFG (19\%).

Comparing the SED classification with the spectroscopic one, we find that
all the 16 objects classified as type 1 AGN from the fitted template
indeed show broad emission lines in their optical spectra. Thus a photometric type 1 AGN
classification appears unambiguous.

On the other hand, there are 10 objects spectroscopically classified as type 1
AGN, which are instead not recognized as such by the SED fitting procedure,
indicating that our method systematically underestimates the fraction of 
broad line AGN. Specifically of the 10 misclassified objects 8 are fitted by a 
Seyfert 1.8 template (all with A$_{\mathrm{V}}$ close to 0), one by a QSO 2 template and 1 
by a SFG template.
These objects appear to be dominated by star-light emission in the 
optical and near-IR where the AGN continuum does not emerge clearly, although 
broad emission lines are visible in the optical spectrum.
Of the remaining 23 objects without broad lines in their optical spectra
only 5 are fitted 
with a Seyfert 1.8 template, in three cases with significant extinction. 
We conclude that SEDs fitted by Seyfert type 1.8 templates are intermediate
between type 1 and type 2 objects and that our method systematically
underestimates the objects spectroscopically classified as type 1.

The sources photometrically
classified as SFGs do not show any AGN signature at optical and IR wavelengths,
however the X-ray to optical and X-ray to IR ratios and the X-ray luminosity
unambiguously point to the
presence of AGN activity also in these objects. 

In the following we will define optically ``unobscured'' AGN all sources 
fitted by a type 1 AGN template.
These sources are expected to unambiguously
correspond to broad line AGN. We will define all other sources (i.e. having
either type 2 AGN or SFG like SEDs) as optically ``obscured'' AGN. As shown above, 
the latter group may include some AGN with broad emission lines, but with a SED
dominated by the host galaxy in the near-IR. We will take into account where relevant that 
the number of unobscured objects should be corrected upwards by a factor 1.6 
(and the number of obscured objects reduced accordingly).

\begin{figure}
\centering
\includegraphics[width=8.5cm]{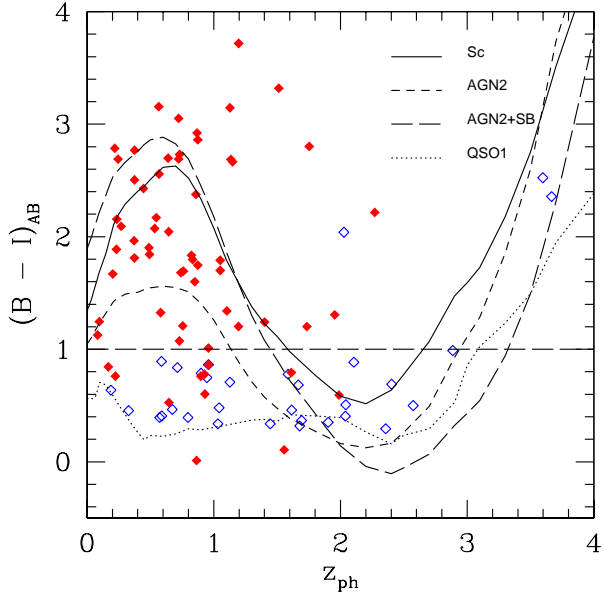}
\caption{$B - I$ distribution of sources in the 3 $\sigma$ sample as a
function of photometric redshift. Empty diamonds are unobscured AGN, filled
diamonds are obscured AGN.
Overimposed to points are the evolutionary tracks for an Sc galaxy (solid line), 
for a type 1 QSO (dotted line), for a type 2 AGN (short-dashed line) and a type
2 AGN plus a starburst component (long-dashed line).}
\label{bizphot}
\end{figure}

In Fig.~\ref{bizphot}, we compare the classification based on the SED shape
with the optical color $B - I$ as a function of redshift. The horizontal
dashed line corresponds to the threshold between blue and red sources ($B -
I = 1$). 90\% of the unobscured AGN have blue optical color and 84\% of
obscured AGN are red. Thus the simple 
classification based on observed color appears in retrospect rather
successful when compared with the more sophisticated template fitting
procedure. However, while at $z < 1.6$ the SED and ``color'' classifications
practically coincide (except for 2 obscured objects near the borderline), at
larger redshifts there is a degeneracy among the different evolutionary
tracks so that the optical color alone is not indicative of a spectral type.

Based on the photometric classification,
the redshift distributions of unobscured and obscured AGN can be derived. They are shown in the left and right panels of
Fig.~\ref{istozphot}, respectively. The two are clearly different,
the first being broader and reaching higher redshifts, while the second is
more concentrated at $z < 1$ (72\%).
Several authors \citep[e.g.][]{Eckart etal 06, Steffen etal 04, Treister etal 05, Lafranca etal 05} find different
redshift distributions for type 1 and non type 1 AGN. An analysis of the
fraction of unobscured and obscured AGN as a function of redshift will be discussed in
Section~\ref{fractions}.

\section{X-ray spectral properties} \label{xray}
We studied the X-ray spectral properties of our sample performing spectral fitting
for individual sources with a sufficient number of counts
(subsection~\ref{singlespec}) and a hardness ratio analysis for for faint
sources to obtain individual values of N$_H$ (subsection~\ref{xcolors}). A
stacking technique was then used to study systematic trends
in the whole sample (subsection~\ref{stack}). The general hardness
ratio definition is
\begin{equation} \label{hreq}
\mbox{HR} = \frac{CR_H - CR_S}{CR_H + CR_S}
\end{equation}
where $CR_H$ and $CR_S$ are the count rates in the hard and in the soft band,
respectively. In subsection~\ref{xcolors} we consider three hardness ratio values for each sources,
HR$_{CB}$, calculated using the energy bands $0.5 - 2$ (B) and $2 - 4.5$ keV
(C), HR$_{DC}$, using the energy bands $2 - 4.5$ and $4.5 - 10$ keV (D) and HR,
computed between the energy bands $0.5 - 2$ and $2 - 10$ keV.

\subsection{Individual sources} \label{singlespec}
We extracted X-ray spectra for all sources in the 3 $\sigma$ hard
sample having at least 50 net counts in the MOS + pn merged image in the $2 -
10$ keV band. There are 55 X-ray sources in our catalogue which
satisfy this criterion; 24 of them are optically unobscured AGN, and 31 are optically obscured AGN. 
  
Counts were extracted for each source using the
\textit{XMM - Newton}
Science Analysis System (SAS)
{\tt evselect} task in a circular region with
a radius of 20\arcsec, corresponding, for a point-like source, to an encircled energy fraction of
$\sim 70 - 75$\%  (off-axis angles between 0 and
10\arcmin). The pn data were used, unless the source was close to a CCD gap, in which
case we used the MOS data, fitting 
simultaneously MOS1 and MOS2. Background counts were
extracted from the nearest source free region, excluding areas near gaps in the
CCD array. We used the SAS
{\tt rmfgen} task to create response matrices (one for each camera and each XMDS pointing)
and {\tt arfgen} to generate ancillary response files (one for each source).

X-ray spectra were analyzed using the XSPEC package (v. 11.3.1). We considered the
energy range $0.3 - 10$ keV. When the number of
counts was large enough, data were binned in order to have
at least 15 or 20 counts for each energy channel and $\chi^2$ statistics was
used, otherwise we used Cash
statistics \citep{Cash 79}, which, however, does not give a ``goodness of fit'' evaluation, like the $\chi^2$.
In order to better match the spectral
resolution of the instruments, we binned the data of these sources with few counts
using a fixed number of PHA channels before fitting using the Cash statistics.  

We first fitted the spectra using a simple power law model with galactic
absorption computed at the XMDS position \citep[N$_H = 2.6 \times 10^{20}$
cm$^{-2}$,][]{Dickey&Lockman 90},
plus a component for intrinsic absorption at $z = 0$ \citep[XSPEC model: {\tt
phabs*zphabs*pow} with abundance table of][]{Wilms etal 00}. 

For all spectra for which $\chi^2$ statistics
can be used in the fit (22 sources), we set both intrinsic 
column density and photon index as free
parameters. Spectral fit results are reported in Table~\ref{z0tab} in Appendix
\ref{appspectra}. The errors in Tables and Figures correspond to the 90\%
confidence level for one interesting parameter. The average photon
index is $\Gamma \sim 2.1$ and N$_H < 21$ cm $^{-2}$.  This is
consistent with their location in the hardness ratio plot,
where they cluster 
around HR$_{CB} = -0.5$ and
HR$_{DC} = -0.5$ (see Table~\ref{z0tab} and subsection~\ref{xcolors}). 
Therefore they cannot be considered representative of the whole sample. 

Since more than half of the X-ray spectra in the sample do not have
a sufficient number of counts to perform a fit with both $\Gamma$
and N$_H$ free parameters, we fixed the photon index for all objects,
in order to obtain an estimate of the column density. We used two
different values of the photon index, $\Gamma = 2.0$ and $\Gamma =
1.7$, both appropriate for AGN \citep{Turner&Pounds 89, Nandra&Pounds
94}. Spectral fit results for the simple absorbed power law model for
each source with both $\Gamma = 2.0$ and $\Gamma = 1.7$ are reported in
Table~\ref{alltab} in Appendix~\ref{appspectra}, where we also list the
sources with peculiar fits. The best fit values of N$_H$ obtained with
$\Gamma = 2.0$ are higher than those obtained with photon index frozen
to 1.7, by about $\Delta \mbox{log(N}_H) = 0.2$.
The two column density estimates are consistent within
errors in 90\% of cases.

We will
consider in the following only the distribution obtained fixing $\Gamma =
2.0$, except for two sources 
(XMDS 124 and 779): in these two cases
we were able to find a stable solution only fixing the photon index to $\Gamma =
1.7$.   
The column density distributions of optically obscured and unobscured AGN turn out
to be different:
16\% of unobscured AGN (3 out of 19) have N$_H > 10^{21}$ cm$^{-2}$, while more than
55\% of obscured AGN (19 out of 34) have N$_H > 10^{21}$ cm$^{-2}$. We recall that these are
lower limits to the column density values, since we did not yet introduce the redshift dependence.

Finally, we introduced the photometric or spectroscopic 
redshift, when available, in order to compute the intrinsic column density. The photon index
was left free when the $\chi^2$ statistics could be used, otherwise it was
fixed to 2.0. 
Spectral fit results, along with redshifts, are reported in Table~\ref{alltab} in Appendix
\ref{appspectra}.
Again, the distributions for
optically obscured and unobscured AGN are different, with 63\% of obscured AGN
(19 out of 30) having
N$_H > 10^{21}$ cm$^{-2}$ and 36\% (11 out of 30) with N$_H > 10^{22}$ cm$^{-2}$.  For
comparison, only $\sim 20$\% of unobscured AGN (4 out of 21) have N$_H > 10^{21}$
cm$^{-2}$ and 10\% (2 out of 21) have N$_H > 10^{22}$ cm$^{-2}$.

\begin{figure}
\centering
\includegraphics[width=8.5cm]{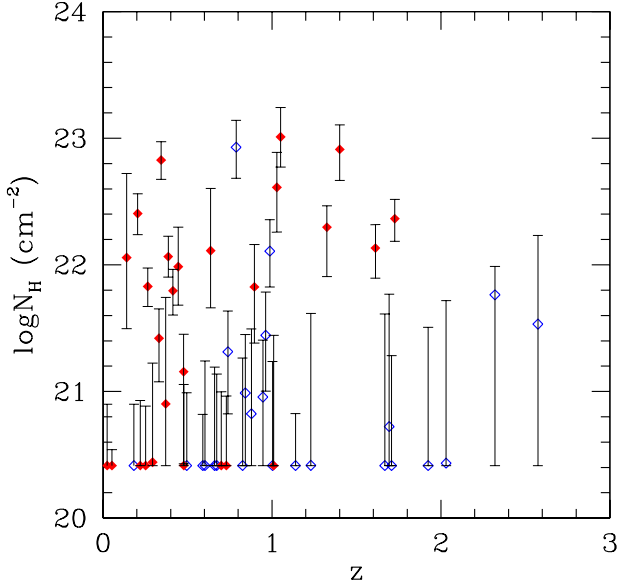}
\caption{Intrinsic column density vs photometric or spectroscopic redshift.
Empty diamonds are optically unobscured AGN, filled diamonds are
optically obscured AGN.}
\label{nhz}
\end{figure}

In Fig.~\ref{nhz} 
the intrinsic column density is shown
as a function of redshift; this figure is qualitatively consistent with those presented
in other surveys \citep[e.g.][]{Eckart etal 06}  and shows no obvious trend with
$z$, although we also notice the paucity of high redshift sources with well
constrained measures of N$_H$.

\subsection{X-ray absorption} \label{xcolors}
\begin{figure}
\centering
\includegraphics[width=8.5cm]{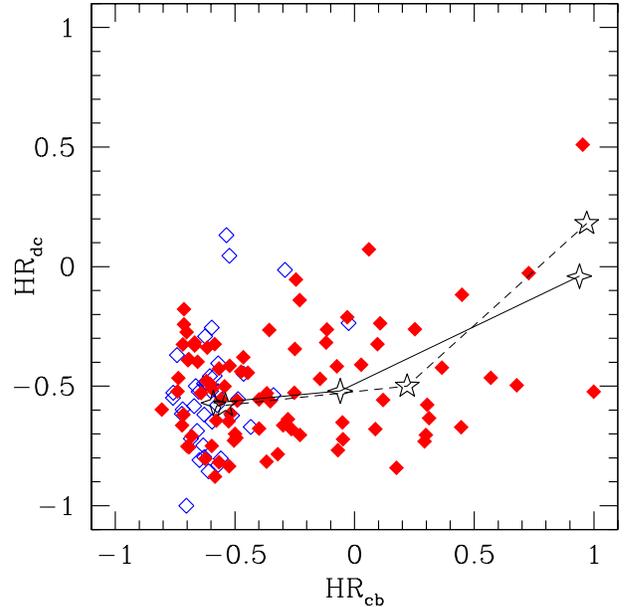}
\caption{X-ray color - color plot for sources in the 3 $\sigma$ hard sample.
Energy bands involved are: $0.5 - 2$ $(B)$, $2 - 4.5$ $(C)$ and $4.5 - 10$
keV $(D)$. 
Sources are
classified as optically unobscured AGN (empty diamonds) or obscured AGN 
(filled diamonds). We also mark the hardness ratios computed for a
simple absorbed power law spectral model, with $\Gamma = 2$ and logN$_H$ = 21,
22, 23 for $z = 0$ (four point stars connected by the solid line, from left to right) and logN$_H$ = 22, 23,
24 for $z = 1$ (five points stars connected by the dashed line, from left to right).}
\label{hrplot}
\end{figure}

The two hardness ratios HR$_{CB}$ and HR$_{DC}$ defined above are compared in
Fig.~\ref{hrplot}. 
As expected, most sources lie along the values expected for a single power law
model. We further distinguish obscured/unobscured sources with different symbols.
Less than
10\% of the objects classified as optically unobscured AGN have HR$_{CB} > -0.3$
(N$_H \sim 10^{21.5}$ at $z = 0$), 
while more than 40\%
of the sources classified as obscured AGN  have HR$_{CB} > -0.3$ indicating
that X-ray absorption and an obscured classification are often 
associated.

\begin{figure*}
\centering
\includegraphics[width=6.5cm]{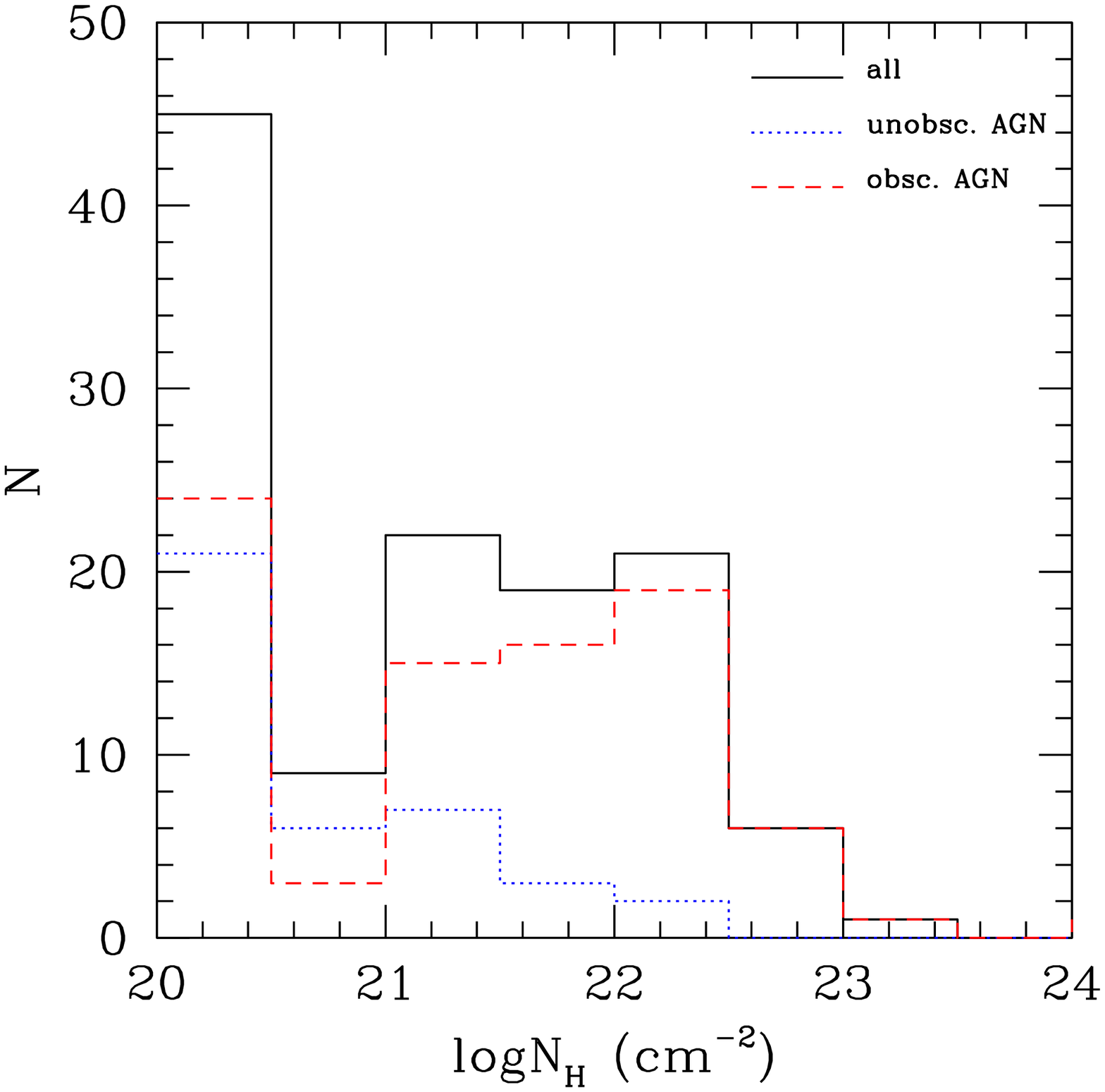}
\includegraphics[width=6.5cm]{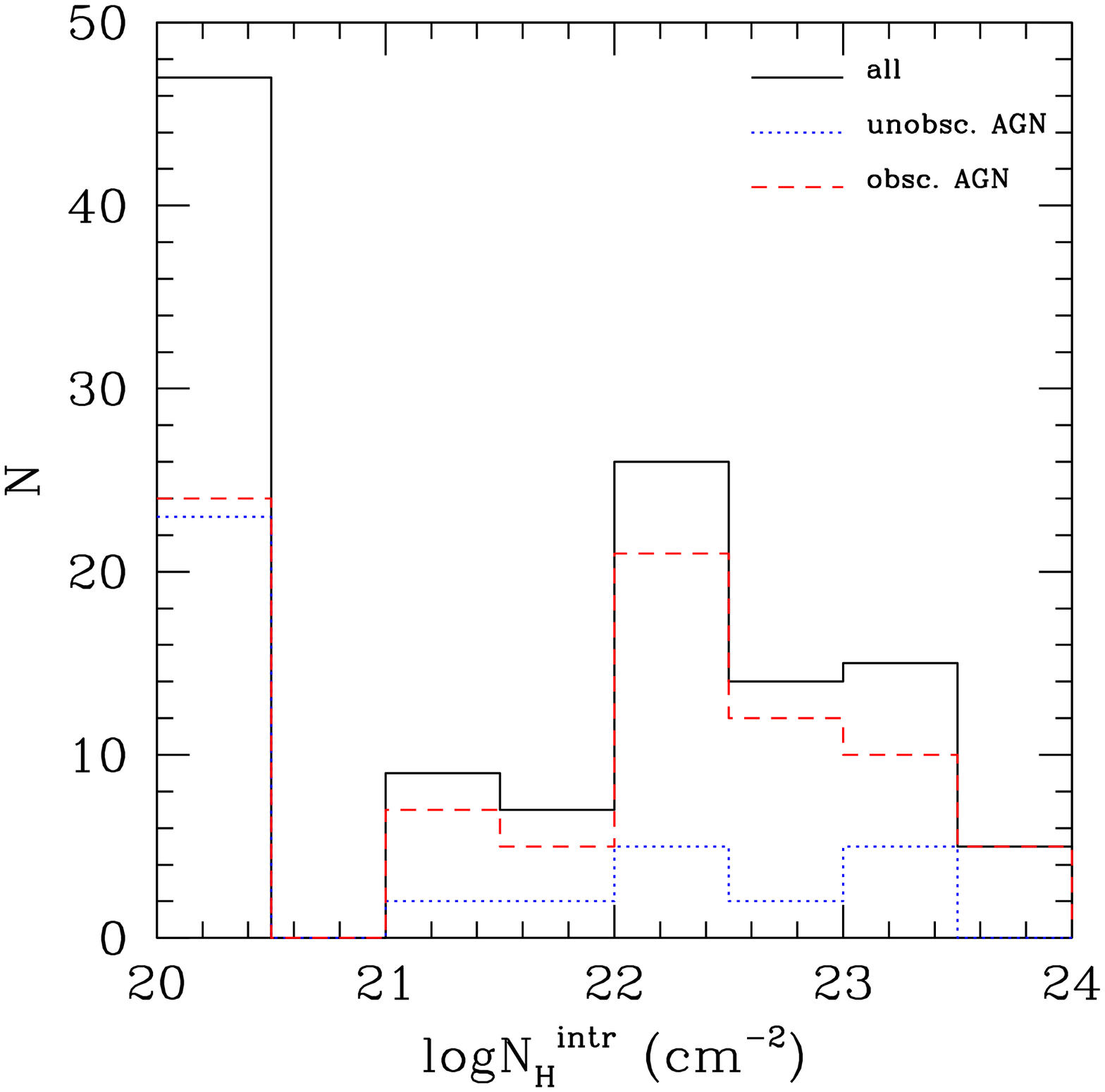}
\caption{Distribution of observed (left panel) and intrinsic (right panel) column
densities. Solid histogram refers to the whole sample, dotted and dashed
histograms refer to optically unobscured AGN and optically
obscured AGN, respectively.}
\label{istonhhr}
\end{figure*} 

For a quantitative estimate of the absorbing column N$_H$
we used the results of the spectral fits described in Section~\ref{singlespec} with $\Gamma$ fixed to 2.0 
for the 51 brightest sources and computed the column density from the X-ray hardness
ratios in the remaining cases in the following way.

We used the standard hardness ratio HR, computed between the $2 - 10$ and the $0.5 -
2$ keV bands. 
We made simulations using XSPEC to obtain hardness ratios corresponding to
typical values of N$_H$ ranging from $10^{20}$ to $10^{24}$ cm$^{-2}$. A simple power law model with photon index $\Gamma = 2.0$ was
assumed, consistently with the model used for the X-ray spectral analysis. 
The simulations and the objects for which a spectrum could be extracted define a clear relationship between N$_H$ and HR for
HR $> -0.5$, while, below these values, the  HR - N$_H$
relation degenerates. We therefore  
fixed the latter value, corresponding to N$_H \sim 10^{21}$ cm$^{-2}$, as a
threshold below which all column densities are fixed to the
galactic value. 
By interpolation we computed 
the observed N$_H$ corresponding
to the hardness ratio of each source. The intrinsic column density was then obtained
from the observed one using the photometric (or spectroscopic, when available) redshift and the expression
N$_H^{intr} = \mbox{N}_H^{obs} (1 + z)^{2.6}$ \citep{Barger etal 02}, also when the observed
column density was estimated from the spectrum (i.e. in this cases we did not use the
N$_H^{intr}$ obtained by the XSPEC and reported in Table~\ref{alltab}, but we recomputed
it from the observed value, in order to be more
consistent with the column density estimates obtained from HR).
N$_H^{intr}$ was not computed when the observed column density was fixed at the
galactic value.

The observed and intrinsic column density distributions are reported in
Fig.~\ref{istonhhr}. Different lines (dotted/dashed) refer to the N$_H$
distribution for optically unobscured and obscured AGN respectively. Unfortunately, our
choice of setting a fixed value for low N$_H$ creates an artificial gap in the
distribution. The majority of optically obscured AGN are X-ray absorbed
(N$_H^{intr} > 10^{22}$ cm$^{-2}$), as expected.
We find that also 12 unobscured AGN (more than 30\%) have N$_H^{intr} > 10^{22}$ cm$^{-2}$.
It is well known that N$_H$ values are less well
constrained with increasing redshifts \citep[see e.g.][]{Eckart etal 06, Akylas
etal 06, Tozzi etal 06}, since the absorption cut - off shifts to lower
energies and becomes comparable to the galactic values or even drops out of the
observed band. For $z \ga 1.5$, an intrinsic N$_H$
of $10^{22}$ cm$^{-2}$ corresponds to an observed column density $\la 10^{21}$
cm$^{-2}$. Unobscured AGN, whose hardness ratios generally cluster around
$-0.5$ (which corresponds to N$_H^{obs} \sim  10^{21}$ cm$^{-2}$), are more
severely affected by this problem
than obscured AGN, which have a broader HR distribution. We have partially compensated for this effect with
our choice of fixing intrinsic columns to 0 when the observed hardness
ratio HR $\sim -0.5$.  Moreover, we have at least 4 examples of optically unobscured
AGN (XMDS 12, 258, 280 and 406) 
in which the $observed$  N$_H$ is already larger than N$_H = 5 \times 
10^{21}$ cm$^{-2}$, ensuring that the large columns are not all to be
attributed to the redshift effects.  

\begin{figure}
\centering
\includegraphics[width=8.5cm]{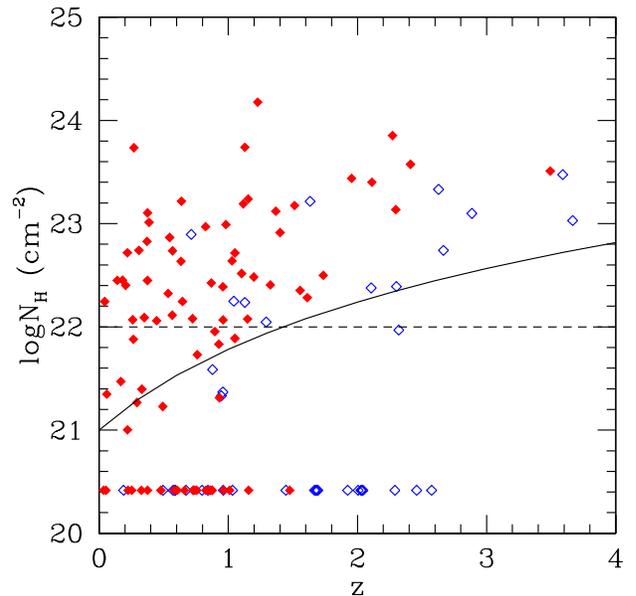}
\caption{Intrinsic column density vs photometric (or spectroscopic, when
available) redshift. Empty diamonds are optically unobscured AGN and filled diamonds are optically
obscured AGN. The dashed line marks the threshold between X-ray absorbed (N$_H
> 10^{22}$ cm$^{-2}$) and
unabsorbed sources. The solid line shows the intrinsic
column density that would be derived at a given redshift, for a measured column density of $10^{21}$
cm$^{-2}$ in the observer frame.}
\label{nhzhr}
\end{figure}

On the other hand, among the objects for which we are not able to estimate the
column density (those with N$_H$ fixed at the galactic value), there could be
some which could be really X-ray absorbed. In
Fig.~\ref{nhzhr} we show the intrinsic column density of our objects as a
function of redshift. The solid line shows the intrinsic N$_H$
values that would be derived at a given redshift, for an observed column density of $10^{21}$
cm$^{-2}$.
Since the objects with HR $< -0.5$ should have N$_H^{obs}
< 10^{21}$ cm$^{-2}$, their intrinsic column density should lie below the solid
line in Figure. It is therefore possible that we underestimate the number of
X-ray absorbed objects for redshift $z \ga 1.5$ (where the solid and dashed
line cross). The column densities are therefore difficult to estimate at high
redshift, but this should not affect our results.

\subsection{Stacking analysis} \label{stack}
Given that less than half of the sources in our sample have a sufficient number of
counts to perform a spectral analysis, we used a stacking technique to measure
the spectral properties, averaged over the whole redshift range, of sources of different classification and in different
flux intervals.

For the stacking analysis we used only pn data, because of the pn larger
effective area; we selected only sources which are not in or near a pn CCD
gap or bad column.  Moreover, since the pn point spread function (PSF)
and the vignetting are not well determined for large off-axis angles
\citep{Ghizzardi 02, Kirsh 06} we only used sources with off-axis angle
$\theta < 11'$. This value allows us to obtain a significant number of
sources (83), for which calibration should be still reliable. 30 sources
are optically unobscured AGN and 53 are optically obscured AGN.

We restricted the source area to a fixed radius of 15\arcsec, which, on-axis,
corresponds to 67\% of the encircled energy for a point-like
source. Since with this radius we sample the PSF core and not the wings,
the encircled energy fraction has a weak dependence on the off-axis angle (at
$\theta = 10'$ 67\% of the encircled energy is within a 16\arcsec{} radius) and
the energy dependence can also be neglected. Therefore, we can consider
all sources together regardless of their position in the field. 
A background spectrum was
extracted for each source in a circle of radius 80\arcsec{} in the nearest source free region. Ancillary response
files were also produced for each source, while, as for single spectra analysis,
we used one response matrix for each \textit{XMM - Newton} pointing.

The sample used in the stacking analysis covers a flux range from about $10^{-14}$ to
$1.2 \times 10^{-13}$ erg cm$^{-2}$ s$^{-1}$;
we divided it in 5 flux bins, chosen to have a sufficient number of counts in
each bin (see Table~\ref{stackall}; on average, brighter sources have a greater number of counts, so in
the higher flux bins a smaller number of sources is included). 

The spectra within the same flux bin were added using the {\tt mathpha} task of {\tt ftools} to produce a
single spectral file. The same was done for background files.
The auxiliary and response files were combined using the {\tt addarf} and
{\tt addrmf} tasks of {\tt ftools}, respectively. The combined spectra were grouped
to a minimum of 20 counts per bin and were analyzed using XSPEC.

\begin{table}
\begin{center}
\begin{tabular}{lrc}
\hline \hline
Flux bins	& N. of sources	& $\Gamma$ \\
\hline
$-14.1 < \mbox{log}F_X < -13.7$	& 25 & $1.72^{+0.09}_{-0.09}$\\
$-13.7 < \mbox{log}F_X < -13.6$ & 22 & $1.81^{+0.09}_{-0.08}$ \\
$-13.6 < \mbox{log}F_X < -13.5$ & 15 & $1.83^{+0.10}_{-0.09}$ \\
$-13.5 < \mbox{log}F_X < -13.3$ & 14 & $1.76^{+0.08}_{-0.08}$ \\
$-13.3 < \mbox{log}F_X < -12.9$ & 7 & $1.75^{+0.12}_{-0.12}$ \\
\hline
\end{tabular}
\caption{Mean spectral properties from the stacking analysis of sources in the $3 \sigma$
hard sample detected in the pn at off-axis angle $< 11$\arcmin. N$_H$ is fixed to the galactic value.}
\label{stackall}
\end{center}
\end{table}

\begin{table*}
\begin{center}
\begin{tabular}{lrclrc}
\hline \hline
\multicolumn{3}{c}{Optically unobscured AGN} & \multicolumn{3}{c}{Optically obscured AGN} \\
Flux bins	& N. of sources	& $\Gamma$ & Flux bins	& N. of sources	& $\Gamma$ \\
\hline
$-14.1 < \mbox{log}F_X < -13.6$ & 12 & $1.91^{+0.12}_{-0.12}$ & $-14.1 < \mbox{log}F_X < -13.7$ & 16 & $1.60^{+0.10}_{-0.09}$ \\
$-13.6 < \mbox{log}F_X < -13.5$ & 8 & $2.04^{+0.09}_{-0.09}$ & $-13.7 < \mbox{log}F_X < -13.6$ & 19 & $1.75^{+0.07}_{-0.07}$ \\
$-13.5 < \mbox{log}F_X < -13.3$ & 7 & $1.85^{+0.11}_{-0.10}$ & $-13.6 < \mbox{log}F_X < -13.4$ & 12 & $1.57^{+0.09}_{-0.08}$ \\
$-13.3 < \mbox{log}F_X < -12.9$ & 3 & $2.13^{+0.21}_{-0.20}$ & $-13.4 < \mbox{log}F_X < -12.9$ & 6 & $1.49^{+0.10}_{-0.09}$ \\
				&	&			& \multicolumn{3}{c}{SFGs} \\
\hline
				&	&			& $-14.1 < \mbox{log}F_X < -13.6$ & 7 & $1.33^{+0.17}_{-0.16}$ \\
				&	&			& $-13.6 < \mbox{log}F_X < -12.9$ & 6 & $1.21^{+0.11}_{-0.10}$ \\
\hline
\end{tabular}
\caption{Mean spectral properties from the stacking analysis of optically unobscured,
optically obscured AGN and of the subclass of SFG objects. N$_H$ is fixed to the galactic value.}
\label{stacktype}
\end{center}
\end{table*}

\begin{figure}
\centering
\includegraphics[width=8.5cm]{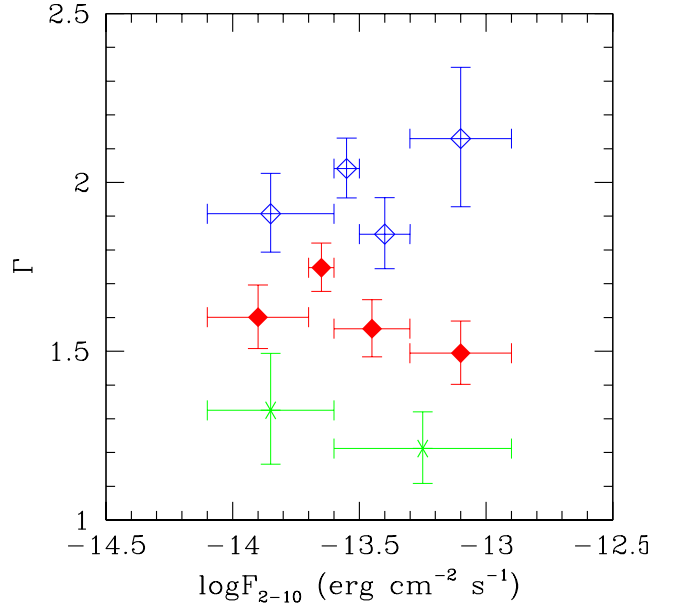}
\caption{Photon index obtained from the fit of stacked spectra as a function of
X-ray flux 
for the optically unobscured AGN (empty diamonds), optically obscured AGN
(filled diamonds) and SFGs (crosses). Vertical bars are errors on $\Gamma$, while horizontal
bars show the flux bin width.}
\label{gammastack}
\end{figure} 

We fitted the stacked spectra in the $0.3 - 10$ keV range using a single power law model with column density
fixed to the galactic value. The fit results are reported in
Table~\ref{stackall}. 
The $\Gamma$ values obtained for the whole sample  are consistent within errors 
with $\Gamma = 1.7 - 1.8$ in all the flux bins.

We then considered the optically unobscured and obscured AGN separately. 
We divided them in 4 bins, using two slightly different binnings for the
two subsamples dictated by the available statistics. The results are reported in
Table~\ref{stacktype} and the photon index as a
function of flux is shown in Fig.~\ref{gammastack}. The difference between the
optically unobscured and obscured AGN
populations is evident: for the unobscured AGN the measured photon index is consistent
with $\Gamma = 2$ over the whole flux range, while for the optically obscured it is
consistent with $\Gamma = 1.5 - 1.6$. Therefore the observed average spectral slope of unobscured AGN is consistent
with that of typical broad line AGN \citep{Turner&Pounds 89, Nandra&Pounds 94}, while that of optically obscured AGN is 
significantly harder. No significant dependence of the spectral index with flux is found for
optically unobscured or optically obscured AGN.

\citet{Georgakakis etal 06} merged the X-ray spectra of hard X-ray sources detected by
\textit{XMM - Newton} at $F_{2 - 8} > 2 \times 10^{-14}$ erg cm$^{-2}$ s$^{-1}$ and having an
optical counterpart in the Sloan Digital Sky Survey \citep{York etal 00} with red color ($g -
r > 0.4$). They found that the stacked spectrum of these sources has a spectral
slope of $\Gamma = 1.47 \pm 0.04$ (sources observed with the THIN filter),
consistent with that of the XRB \citep[$\Gamma \sim 1.4$,][]{Gendreau etal 95, Chen etal 97, Vecchi etal 99}.
As shown in the previous Sections, optically obscured AGN generally have red optical color
and the average spectral slope obtained for obscured AGN is only slightly higher than that of red
objects of \citet{Georgakakis etal 06}, showing that we are sampling similar populations.

According to \citet{Worsley etal 05}, whilst the XRB is $\sim 85$\% and $\sim 80$\%
resolved in the $0.5 - 2$ and $2 - 10$ keV bands respectively, it is only $\sim 60$\%
resolved above $\sim 6$ keV and $\sim 50$\% resolved above $\sim 8$ keV. The missing
population should be made of faint, heavily obscured AGN located at redshift of $\sim 0.5
- 1.5$, and with intrinsic absorption of $\sim 10^{23} - 10^{24}$ cm$^{-2}$. As noted in
Section~\ref{photoz}, the
sources classified as SFG do not show any AGN signature in the optical and IR.
We also find that the fraction of
X-ray absorbed sources in the SFG class ($\sim 67$\%, 16 out of 24) is larger than that of X-ray absorbed
sources in the type 2 AGN class ($\sim 54$\%, 33 out of 61).
Thus sources belonging to this class appear to be good candidates to be responsible for the XRB in the
harder X-ray range. We therefore applied
the stacking analysis to study separately the spectral properties of 
the SFG population in our sample.

Only 13 SFGs are detected in the pn at off-axis angle $< 11$\arcmin, so we grouped
them in two flux bins. The spectral fits of the SFG stacked spectra give $\Gamma \sim
1.2 - 1.3$, with no significant differences in the two bins (Table~\ref{stacktype} and
crosses in Fig.~\ref{gammastack}).
Therefore the average spectra of the SFGs are harder than those
of optically obscured AGN (type 2 + SFGs) and even harder than the XRB spectrum
in the same band. If the population responsible for
the high energy XRB has the same SED properties as the SFG objects discussed in this work,
they might go unidentified as AGN even in the IR, where they look like star forming
galaxies. A more detailed discussion about this topic is presented in
\citet{Polletta etal 07}.

\section{The surface density of optically obscured and unobscured AGN}
\label{lognlogs}

\begin{figure}
\centering
\includegraphics[width=8.5cm]{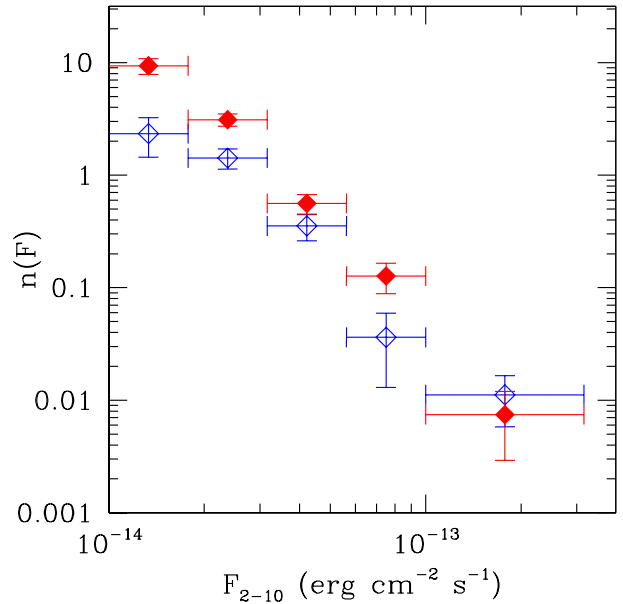}
\caption{Differential logN-logS for unobscured (empty diamonds) and obscured AGN
(filled diamonds). The units of $n(F)$ are number per 10$^{-15}$ erg cm$^{-2}$
s$^{-1}$ deg$^{-2}$. Vertical bars show the error on the number of sources,
while the horizontal bars show the flux bin width.}
\label{lognlogsfig}
\end{figure}

In Paper I we computed the logN-logS  distribution for all the sources detected with a probability of false detection $P < 2
\times 10^{-5}$ in the $0.5 - 2$ and $2 - 10$ keV bands; for the $2 - 10$ keV
band, this probability threshold is slightly lower than the 3 $\sigma$ threshold
chosen for the present sample, therefore all the sources of the 3
$\sigma$ hard sample were included in the logN-logS.
We used the differential
logN-logS for the $2 - 10$ keV band reported in Paper I, computed
the fraction of optically unobscured and obscured AGN in each flux bin from the
present sample and rescaled
the logN-logS relationship accordingly. We made the reasonable assumption that the
fraction of optically obscured and unobscured AGN should be the same in the area covered by the VVDS as well as in the whole XMDS area. 

The differential logN-logS relationships for optically obscured and unobscured
AGN are shown in Fig.~\ref{lognlogsfig}. The errors are the
combination of the errors on the original logN-logS with those on the fractions,
according to the error propagation formula. 
The two logN-logS are quite similar,
except for the faintest fluxes, where the density of the optically unobscured AGN is
significantly lower (by a factor of $\sim 4.6$) than that of optically
obscured AGN. Considering the cumulative
logN-logS instead of the differential one, we can give an estimate of the
integral surface density of optically obscured and unobscured AGN at $F >
10^{-14}$ erg cm$^{-2}$ s$^{-1}$. We find 138
and 59 sources deg$^{-2}$, respectively, and the ratio between optically
obscured and unobscured AGN is $\sim 2.3$ for the whole flux range
covered. The ratio would decrease from 2.3 to 1.1 if we assume that the
fraction of unobscured AGN should be corrected by a factor of 1.6 (see
subsection~\ref{photoclass}).

We compared these values with the surface densities of broad line and non broad line
AGN, estimated by \citet{Bauer etal 04} in
their study of the \textit{Chandra} Deep Fields. We obtained values from their
Fig.~4 and 8. At  $F_{2-10} > 10^{-14}$ erg cm$^{-2}$ s$^{-1}$, the surface densities
differ by a factor of $\sim 2$.
This discrepancy is a consequence
of the fact that the logN-logS computed in Paper I is lower than that of
\citet{Bauer etal 04}. As discussed in Paper I, the XMDS logN-logS
is slightly lower than those of \citet{Baldi etal 02} and \citet{Moretti etal
03} for $F_{2-10} > 2 \times 10^{-14}$ erg cm$^{-2}$ s$^{-1}$, but consistent within the
errors. Instead, the \citet{Bauer etal
04} surface density for fluxes $F_{2-8} \sim 10^{-14}$ erg cm$^{-2}$ s$^{-1}$ is
slightly higher than that obtained by \citet{Moretti etal 03}. Since all these
surveys refer to small connected areas in different parts of the sky, it is
possible that the differences in the derived number counts may be due to cosmic
variance. 

On the other hand, the \citet{Bauer etal 04} ratio of non broad line and broad line AGN at $F
\sim 10^{-14}$ is $\sim 1.5 - 2$, intermediate
between our values of 2.3 and 1.1. These authors describe some caveats
about their AGN classification criteria, and point out that their number counts
of broad line AGN must be considered a lower limit and that of non broad line AGN
an upper limit, so that their non broad line/broad line AGN ratio could approach
our lower estimate. 

\subsection{Type 2 QSO candidates} \label{qso2}

\begin{figure}
\centering
\includegraphics[width=8.5cm]{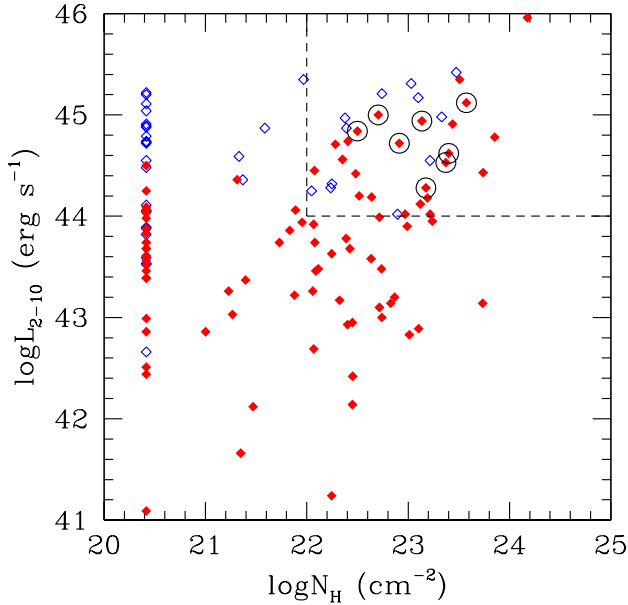}
\caption{X-ray ($2 - 10$ keV) absorption corrected luminosity vs intrinsic column
density. Empty diamonds are optically unobscured AGN and filled diamonds are optically
obscured AGN. 
Encircled points are objects with $F_X/F_R > 40$. 
The dashed lines
mark the region where type 2 QSOs should be found.}
\label{lxnh}
\end{figure}

We searched for type 2 QSO candidates in the 3 $\sigma$ hard sample. In the X-ray domain the type 2 QSO population is characterized by high intrinsic
absorption (N$_H > 10^{22}$ cm$^{-2}$) and high X-ray luminosity ($L_X
> 10^{44}$ erg s$^{-1}$). Since locally (in the Seyfert luminosity regime) X-ray
absorbed AGN are 4 times more numerous than unabsorbed ones \citep{Maiolino&Rieke 95, Risaliti etal 99}, 
according to the unified AGN model, one would expect that the same should be
true at high luminosities and redshifts, i.e. in the QSO regime. A still 
undiscovered large population of obscured AGN is indeed predicted by X-ray 
background synthesis models \citep[e.g.][]{Gilli etal 01, Franceschini etal 02,
Gandhi&Fabian 03, Ueda etal 03, Worsley etal 05}. 
Before the advent of \textit{Chandra} and \textit{XMM - Newton}, only a few type 2 QSOs were 
known \citep[see e.g.][]{Akiyama etal 98, Della Ceca etal 00, Franceschini etal
00}. 
Deep X-ray surveys found indeed a large fraction of the objects to be obscured
\citep[e.g.][]{Barger etal 03, Mainieri etal 02}, but the
number of identified type 2 QSOs is still very low 
compared
to the model predictions. A significant fraction of high - $z$,
obscured QSOs may have escaped optical spectroscopic identification due to the
weakness of their optical counterparts and misclassification due to the lack of AGN signature. On the other hand, medium deep X-ray 
surveys, covering
relatively large sky areas at a higher flux limit, proved to be effective to 
select significant samples
of type 2 QSO candidates among objects with high values of the X-ray to
optical ratio \citep[$F_X/F_R > 10$, see][]{Fiore etal 03}. Recent findings 
also suggest a connection between Extremely Red Objects (EROs, $R - K
> 5$ in the Vega system) and type 2 QSOs 
\citep[see e.g.][and references therein]{Brusa etal 05, Severgnini etal 05}. \citet{Maiolino etal 06} suggest that by selecting
extreme values of $F_X/F_R (> 40)$ and extreme values of $R - K (> 6)$, the type
2 QSO selection efficiency may approach 100\%.

The absorption corrected X-ray luminosity is shown as a function of
the intrinsic column density in Fig.~\ref{lxnh}.
There is a significant number of objects (34 out of 124) having
$L_X > 10^{44}$ erg s$^{-1}$ and N$_H > 10^{22}$ cm$^{-2}$ in our sample. 12 of
them (35\%) are classified as unobscured AGN, while the remaining 22 are classified
as obscured AGN, based on the SED classification. We verified that these
objects generally have high X-ray to optical ratios, in
particular 21 of the 25 X-ray sources in the 3 $\sigma$ sample having $F_X/F_R >
10$ have N$_H > 10^{22}$ cm$^{-2}$ and $L_{2-10} > 10^{44}$ erg s$^{-1}$. On the
other hand, while for the optically unobscured AGN the fraction of high luminosity, X-ray absorbed sources having $F_X/F_R > 10$ is
only 25\% (3 out of 12), this
fraction is 77\% (17 out of 22) for the obscured ones.

All the 8 objects with $F_X/F_R > 40$ \citep[the threshold used by][]{Maiolino
etal 06} satisfy the X-ray definition of a type 2 QSO (see encircled objects in
Fig.~\ref{lxnh}). All are
optically obscured. 
These results confirm that type 2 QSO candidates are found between
the high X-ray to optical ratio population and that the threshold proposed by
\citet{Maiolino etal 06} is highly efficient in finding type 2 QSOs
but it is far from 
exhaustive (i.e. many type 2 QSOs have $F_X/F_R < 40$).

\begin{figure}
\centering
\includegraphics[width=8.5cm]{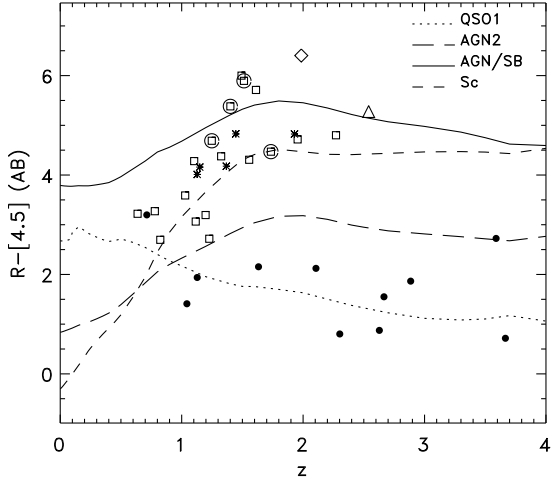}
\caption{Optical-infrared color (in the AB system) as a function of redshift for the
type 2 QSO candidates compared with the expected colors for various types of
galaxy and AGN templates. Filled circles are sources fitted by a type 1 AGN
template, squares are sources fitted by a type 2 AGN template,
asterisks are object fitted by a SFG template.
Encircled points are objects having $F_X/F_R > 40$. 
Triangle is
SWIRE\_J104409.95+585224.8  in \citet{Polletta etal 06}, diamond is XBS
J0216-0435 in \citet{Severgnini etal 06}.}
\label{qso2colors}
\end{figure}

In Fig.~\ref{qso2colors} the color between VVDS $R$ band and SWIRE 4.5 $\mu$m
band is shown as a function of redshift for the type 2 QSO candidates. Objects
fitted by a type 1 AGN template have generally blue colors. About 70\% of the 
candidates fitted by a type 2 AGN or a SFG template have extremely red
infrared/optical flux ratios, 
as observed in extremely obscured AGN and similar to those
observed in spectroscopically confirmed type 2 QSOs at high redshift
\citep[$z = 1.5 - 2.5$, see][]{Severgnini etal 06,
Polletta etal 06}. 
We have $K$ magnitudes from the UKIDSS Early Data Release or from
the VVDS for 18 objects: 4 of them are EROs, all having $F_X/F_R > 40$ and all fitted by a type 2 AGN or
a SFG template. Given 
the blue optical/IR colors of the optically unobscured AGN,
we exclude them from the type 2 QSO candidates. Therefore the
sample of type 2 QSOs reduces to 22 objects. One of them (XMDS 55) has been
indeed spectroscopically confirmed as a type 2 object (see Garcet et al. in
prep.). The type 2 QSO candidates represent ($18 \pm 4$)\% of the sources in the 3 $\sigma$ hard sample and have X-ray fluxes in the
range $1 - 5 \times 10^{-14}$ erg cm$^{-2}$ s$^{-1}$. 

As done before for optically obscured and unobscured AGN, we rescaled the surface density of X-ray sources
at $F_{2 - 10} > 10^{-14}$ erg cm$^{-2}$ s$^{-1}$ 
to the type 2 QSO fraction to estimate the type 2 QSOs surface density.
It results ($35 \pm 8$) deg$^{-2}$, lower but consistent, within errors, with
($45 \pm 15$) deg$^{-2}$ found by \citet{Cocchia etal 07} in the
HELLAS2XMM at the same flux level. 

The type 2 QSO population represents about 35\% of all high luminosity sources ($L_X >
10^{44}$ erg s$^{-1}$) in our sample. For
comparison, according to \citet{Perola etal 04}, in the HELLAS2XMM the fraction of X-ray absorbed
sources
(N$_H > 10^{22}$ cm$^{-2}$) in the high luminosity ($L_X > 10^{44}$ erg
s$^{-1}$) AGN population would be between 28\% and 40\%.

\section{Discussion} \label{fractions}
\subsection{Optical Obscuration vs X-ray Absorption} 

\begin{figure*}[!h]
\centering
\includegraphics[width=6.5cm]{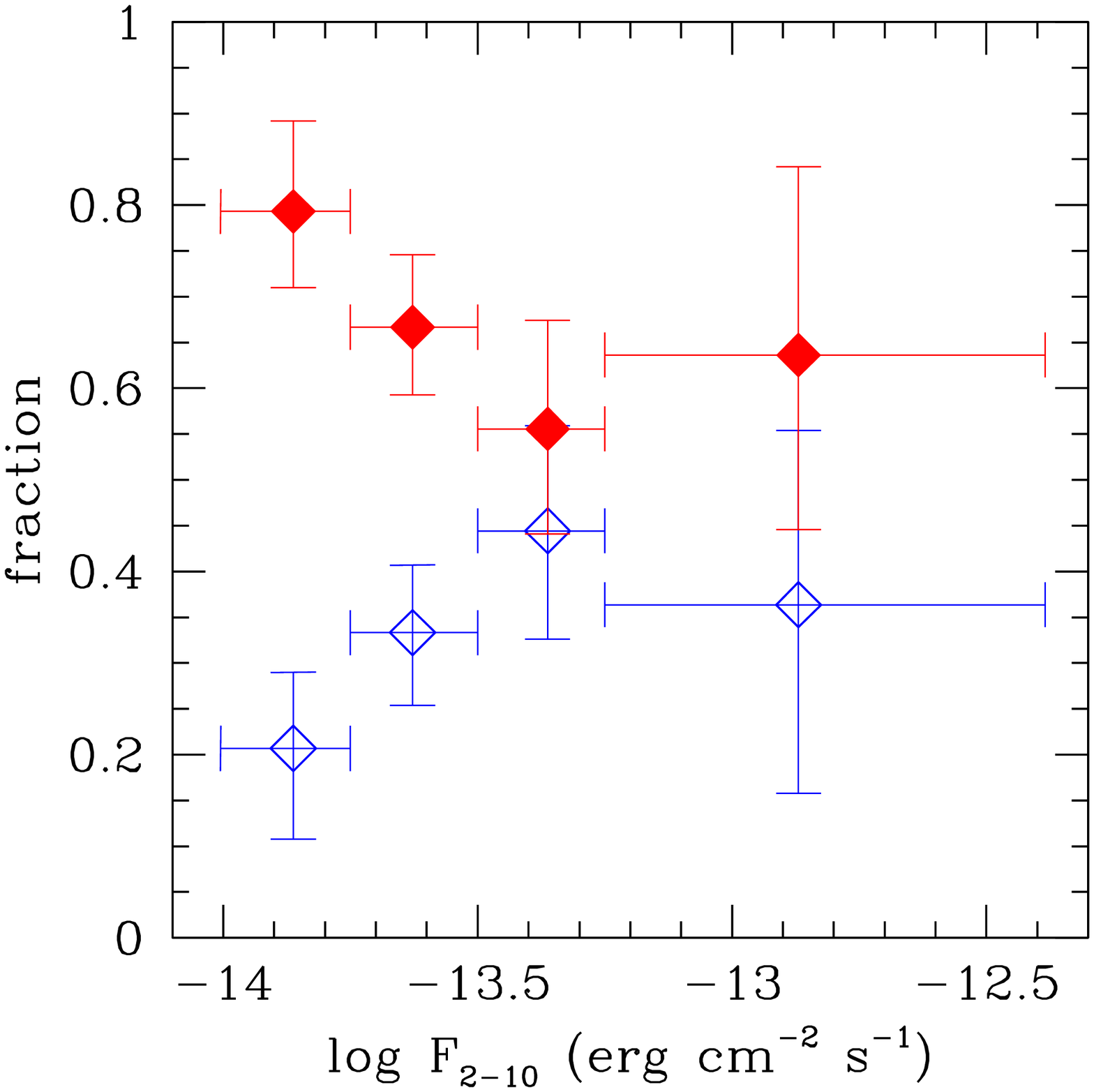}
\includegraphics[width=6.5cm]{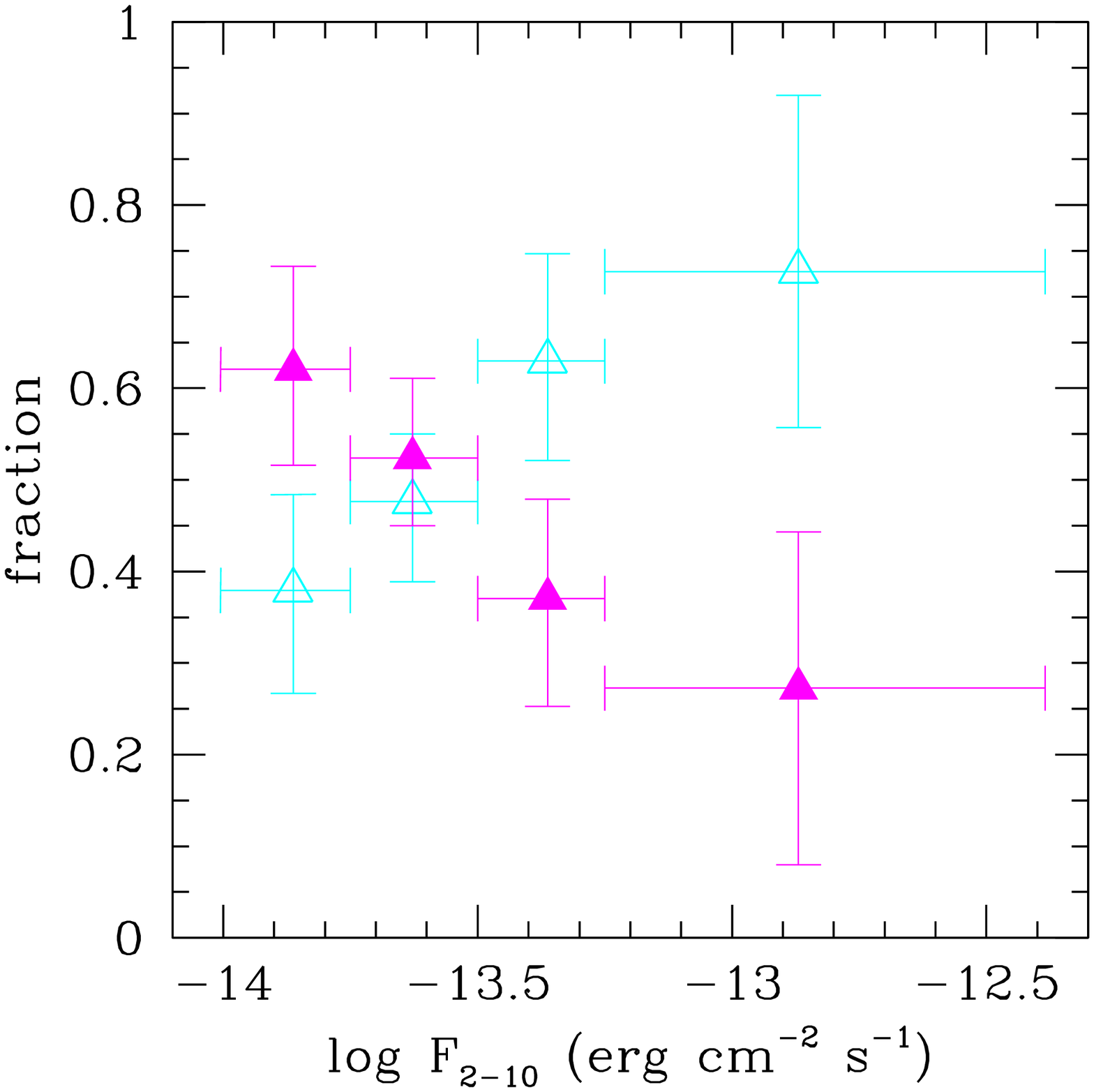}
\includegraphics[width=6.5cm]{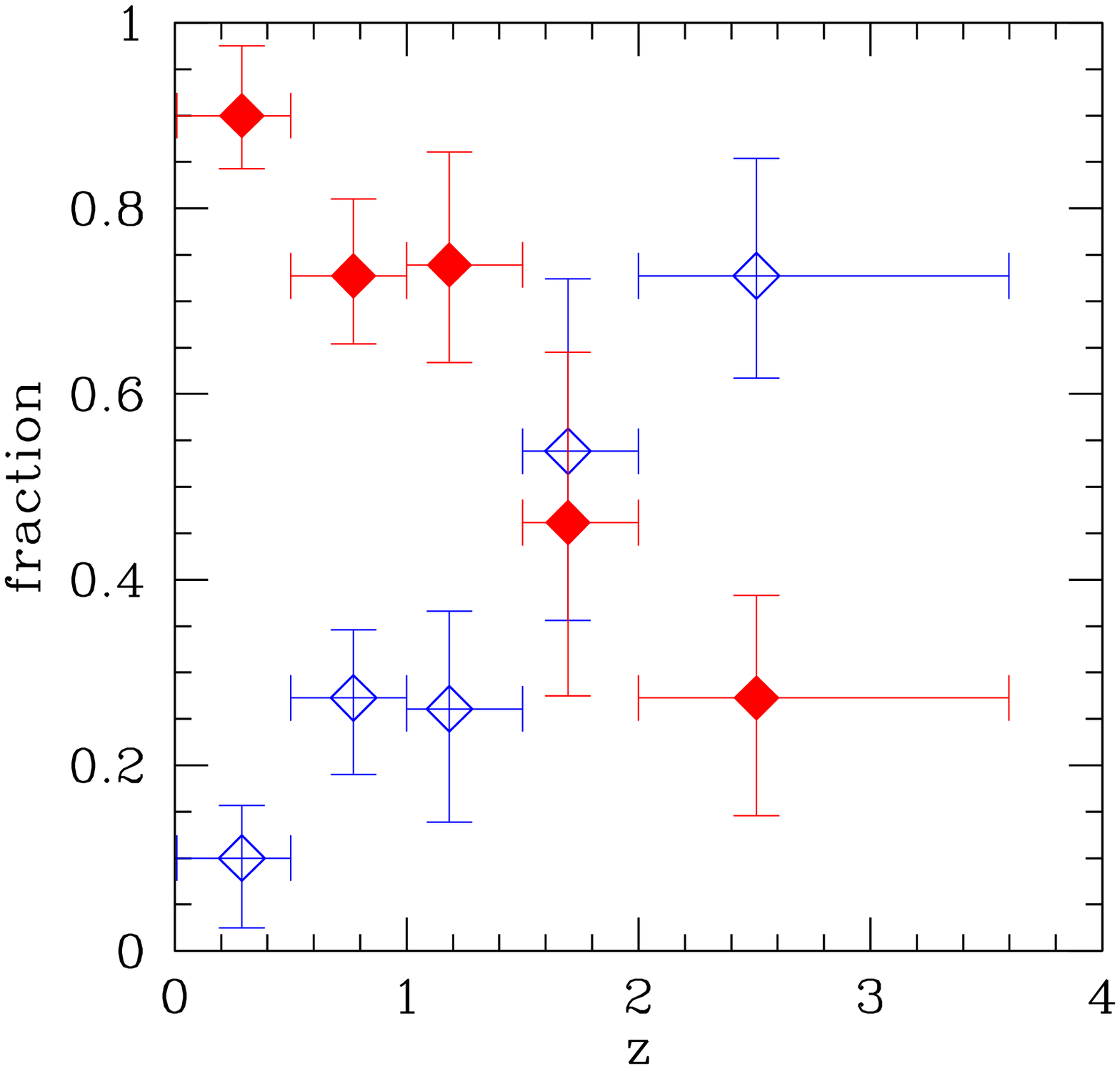}
\includegraphics[width=6.5cm]{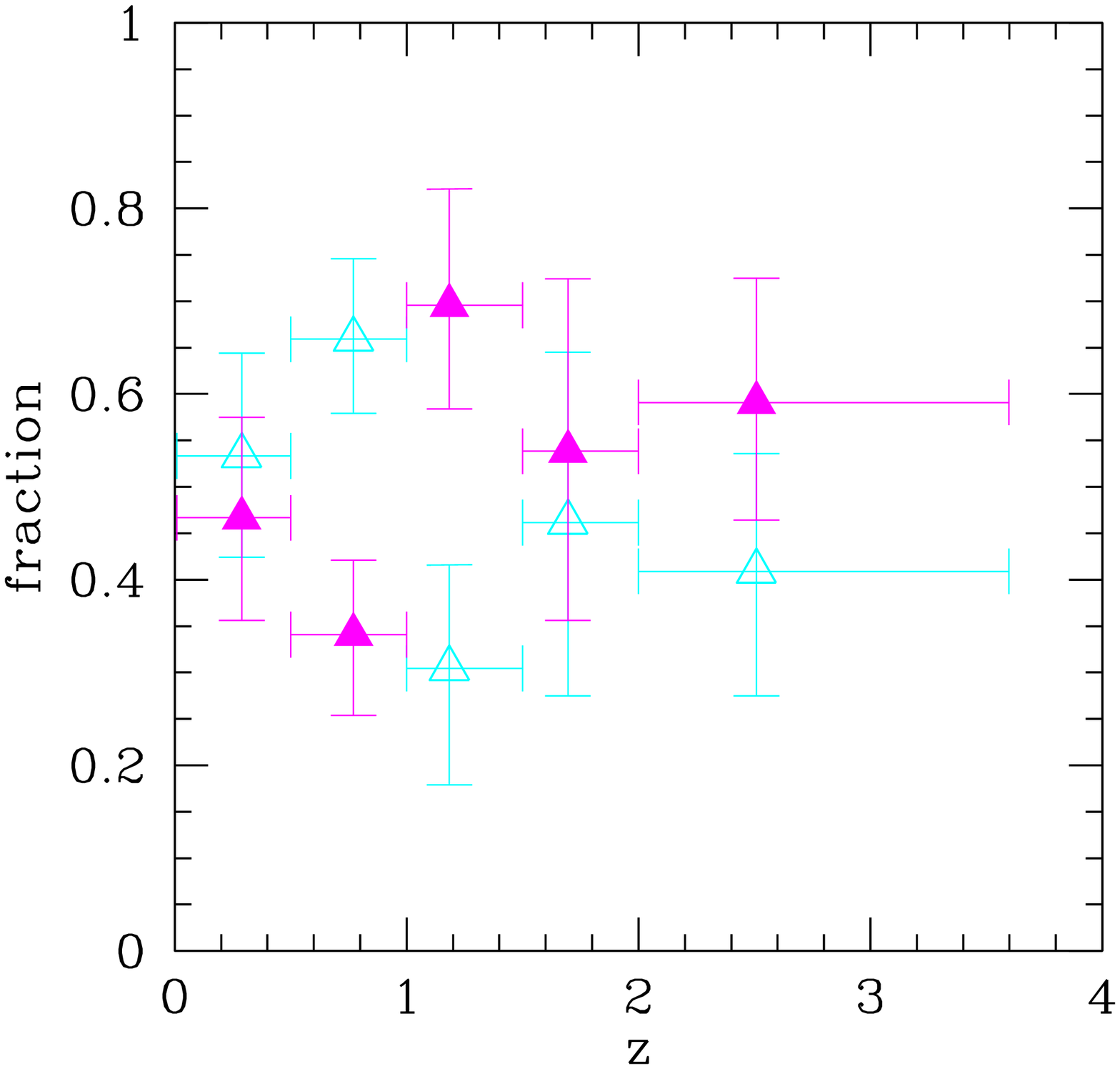}
\includegraphics[width=6.5cm]{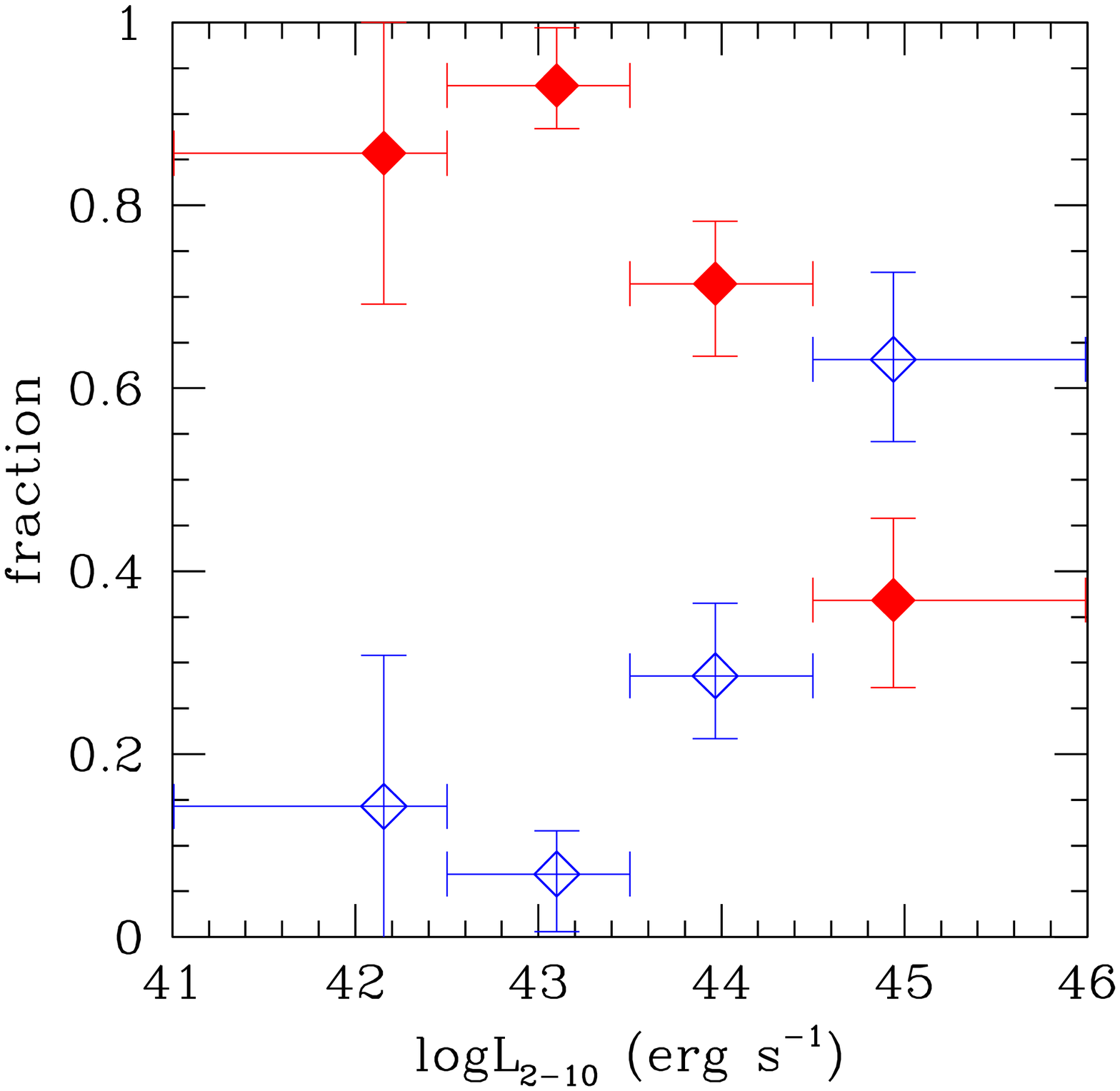}
\includegraphics[width=6.5cm]{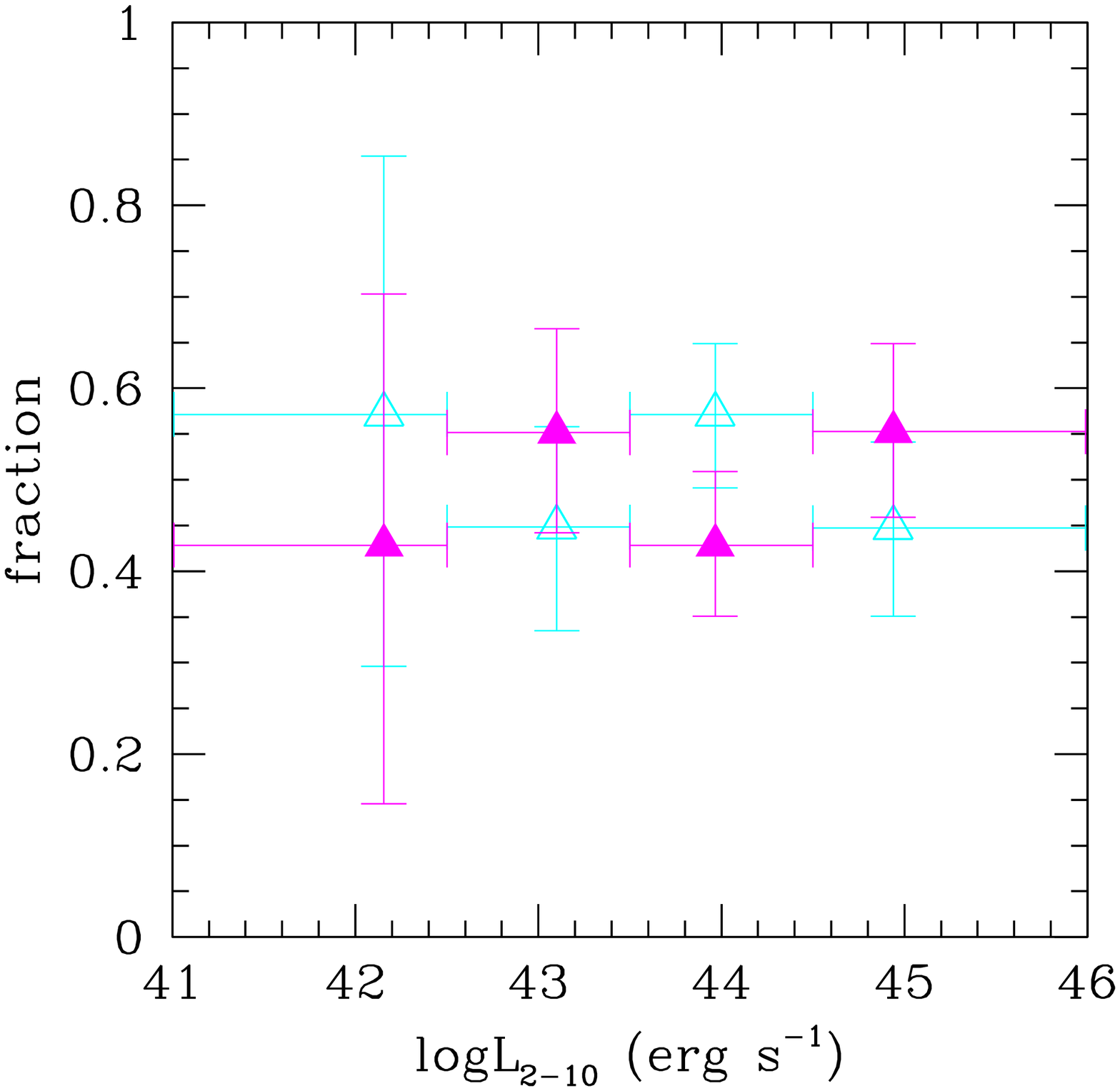}
\caption{Left panels: fraction of optically obscured (filled diamonds) and unobscured
(empty diamonds) AGN as a
function of X-ray flux (upper panel), redshift (middle panel) and
luminosity (lower panel). Right panels: fraction of X-ray absorbed (N$_H >
10^{22}$ cm$^{-2}$, filled
triangles) and unabsorbed
(empty triangles) AGN as a
function of X-ray flux (upper panel), redshift (middle panel) and
luminosity (lower panel).}
\label{fracfig}
\end{figure*}

The wide and well sampled multiwavelength coverage from the optical  
through the mid-IR allowed us to use the photometric approach to redshift
determination and classification in a very effective way. 
The comparison with a spectroscopic sample gives us confidence in the estimated
redshifts and in the fact that a photometric type 1 classification is
unambiguous, but reveals a bias against the recognition of a number of 
broad line objects as type 1 SEDs. This is due to the coexistence  
of Seyfert 1.8 type SEDs with the presence of broad line emission. 
Since we adopt here the SED classification,
these objects will be considered ``optically obscured''. 
In optically obscured objects
the optical-UV emission from the AGN may be either dimmed by intervening 
dust or be weaker than that of the host. 
In this sense, ``red quasars'' \citep{Wilkes etal 02, Gregg etal 02, 
Glikman etal 04, Urrutia etal 05, Wilkes etal 05} 
would be classified here as obscured objects.

The X-ray data give us an independent and complementary information 
essential to identify AGN where the nuclear X-ray emission is heavily
absorbed, and AGN features may be completely hidden
both in the optical and IR bands, as in the case of sources identified 
with SFGs.  
In the following we discuss the trends of optical obscuration and X-ray
absorption within our sample separately, in order to explore to what extent
the two are associated.

\begin{table*}
\begin{center}
\begin{tabular}{lr|rr|rrr}
\hline \hline
\multicolumn{1}{c}{Bins} & & & & \multicolumn{3}{c}{N$_H > 10^{22}$} \\ 
	& N$_{tot}$ & Opt. Unobsc. & Opt. Obsc. & N$_{abs}$	& Opt. Unobsc. & Opt. Obsc. \\
\hline
$-14 < \mbox{log}F_{2-10} < -13.75$	& 27	& 5	& 22 	& 17	& 2	& 15 \\
$-13.75 < \mbox{log}F_{2-10} < -13.5$	& 61	& 20	& 41 	& 32	& 8	& 24 \\
$-13.5 < \mbox{log}F_{2-10} < -13.25$	& 25	& 11	& 14 	& 9	& 2	& 7 \\
$-13.25 < \mbox{log}F_{2-10} < -12.5$  & 9	& 3	& 6 	& 2	& 0	& 2 \\
	&	&	& 	&	&	&	\\
$z < 0.5$			& 28	& 2	& 26 	& 13	& 0	& 13 \\
$0.5 < z < 1$			& 42	& 11	& 31 	& 14	& 1	& 13 \\
$1 < z < 1.5$			& 21	& 5	& 16 	& 15	& 3	& 12 \\
$1.5 < z < 2$			& 11	& 6	& 5 	& 6	& 1	& 5 \\
$z > 2$				& 20	& 15	& 5 	& 12	& 7	& 5 \\
	&	&	& 	&	&	&	\\
log$L_{2-10} < 42.5$			& 5	& 0	& 5 	& 2	& 0	& 2 \\
$42.5 < \mbox{log}L_{2-10} < 43.5$	& 27	& 1	& 26 	& 15	& 0	& 15 \\
$43.5 < \mbox{log}L_{2-10} < 44.5$	& 54	& 15	& 39 	& 23	& 4	& 19 \\
log$L_{2-10} > 44.5$			& 36	& 23	& 13 	& 20	& 8	& 12 \\
\hline 
\end{tabular}
\caption{Total number of sources, number of optically unobscured and optically obscured AGN, number of
X-ray absorbed AGN (N$_H > 10^{22}$ cm$^{-2}$), and number of optically
unobscured and obscured AGN among X-ray
absorbed AGN in each X-ray flux, redshift and absorption corrected X-ray luminosity bin
(see text and Fig.~\ref{fracfig}). Fluxes are in erg cm$^{-2}$ s$^{-1}$,
luminosities are in erg s$^{-1}$.}
\label{fractab}
\end{center}
\end{table*}

The sample contains 39 optically unobscured AGN and 83 optically obscured
AGN (of which 22 best fitted by a SFG template). The two sources with X-ray to optical ratios
and luminosities typical of normal galaxies are excluded from this analysis.
The X-ray absorbed AGN are 60. X-ray absorption occurs in 48 of the 83 
obscured AGN (of which 16 are SFGs). 
The remaining 12 X-ray absorbed AGN are
classified as unobscured on the basis of their SEDs. We conclude that
X-ray absorption is commonly but not exclusively associated with obscuration 
since 30\% of the unobscured AGN are X-ray
absorbed.

The numbers of optically unobscured, optically obscured and
X-ray absorbed AGN in different flux, redshift and luminosity bins are given
in Table~\ref{fractab}, where we also
report separately the number of X-ray absorbed sources within the optically unobscured and 
obscured AGN respectively. 
   
The fractions of obscured/absorbed AGN in our sample are shown 
in Fig.~\ref{fracfig}
as a function of observed flux, redshift and absorption corrected
hard X-ray luminosity. The left panels refer to
optically obscured and unobscured AGN,
while the right panels refer to X-ray absorbed (N$_H^{intr} > 10^{22}$ cm$^{-2}$) 
and unabsorbed  AGN, irrespectively of their SED classification. 

We used the Bayesian
statistics to estimate the ``true value'' of the fractions and their errors 
\citep[68\% credible interval, see][and references therein]{Andreon etal 06} and
to evaluate the reliability of the suggested correlations. 
We tested whether existing data support a
model in which the fraction 1) is constant or 2) has a linear dependence with
redshift, flux or luminosity, by computing the Bayesian Information
Criterion (BIC, Schwartz 1978; an astronomical introduction to it
can be found in Liddle 2004) for both models and then the difference $\Delta$BIC
between the BICs of the two models. A $\Delta$BIC of 6 or more can be
used to reject the model with the largest value of BIC,
whereas a value between 2 and 6 is only suggestive \citep{Jeffreys 61}. We also
compared the trends of optically obscured and X-ray absorbed AGN using the same
criterion.

Obscured sources are dominant in the 
lowest flux bin ($79^{+0.10}_{-0.08}$\%, upper left panel), although a systematic trend of optically obscured AGN to
decrease with X-ray flux is not established ($\Delta$BIC = 0.7). This result was already apparent
from the logN-logS curves, where the surface density of 
optically obscured AGN largely exceeds that of unobscured objects in the
lowest flux interval ($F < 2 \times 10^{-14}$ erg cm$^{-2}$ s$^{-1}$),
see Fig.~\ref{lognlogsfig}.
The fraction of X-ray absorbed AGN (upper right panel) is also higher at the 
faintest fluxes and there is a positive indication that it changes
systematically with flux ($\Delta$BIC = 4.9). 

The fraction of optically obscured AGN shows instead a steep decrease as a function 
of redshift (middle left panel), from $\sim 90$\% at $z < 0.5$ to
$\sim 30$\% at $z > 2$, and the trend is highly significant ($\Delta$BIC = 26.3).
Similarly, the data strongly support a decrease of the fraction of optically obscured
AGN with luminosity (lower left panel, $\Delta$BIC = 23.9). The two trends are
not independent, since in flux limited surveys
a correlation between luminosity and redshift is expected.
For X-ray absorbed AGN, data suggest constancy with both redshift and luminosity.
In summary there is 
evidence that the trends of optically obscured and X-ray absorbed AGN are different
both as a function of redshift ($\Delta$BIC = 17.5) and luminosity ($\Delta$BIC =
20.1), the former showing a decrease with redshift and luminosity, the latter
being essentially constant.
 
\subsection{Comparison with the literature} \label{fraclett}
Several authors compute the fraction of X-ray absorbed or optically obscured 
AGN as a function of all or some of the quantities described above, however 
in the literature a detailed comparison between optical obscuration and 
X-ray absorption seems to be lacking.

The trends of broad line AGN as a function of redshift and X-ray
luminosity are explored e.g by \citet{Steffen etal 03},
\citet{Barger etal 05}, \citet{Treister&Urry 05}, \citet{Tozzi etal 06}, who
use data from 
the \textit{Chandra} Deep Fields, in
some cases complemented by shallower 
\textit{Chandra} observations. 
All these samples reach flux levels significantly deeper than ours
thus probing the AGN population in more depth; 
given this, we limit ourselves to a qualitative comparison.
Assuming that unobscured AGN 
correspond to broad line AGN, the trends presented by the above authors 
are in agreement with those shown in the middle and lower left panels 
of Fig.~\ref{fracfig}. 

We examined in more detail the data of the HELLAS2XMM 1df 
\citep{Fiore etal 03, Perola etal 04} and of the Serendipitous Extragalactic 
X-ray Source Identification \citep[SEXSI,][]{Eckart etal 06},
whose flux limits and areas are comparable to those of the XMDS.
For the HELLAS2XMM, we computed the fraction of optically obscured AGN 
using all the
sources for which a spectroscopic classification is available (their samples S1,
S2 and S4). Consistently with our classification scheme, we grouped together the
objects spectroscopically classified as type 2 AGN, emission line galaxies
(ELGs) and early type galaxies (ETGs), considering them as optically obscured
AGN. We did the same for the SESXI data, considering as optically obscured AGN
all sources spectroscopically classified as narrow line AGN (NLAGN), ELGs and
absorption line galaxies (ALGs). Broad line AGN are instead classified as
optically unobscured. 

We computed the fractions 
in the HELLAS2XMM and SEXSI using the Bayesian 
statistics, as done for our sample. We find that also in the HELLAS2XMM and 
SEXSI cases the fraction of optically
obscured AGN decreases (and conversely the fraction of optically unobscured AGN increases)
with redshift and luminosity. We notice that in the HELLAS2XMM survey the fraction of
unobscured AGN is larger by a factor of $\sim 3$ than ours, for redshifts $z < 1.5$, while it is consistent with ours at higher
redshifts. The agreement with the SEXSI survey is instead
better. The spectroscopic completeness is however about 90\% for the HELLAS2XMM,
while it ranges from 40\% to 70\% for the SEXSI. The larger fractions of type 1 AGN in
spectroscopic samples are expected, given the
differences between photometric and spectroscopic classifications discussed above, however the
fraction of type 1 AGN in the HELLAS2XMM is still larger than
ours even when the correction factor computed in
subsection~\ref{photoclass} is taken into account.
Nevertheless, the trends observed in spectroscopic samples are consistent with
ours. 

The X-ray properties (fraction of AGN with N$_H > 10^{22}$ cm$^{-2}$) are
explored by a number of authors who use \textit{XMM - Newton} or \textit{Chandra} 
data of similar depth \citep[flux limit of $\sim 10^{-14}$ erg cm$^{-2}$ s$^{-1}$, 
see e.g.][]{Piconcelli etal 03, Perola etal 04, Eckart etal 06,
Akylas etal 06} and a quantitative comparison with their results is possible. 
Taking into account the different selection criteria and corrections that the
different authors apply to
the data, we find agreement within the errors with the results reported here. 

Again, we concentrated in particular on the results obtained 
in the HELLAS2XMM and in the SEXSI surveys.  
Both \citet{Perola etal 04} and \citet{Eckart etal 06} show that the fraction of
X-ray absorbed AGN increases with decreasing X-ray flux, even if the trends
are significant only when X-ray fluxes as faint as $10^{-15}$ erg cm$^{-2}$ s$^{-1}$ are
considered.  In our flux range, we are consistent with the HELLAS2XMM and SEXSI values.
\citet{Perola etal 04} and \citet{Eckart etal 06} also find that there is no evidence of a dependence
of the fraction of X-ray absorbed AGN on luminosity. Again, these results are
consistent with ours, with a better quantitative agreement with the SEXSI than with the
HELLAS2XMM survey (for example, the fraction of X-ray absorbed AGN is between
$\sim 0.4$ and $\sim 0.6$ in our case and in the SEXSI, while it is between
$\sim 0.2$ and $\sim 0.4$ in the HELLAS2XMM).

In conclusion, the analysis of our, the HELLAS2XMM and the SEXSI data indicates 
that the percentages of optically obscured and X-ray absorbed AGN within the same sample show different dependences on redshift and X-ray luminosity. Recent models which
describe  the cosmological evolution of the AGN space density, such as
those of \citet{Ueda etal 03} and \citet{Lafranca etal 05}, predict that the
fraction of X-ray absorbed AGN decreases with luminosity, and increase with
redshift, in the case of the \citet{Lafranca etal 05} model.
The combination of a decrease in luminosity and an increase with redshift
within a single flux limited sample, where the redshift and luminosity 
dependences tend to compensate each other, may well be the explanation
underlying the observed ``constancy'' 
with redshift and luminosity 
of the absorbed AGN fraction in our data. 
In fact, \citet{Lafranca etal 05} point out that the result 
of the opposite trends in $L_X$ and $z$ leads to an apparent constancy
of the X-ray absorbed fraction of AGN.
Only by combining several samples and thus
covering wide strips of the $L_X - z$ plane with almost constant redshift or
luminosity is it possible to disentangle the true dependences.

On the other hand, in a recent analysis of the \textit{XMM - Newton} observation of the \textit{Chandra} Deep Field South,
\citet{Dwelly&Page 06} do not find any dependence of the X-ray absorbed AGN
fraction on redshift and luminosity and suggest that the trends observed 
by other authors
could be the result of using deep X-ray
data from \textit{Chandra}, which could be biased against high redshift X-ray
absorbed AGN. Therefore the redshift and luminosity dependence of X-ray
absorption in the AGN population is still an open issue. 

In any case, the different redshift and luminosity dependences observed for optically
obscured and X-ray absorbed AGN imply that in a significant number of objects 
obscuration and absorption are not strictly related, moreover the relation depends on
redshift/luminosity in a systematic way.

\subsection{Objects with different absorption properties}
There is a number of examples in the literature of objects
that have opposite X-ray and optical properties, such as X-ray absorbed type 1
AGN \citep[e.g.][]{Comastri etal 01, Brusa etal 03, Akiyama etal 03} or X-ray
unabsorbed type 2 AGN \citep[e.g.][]{Panessa&Bassani 02, Caccianiga etal 04, Wolter etal 05}, 
however it is not clear so far how
common these exceptions are and how they can be reconciled
with the unified model \citep{Antonucci 93}. \citet{Perola etal 04} find that about 10\% of broad
line AGN are X-ray absorbed, while \citet{Tozzi etal 06} estimate that the
correspondence of unabsorbed (absorbed) X-ray sources to optical type 1 (type 2)
AGN is accurate for at least 80\% of the sources. We address this question in
the following subsections, where
instead of type 1 and type 2 AGN as derived from the optical spectra,
we will consider optically obscured and unobscured AGN as derived from
the SED classification.

\subsubsection{X-ray absorption without optical obscuration}
We find 12 X-ray absorbed, optically unobscured AGN, 31\% of all unobscured AGN. 
In 7 cases, the fit requires additional extinction,
A$_{\mathrm{V}}$=0.40--0.55, but lower than that would be derived from the column
density in the X-ray spectral fits assuming the standard dust-to-gas ratio. The discrepancy between optical and X-ray properties
can be explained e.g. by a dust-to-gas ratio lower than
the Galactic value \citep{Maiolino etal 01}
or by a different path of
the line of sights to the X-ray and the optical sources (e.g. see dual
absorber model in Risaliti et al. 2000
or torus clumpy model, Hoenig et al. 2006).
Another possibility is that in some objects the absorbing gas is ionized
rather than neutral: in that case, dust would likely be absent
and the intrinsic continuum plus broad
emission lines would be observed in the optical spectrum of the AGN. 
Examples of broad line
AGN whose X-ray spectra show absorption well fitted by an
ionized absorber model are reported in \citet{Page etal 06}. 

\subsubsection{Optical obscuration without X-ray absorption}
35 out of 83 optically obscured sources do not
show high X-ray absorption. 
We note however that 22 of them (63\%) are indeed fitted by the Seyfert 1.8
template, which in a number of cases corresponds to objects with broad emission
lines (subsection~\ref{photoclass}). It is therefore likely that an intermediate
class between unobscured and truly obscured AGN exists, made of objects which
are dominated by star-light emission in the near-IR, and X-ray unabsorbed.
In a future work we will extend our analysis to the whole XMDS area; the
larger statistics will allow us to refine the photometric classification using
a wider set of templates and better explore the relation between X-ray 
absorption and the optical to mid-IR SED. However, we do not expect all 
obscured AGN that are unabsorbed in the X-rays to be 
misclassified.

Another plausible explanation for (apparent) obscuration without
X-ray absorption
is a larger relative luminosity of the host galaxy compared to the 
AGN optical light. In this scenario, the AGN optical
light might simply be fainter than that of the host galaxy and not
extincted. 

Our analysis shows that about 40\% of our sources have opposite optical and
X-ray properties (12 X-ray absorbed, optically unobscured AGN
and 35 X-ray unabsorbed, optically obscured AGN). 
The uncertainties related to this fraction are linked, on the one side,
to the large errors involved in the
computation of intrinsic column densities for high redshift AGN (see
subsection~\ref{xcolors}), and on the other side, to our classification of 
sources based on SED fitting templates. 
However, these effects cannot fully account for the large number of objects with
discrepant optical and X-ray properties and the very different trends that we
observe in Fig.~\ref{fracfig}. We remind that similar discrepancies are also
apparent in the literature, where spectroscopic
classifications are used (subsection~\ref{fraclett}). These results suggest
that the basic formulation
of the unified model, in which the viewing angle is
the sole factor in determining the AGN type, might be too simplistic. As an example,
\citet{Elitzur 06} propose that the difference between type 1 and type 2 AGN is
instead an issue of
probability for direct view of the AGN through a clumpy, soft - edge torus.
Moreover, he suggests that the ``X-ray torus'' does not coincide with the
``dusty torus'' and that the bulk of the X-ray absorption likely comes in most
cases from clouds located in the inner, dust free portion of the X-ray torus.
The trends with luminosity described above agree with this kind of picture, 
where at high luminosities dust can be evaporated/expelled, while absorption
by gas can be associated with strong outflowing winds outside the dusty torus. 

Larger samples obtained by combining sub-samples from surveys of various
depths and areas, joined with optical spectroscopic data, are necessary to
minimize selection effects, provide a deeper understanding of the AGN
properties and test this scenario. 

\section{Summary and conclusions} \label{conclusions}
We have selected 136 X-ray sources detected at $\geq 3 \sigma$ in the $2 - 10$
keV band in the XMDS area also covered by the VVDS. More than 90\% of the
sources have been identified with an optical and/or infrared counterpart
(Section~\ref{id}) and 98\% of
them have X-ray to optical and X-ray to infrared ratios
typical of AGN (Section~\ref{xottir}).

We used the optical and
infrared data to construct SEDs and compute photometric redshifts. The comparison with the
spectroscopic redshifts available shows that in 90\% of cases there is agreement
between photometric and spectroscopic estimates (Section~\ref{photoz}). 
All sources fitted by
a SFG template which have X-ray to optical
ratios typical of AGN (22 out of 24) 
have high hard X-ray luminosities ($\geq 10^{42}$ erg
s$^{-1}$), suggesting that all are indeed AGN with the host galaxy emission
dominating in the optical - IR bands. 
Objects fitted by a type 1 AGN template
have generally blue optical/IR colors and in most cases do not show X-ray
absorption, while those fitted by a type 2 AGN or a SFG template have red
optical/IR colors and most of them are X-ray absorbed
(Section~\ref{photoz} and \ref{xray}). 

Comparison between photometric and spectroscopic classification, when
available, shows that the type 1 AGN photometric classification is
unambiguous, but we underestimate the fraction of broad line AGN, since 10 out
of 26 are classified as type 2 AGN due to the dominance of star
light emission in the near-IR. 
AGN
fitted by a type 1 template are referred as optically unobscured, while those
fitted by a type 2 AGN or a SFG template are referred as optically obscured
(subsection~\ref{photoclass}). 

We extracted the X-ray spectra of the 55 X-ray sources
having at least 50 net counts in the $2 - 10$ keV band. For
sources with a smaller number of detected counts, we
used hardness ratios to compute the column densities. We find that, when the redshift
dependence is taken into account, 60\% of the optically obscured AGN are X-ray absorbed, but
also 30\% of the optically unobscured AGN have N$_H^{intr} > 10^{22}$
cm$^{-2}$, showing that optical and X-ray
classifications are not strictly related (Section~\ref{xray}).

We constructed stacked X-ray spectra to measure average spectral properties
of our sample and to find differences between optically obscured and unobscured AGN as a function
of X-ray flux. We find that stacked spectra of optically unobscured AGN have a photon index
consistent with $\Gamma = 2$, similar to average values found for X-ray unabsorbed AGN.
On the other hand, the slope of stacked spectra of optically obscured AGN is consistent
with $\Gamma \sim 1.6$. 
The stacked
spectrum of the objects fitted by a SFG template is even harder, with $\Gamma \sim 1.2 - 1.3$. (subsection~\ref{stack}).

Comparing the fractions of optically obscured and X-ray absorbed AGN, we find
that while the fraction of
optically obscured AGN steeply decreases with redshift and luminosity, that of
X-ray absorbed AGN is nearly constant at $\sim 50$\% as
a function of redshift and X-ray luminosity (Section~\ref{fractions}). 
The constancy of the population of X-ray absorbed AGN with redshift and
luminosity observed in our sample can be explained by the
\citet{Lafranca etal 05} predictions that the fraction of X-ray absorbed AGN
should decrease with luminosity and increase with redshift, since these two
dependences tend to compensate each other in a single, flux limited sample.
The different trends of optically obscured and X-ray absorbed AGN are
confirmed also by the analysis of spectroscopic samples from the literature, showing that this
result is not biased by the uncertainties in the photometric classification.

In 39\% sources (47
out of 122) an inconsistency between X-ray absorption and optical
obscuration is observed (12 X-ray absorbed, optically unobscured AGN, and 35
X-ray unabsorbed, optically obscured AGN). 
About 63\% of the optically obscured X-ray absorbed AGN can be indeed
misclassified broad line AGN. On the other hand, the significant fraction of
optically unobscured, X-ray absorbed AGN found in our sample suggests that the
basic formulation of the AGN unification model can be too simplistic.

We also computed the differential logN-logS relationship for the obscured and
unobscured AGN, finding that the optically obscured AGN begin to dominate for $F < 2 \times 10^{-14}$
erg cm$^{-2}$ s$^{-1}$, where the
ratio between obscured and unobscured AGN is $\sim 4.6$
(Section~\ref{lognlogs}). In the whole flux range considered, the surface
density of the optically obscured AGN is higher than that of the optically
unobscured ones by a factor of $\sim 2.4$. However, if a correction is
applied to account for misclassified type 1 AGN, the ratio between optically
obscured and unobscured AGN is $\sim 1.1$.

We find 22 sources that could be classified as type 2 QSO candidates (N$_H > 10^{22}$ cm$^{-2}$, $L_X > 10^{44}$ erg
s$^{-1}$).
They are fitted by a type 2 AGN or a SFG
template and on average their infrared/optical and X-ray/optical flux
ratios are typical of extremely obscured AGN. 4 of the 18 having a measured $K$
magnitude are EROs ($R - K > 5$ in the Vega system). We estimate a surface density of type 2 QSOs at $F_{2-10} > 10^{-14}$
erg cm$^{-2}$ s$^{-1}$ of
($35 \pm 8$) deg$^{-2}$
(subsection~\ref{qso2}).

In this work the full power of multiwavelength observations is exploited to
understand the global properties of AGN. We plan to extend our analysis to the
whole XMDS to improve the statistical significance of our results. 

\begin{acknowledgements}
We thank M. Bolzonella for her suggestions in computing the photometric
redshifts and her comments on the paper and A. Bongiorno for helping us with the
VVDS spectra. We also thank D. Alloin, R. Della Ceca, P. Severgnini and G. Zamorani
for useful comments.\\
M.T. acknowledges financial support from MIUR Cofin 2004-023189-005.
The INAF members of the team 
acknowledge financial contribution from contract ASI-INAF I/023/05/0. O.G. and
J.S. would like to acknowledge support by contract Inter - University Attraction
Pole P5/36 (Belgium), the ESA PRODEX Programme (XMM - LSS) and the Belgium
Federal Science Policy Office. This work is based on observations obtained with \textit{XMM-Newton}, an ESA science
mission with instruments and contributions directly funded by ESA Member
States and the USA (NASA). Also based on observations obtained with MegaPrime/MegaCam, a joint project of 
CFHT and CEA/DAPNIA, at the Canada-France-Hawaii Telescope (CFHT) which is 
operated by the National Research Council (NRC) of Canada, the Institut 
National des Science de l'Univers of the Centre National de la Recherche 
Scientifique (CNRS) of France, and the University of Hawaii. This work is based 
in part on data products produced at TERAPIX and the Canadian Astronomy Data 
Centre as part of the Canada-France-Hawaii Telescope Legacy Survey, a 
collaborative project of NRC and CNRS. This work is in part based on observations made with the {\it Spitzer
Space Telescope}, which is operated by the Jet Propulsion Laboratory,
California Institute of Technology under NASA. Support for this work,
part of the {\it Spitzer Space Telescope} Legacy Science Program, was
provided by NASA through an award issued by the Jet Propulsion
Laboratory, California Institute of Technology under NASA contract 1407.

\end{acknowledgements}

\Online
\appendix
\section{Properties of the 3 $\sigma$ hard sample}
We report in Table~\ref{gentab} the X-ray, optical and infrared properties of the 3
$\sigma$ hard sample. Column 1 list the XMDS identifier, column 2 the XMDS name,
columns 3 and 4 the X-ray coordinates (astrometrically corrected), column 5 the
$2 - 10$ keV flux, column 6 the hardness ratio between the $2 -10$ and $0.5 - 2$
keV bands, column 7 the distance between the X-ray and the optical VVDS
position, column 8 the VVDS $B$ magnitude, column 9 the VVDS $I$ magnitude,
column 10 the color classification (B for blue, i.e. $B - I \leq 1$ and R for
red, i.e. $B - I > 1$), column 11 the CFHTLS $i'$ magnitude, column 12 the SWIRE
3.6 $\mu$m flux, column 13 the SWIRE 24 $\mu$m flux, column 14 the spectroscopic
redshift, column 15 the photometric
redshift, column 16 the photometric classification based on SEDs (see text) and
column 17 the absorption corrected $2 - 10$ keV luminosity. 

A ``...'' means no
data available (i.e. source outside the field of view or not measured flux
because of instrumental problems). For sources undetected at 24 $\mu$m,
a 5 $\sigma$ upper limit (241 $\mu$Jy) was used.
For sources with
ambiguous optical - IR identification, two or more rows are associated with the
X-ray source, one for each candidate counterpart, with the one having the best
probability (see Section~\ref{id}) listed as first. In these cases, X-ray data are
reported only in the first row.
 All magnitudes are in the AB system. The systematic error on SWIRE fluxes is
5\%.

\begin{landscape}
\begin{center}
\scriptsize{%
\tablefirsthead{%
\hline \hline
ID & XMDS Name & RA & Dec. & \multicolumn{1}{c}{$F_{2-10}$} & HR & X -- 0 & mag. $B$ & mag. $I$ & color & mag. $i'$ & $F_{3.6}$ & $F_{24}$ & $z_{sp}$ & $z_{ph}$ & SED class. & log$L_{2 - 10}^{corr}$ \\
   &      &    &      & ($10^{-14}$ cgs) & & \multicolumn{1}{c}{($''$)} &         &          &       &           & ($\mu$Jy) & ($\mu$Jy) & & & & (erg s$^{-1}$) \\}
\tablehead{%
\multicolumn{17}{l}{\textbf{Table \ref{gentab}} continued} \\
\hline \hline
ID & XMDS Name & RA & Dec. & \multicolumn{1}{c}{$F_{2-10}$} & HR & X -- 0 & mag. $B$ & mag. $I$ & color & mag. $i'$ & $F_{3.6}$ & $F_{24}$ & $z_{sp}$ & $z_{ph}$ & SED class. & log$L_{2 - 10}^{corr}$ \\
   &      &    &      & ($10^{-14}$ cgs) & & \multicolumn{1}{c}{($''$)} &         &          &       &           & ($\mu$Jy) & ($\mu$Jy) & & & & (erg s$^{-1}$) \\
\hline}
\tablecaption{The 3 $\sigma$ hard sample: XMDS identifier and name, X-ray
coordinates, X-ray flux ($2 - 10$ keV), hardness ratio between the $2 -10$ and $0.5 - 2$
keV bands, distance between the X-ray and optical VVDS positions, VVDS $B$
and $I$ magnitudes, color classification, CFHTLS $i'$ magnitude, SWIRE 3.6 and
24 $\mu$m fluxes, spectroscopic and photometric redshifts, SED classification
(see text) and absorption corrected X-ray luminosity. All magnitudes are in the AB system.}
\label{gentab}
\tabletail{%
\hline}
\tablelasttail{%
\hline}
\begin{supertabular}{llllrrrrrrrrrrrll}
\hline
4 & XMDS J022521.0$-$043949 & 02:25:21.079 & $-$4:39:48.476 & 7.45 & $-$0.07 & 0.8 & 21.08 & 18.39 & R & 18.56 & 259 & 410 & 0.265 & 0.245 & SFG & 43.122 \\
12 & XMDS J022544.9$-$043735 & 02:25:44.988 & $-$4:37:35.230 & 2.13	& $-$0.23	& 1.0 & 24.66 & 22.14 & R & 21.96 & 32 & $< 241$ & 3.5892 & 3.597 & AGN1 & 45.42 \\
13 & XMDS J022504.5$-$043707 & 02:25:04.546 & $-$4:37:07.416 & 2.30	& $-$0.28	& 0.4 & 24.04 & 22.69 & R & 22.91 & 59 & 470 & ... & 1.103 & AGN2 & 44.20 \\
16 & XMDS J022506.4$-$043621 & 02:25:06.484 & $-$4:36:20.932 & 1.36	& $-$0.65	& 0.6 & 22.37 & 20.77 & R & 20.87 & 109 & 362 & ... & 0.850 & AGN2 & 43.68 \\
18 & XMDS J022510.6$-$043549 & 02:25:10.665 & $-$4:35:49.283 & 2.97	& $-$0.44	& 0.3 & 25.65 & 24.86 & B & 25.31 & 27 & 300 & ... & 1.612 & AGN2 & 44.71 \\
29 & XMDS J022521.0$-$043228 & 02:25:21.067 & $-$4:32:27.837 & 1.25	 & $-$0.25	 & 1.3 & 24.51 & 21.59 & R & 21.93 & 52 & $< 241$ & ... & 0.868 & SFG & 43.68 \\
36 & XMDS J022449.8$-$043026 & 02:24:49.888 & $-$4:30:26.192 & 3.54	 & $-$0.22 & 0.6 & 20.42 & 19.94 & B & 20.10 & 110 & 895 & ... & 1.043 & AGN1 & 44.32 \\
40 & XMDS J022510.6$-$042928 & 02:25:10.689 & $-$4:29:28.264 & 2.65	 & 0.23 & 1.9 & 24.55 & 23.25 & R & 23.71 & 40 & 776 & ... & 1.955 & AGN2 & 44.91 \\
55 & XMDS J022522.8$-$042648 & 02:25:22.870 & $-$4:26:47.807 & 2.65	 & 0.36 & 0.2 & 23.37 & 20.68 & R & 20.53 & 176 & 1065 & 1.029 & 1.135 & AGN2 & 44.19 \\
58 & XMDS J022436.2$-$042511 & 02:24:36.239 & $-$4:25:10.816 & 2.31	 & $-$0.27 & 0.4 & 25.35 & 23.31 & R & 23.06 & 33 & 357 & ... & 0.645 & AGN2 & 43.63 \\
60 & XMDS J022439.6$-$042401 & 02:24:39.729 & $-$4:24:01.115 & 4.71	 & $-$0.49 & 0.2 & 20.94 & 19.27 & R & 19.36 & 238 & 1477 & 0.478 & 0.203 & AGN2 & 43.61 \\
67 & XMDS J022537.0$-$042132 & 02:25:37.108 & $-$4:21:32.123 & 2.35	 & $-$0.69 & 0.8 & 19.39 & 19.11 & B & 19.42 & 170 & 1114 & ... & 0.961 & AGN1 & 44.05 \\
71 & XMDS J022511.9$-$041911 & 02:25:11.970 & $-$4:19:10.526 & 4.12	 & 0.00 & 1.4 & 24.83 & 23.59 & R & 23.59 & 68 & 805 & ... & 1.401 & AGN2 & 44.72 \\
91 & XMDS J022710.0$-$041649 & 02:27:10.185 & $-$4:16:49.536 & 2.22	 & $-$0.45 & 0.9 & 23.35 & 21.76 & R & 22.24 & 57 & 705 & ... & 1.015 & AGN2 & 44.08 \\
   &                       &        	   &		  &	 &	 & 2.0 & 21.80 & 19.87 & R & 20.02 & 60 & $< 241$ & ... & 0.396 & SFG & 43.09 \\
94 & XMDS J022643.8$-$041626 & 02:26:43.962 & $-$4:16:26.397 & 5.37	 & $-$0.15 & 0.3 & 20.96 & 18.34 & R & 18.52 & 405 & 1109 & ... & 0.349 & AGN2 & 43.46 \\
101 & XMDS J022809.4$-$041524 & 02:28:09.546 & $-$4:15:24.972 & 2.85 & $-$0.53 & 0.9 & 24.60 & 21.74 & R & ... & ... & ... & ... & 0.874 & SFG & 44.03 \\
106 & XMDS J022719.5$-$041407 & 02:27:19.650 & $-$4:14:07.764 & 1.01 & 0.64 & ... & ... & ... & ... & ... & 53 & 1909 & ... & 0.387 & AGN2 & 42.83 \\
111 & XMDS J022735.6$-$041317 & 02:27:35.721 & $-$4:13:17.043 & 3.30 & $-$0.03 & 0.4 & 25.43 & 24.46 & B & ... & ... & ... & ... & 2.360 & SFG & 45.16 \\
    &                       &              &		   &	  &	  & 0.8 & 25.27 & 24.47 & B & ... & ... & ... & ... & 2.514 & SFG & 45.23 \\
112 & XMDS J022809.0$-$041232 & 02:28:09.129 & $-$4:12:32.857 & 19.44 & $-$0.52 & 1.8   & 19.83 & 19.04 & B & ... & ... & 0.878 & ... & 0.900 & AGN1 & 44.90 \\
114 & XMDS J022649.7$-$041240 & 02:26:49.855 & $-$4:12:40.512 & 2.03 & $-$0.55 & 0.4 & 21.70 & 18.93 & R & 19.14 & 161 & 640 & ... & 0.375 & SFG & 42.99 \\
118 & XMDS J022649.3$-$041154 & 02:26:49.466 & $-$4:11:54.248 & 1.49 & $-$0.69 & 1.2 & 22.93 & 21.73 & R & 21.85 & 78 & 361 & 1.1572 & 1.195 & AGN2 & 44.05 \\
120 & XMDS J022735.7$-$041122 & 02:27:35.841 & $-$4:11:22.637 & 9.05 & $-$0.50 & 0.6 & 25.31 & 23.26 & R & ... & ... & ... & ... & 1.043 & AGN2 & 44.72 \\
    &                       &              &		   &	  &	  & 1.6 & 22.54 & 21.32 & R & ... & ... & ... & ... & 1.082 & AGN2 & 44.76 \\
124 & XMDS J022659.7$-$041108 & 02:26:59.823 & $-$4:11:08.500 & 2.46 & 0.46	 & 0.2 & 22.28 & 20.32 & R & 20.45 & 100 & 912 & ... & 0.371 & AGN2 & 43.14 \\
133 & XMDS J022713.1$-$040912 & 02:27:13.216 & $-$4:09:12.800 & 2.28 & $-$0.43 & 0.7 & 25.03 & 22.34 & R & 22.49 & 15 & $< 241$ & ... & 0.723 & AGN2 & 43.74 \\
134 & XMDS J022701.3$-$040912 & 02:27:01.488 & $-$4:09:12.270 & 1.26 & $-$0.54 & 0.4 & 21.39 & 20.18 & R & 20.24 & 117 & 566 & ... & 0.755 & AGN2 & 43.52 \\
138 & XMDS J022656.0$-$040821 & 02:26:56.201 & $-$4:08:21.483 & 1.15 & 0.35	 & 4.6 & 22.47 & 20.30 & R & 20.45 & 119 & 523 & ... & 0.546 & AGN2 & 43.20 \\
139 & XMDS J022727.7$-$040806 & 02:27:27.835 & $-$4:08:06.268 & 2.29 & $-$0.63 & 0.6 & 21.03 & 19.96 & R & ... & ... & ... & ... & 0.729 & AGN2 & 43.74 \\
140 & XMDS J022701.3$-$040751 & 02:27:01.437 & $-$4:07:51.221 & 5.05 & $-$0.38 & 0.4 & 20.20 & 18.04 & R & 18.20 & 270 & 633 & 0.220 & 0.235 & SFG & 42.86 \\
142 & XMDS J022644.1$-$040720 & 02:26:44.262 & $-$4:07:20.192 & 2.53 & $-$0.64 & 0.4 & 19.87 & 19.47 & B & 19.84 & 168 & 1412 & ... & 0.573 & AGN1 & 43.53 \\
143 & XMDS J022655.4$-$040650 & 02:26:55.586 & $-$4:06:50.880 & 1.71 & 0.04	 & 0.7 & 22.35 & 20.54 & R & 20.87 & 87 & 1112 & ... & 0.374 & AGN2 & 42.95 \\
144 & XMDS J022652.0$-$040556 & 02:26:52.138 & $-$4:05:56.361 & 3.33 & $-$0.49 & 0.1 & 20.62 & 20.61 & B & 20.37 & 100 & 388 & ... & 0.864 & AGN2 & 44.09 \\
149 & XMDS J022707.2$-$040438 & 02:27:07.359 & $-$4:04:38.575 & 1.96 & $-$0.53 & 1.2 & 21.54 & 19.69 & R & 19.90 & 180 & 851 & ... & 0.493 & AGN2 & 43.26 \\
161 & XMDS J022700.7$-$042020 & 02:27:00.832 & $-$4:20:20.357 & 41.26 & $-$0.76 & 0.6   & 17.71 & 16.34 & R & 16.42 & 1803 & 27141 & 0.053 & 0.086 & AGN2 & 42.44 \\
178 & XMDS J022544.6$-$041936 & 02:25:44.673 & $-$4:19:35.499 & 5.40 & $-$0.43 & 1.7 & 19.64 & 17.49 & R & 17.79 & 285 & $< 241$ & ... & 0.059 & SFG & 41.66 \\
179 & XMDS J022607.7$-$041843 & 02:26:07.740 & $-$4:18:42.879 & 12.07 & $-$0.56 & 0.4   & 19.41 & 18.95 & B & 19.26 & 347 & 1577 & 0.495 & 0.328 & AGN1 & 44.05 \\
191 & XMDS J022626.5$-$041214 & 02:26:26.535 & $-$4:12:13.575 & 4.38 & 0.65 & 1.5 & 23.34 & 20.42 &	 R & 20.66 & 188 & 2267 & ... & 0.763 & AGN2 & 44.18 \\
    &                       &              &		   &	  &	 & 4.2 & 24.16 & 20.60 & R & 20.82 & 64 & $< 241$ & ... & & & \\
197 & XMDS J022539.0$-$040823 & 02:25:39.020 & $-$4:08:22.499 & 2.87 & 0.26	& 1.6 & 22.08 & 20.25 & R & 20.65 & 186 & 1035 & ... & 0.824 & AGN2 & 44.02  \\
199 & XMDS J022614.5$-$040738 & 02:26:14.540 & $-$4:07:37.479 & 2.61 & 0.15	& ... & ... & ... & ... & ... & 26 & $< 241$ & ... & 2.411 & AGN2 & 45.12 \\
227 & XMDS J022511.4$-$041916 & 02:25:11.497 & $-$4:19:17.389 & 3.51 & 0.00	& 2.8 & 27.41 & 24.82 & R & 24.62 & 22 & $< 241$ & ... & 1.446 & SFG & 44.69 \\
    &                       &              &		   &	  &	 & 3.7 & 24.12 & 23.85 & B & 23.56 & 10 & $< 241$ & ... & 1.683 & AGN1 & 44.85 \\
229 & XMDS J022406.4$-$041830 & 02:24:06.475 & $-$4:18:31.927 & 2.57 & 0.85	& 1.6 & 25.90 & 22.75 & R & 23.45 & 33 & $< 241$ & ... & 1.128 & SFG & 44.43 \\
232 & XMDS J022449.2$-$041800 & 02:24:49.235 & $-$4:18:01.477 & 2.11 & $-$0.65 & 0.4 & 21.72 & 20.39 & R & 20.52 & 66 & 365 & ... & 0.581 & AGN2 & 43.46 \\
233 & XMDS J022456.0$-$041725 & 02:24:56.112 & $-$4:17:26.698 & 1.80 & $-$0.48 & ... & ... & ... & ... & ... & 20 & $< 241$ & ... & 1.293 & AGN1 & 44.25 \\
242 & XMDS J022437.8$-$041520 & 02:24:37.842 & $-$4:15:21.854 & 1.56 & $-$0.12 & 3.4 & 24.10 & 22.31 & R & 22.56 & 37 & $< 241$ & ... & 1.051 & SFG & 43.99 \\
246 & XMDS J022415.6$-$041416 & 02:24:15.678 & $-$4:14:17.652 & 2.80 & $-$0.48 & 1.7 & 21.53 & 20.65 & B & 21.14 & 59 & 533 & ... & 2.106 & AGN1 & 44.97 \\
253 & XMDS J022451.9$-$041209 & 02:24:52.013 & $-$4:12:10.157 & 1.85 & $-$0.54 & 0.5 & 19.77 & 18.91 & B & 19.38 & 154 & 2366 & ... & 1.686 & AGN1 & 44.55 \\
255 & XMDS J022408.4$-$041149 & 02:24:08.469 & $-$4:11:50.595 & 1.66 & 0.24	& 2.1 & 24.61 & 23.91 & B & 23.96 & 45 & $< 241$ & ... & 2.063 & SFG & 44.77 \\
    &                       &              &		   &	  &	 & 4.9 & 22.04 & 20.02 & R & 20.17 & 73 & $< 241$ & ... & 0.340 & SFG & 42.86 \\
258 & XMDS J022447.4$-$041049 & 02:24:47.510 & $-$4:10:50.701 & 1.63 & $-$0.14 & 0.2 & 21.41 & 21.16 & B & 21.17 & 20 & $< 241$ & ... & 2.628 & AGN1 & 44.98 \\
270 & XMDS J022449.2$-$040841 & 02:24:49.300 & $-$4:08:42.809 & 1.25 & $-$0.30 & 0.4 & 22.18 & 21.17 & R & 21.71 & 43 & 341 & ... & 0.958 & AGN2 & 43.78 \\
271 & XMDS J022509.5$-$040836 & 02:25:09.623 & $-$4:08:37.874 & 1.79 & $-$0.57 & 1.3 & 20.38 & 19.87 & B & 19.79 & 95 & 786 & ... & 2.042 & AGN1 & 44.74 \\
272 & XMDS J022501.6$-$040752 & 02:25:01.725 & $-$4:07:53.534 & 2.57 & $-$0.53 & 0.8 & 19.88 & 19.49 & B & 19.42 & 173 & 1099 & ... & 0.797 & AGN1 & 43.89 \\
279 & XMDS J022421.3$-$040607 & 02:24:21.350 & $-$4:06:08.861 & 2.24 & $-$0.19 & 1.1 & 21.26 & 19.39 & R & 19.60 & 301 & 2330 & ... & 0.260 & AGN2 & 42.69 \\
280 & XMDS J022417.9$-$040606 & 02:24:18.033 & $-$4:06:07.171 & 1.86 & 0.08	& 0.7 & 23.84 & 23.01 & B & ... & 6 & $< 241$ & ... & 1.633 & AGN1 & 44.55 \\
281 & XMDS J022503.2$-$040538 & 02:25:03.251 & $-$4:05:39.467 & 5.28 & $-$0.58 & 0.7 & 21.22 & 20.62 & B & 20.88 & 92 & 669 & ... & 0.930 & AGN2 & 44.36 \\
282 & XMDS J022452.1$-$040518 & 02:24:52.146 & $-$4:05:20.053 & 4.53 & $-$0.66 & 0.9 & 20.23 & 19.60 & B & 19.52 & 260 & 2366 & ... & 0.189 & AGN1 & 42.66 \\
288 & XMDS J022421.2$-$040351 & 02:24:21.265 & $-$4:03:52.487 & 2.26 & $-$0.31 & 5.2 & 24.50 & 21.34 & R & 21.57 & 81 & $< 241$ & ... & 0.566 & SFG & 43.48 \\
291 & XMDS J022452.0$-$040258 & 02:24:52.111 & $-$4:02:59.465 & 2.30 & 0.98	& 1.8 & 20.91 & 18.82 & R & 19.01 & 220 & 1052 & ... & 0.269 & AGN2 & 43.14 \\
330 & XMDS J022333.0$-$041525 & 02:23:33.118 & $-$4:15:25.133 & 1.24 & 0.50	& 1.4 & 25.72 & 23.50 & R & 23.39 & 56 & 369 & ... & 2.271 & AGN2 & 44.78 \\
351 & XMDS J022356.5$-$041105 & 02:23:56.652 & $-$4:11:05.848 & 1.66 & $-$0.48 & 0.5 & 22.31 & 21.53 & B & 22.31 & 47 & $< 241$ & ... & 0.927 & AGN2 & 43.86 \\
359 & XMDS J022325.3$-$040922 & 02:23:25.479 & $-$4:09:22.450 & 1.03 & 0.35	& 2.0 & 22.55 & 21.50 & R & 22.13 & 45 & $< 241$ & ... & 1.154 & SFG & 43.95 \\
403 & XMDS J022742.1$-$043607 & 02:27:42.144 & $-$4:36:07.737 & 1.92 & $-$0.19 & ...	& ... & ... & ... & 29.88 & 13 & $< 241$ & ... & 3.493 & SFG & 45.35 \\
406 & XMDS J022732.7$-$043544 & 02:27:32.736 & $-$4:35:44.652 & 4.03 & 0.22	& 1.5 & 22.74 & 21.90 & B & 22.02 & 66 & 895 & ... & 0.713 & AGN1 & 44.02 \\
414 & XMDS J022726.3$-$043327 & 02:27:26.389 & $-$4:33:27.610 & 1.62 & $-$0.43 & 0.9 & 21.39 & 19.03 & R & 19.18 & 113 & 489 & ... & 3.666 & AGN1 & 45.31 \\
416 & XMDS J022812.2$-$043230 & 02:28:12.290 & $-$4:32:30.771 & 2.91 & $-$0.50 & 3.2 & 21.17 & 20.48 & B & ... & ... & $-$ 1 & ... & 1.668 & AGN1 & 44.73 \\
420 & XMDS J022729.2$-$043225 & 02:27:29.235 & $-$4:32:25.984 & 1.91 & $-$0.57 & 1.5 & 19.33 & 19.04 & B & 19.26 & 104 & 1427 & 2.2899 & 2.357 & AGN1 & 44.89 \\
427 & XMDS J022758.6$-$043112 & 02:27:58.647 & $-$4:31:12.027 & 1.33 & $-$0.55 & 0.2 & 25.67 & 23.29 & R & ... & ... & ... & ... & 0.859 & AGN2 & 43.68 \\
430 & XMDS J022737.1$-$043031 & 02:27:37.109 & $-$4:30:31.532 & 2.04 & $-$0.48 & 1.1 & 23.79 & 22.10 & R & 22.03 & 38 & $< 241$ & ... & 0.760 & AGN2 & 43.74 \\
438 & XMDS J022756.3$-$042905 & 02:27:56.391 & $-$4:29:05.354 & 1.18 & 0.74	& 0.4 & 23.36 & 20.85 & R & ... & ... & ... & ... & 0.375 & SFG & 42.89 \\
439 & XMDS J022746.0$-$042853 & 02:27:46.020 & $-$4:28:53.194 & 1.06 & 0.10	& 1.4 & 23.79 & 23.19 & B & 23.28 & 54 & $< 241$ & 1.3679 & 1.988 & SFG & 44.12 \\
440 & XMDS J022748.8$-$042820 & 02:27:48.829 & $-$4:28:20.933 & 3.12 & $-$0.47 & 0.5 & 19.59 & 19.09 & B & ... & ... & ... & ... & 2.574 & AGN1 & 45.22 \\
449 & XMDS J022815.2$-$042617 & 02:28:15.285 & $-$4:26:17.032 & 1.68 & $-$0.10 & 1.8 & 25.29 & 23.94 & R & ... & ... & ... & ... & 2.504 & SFG & 44.95 \\
    &                       &              &		   &	  &	  & 3.6 & 23.89 & 23.58 & B & ... & ... & ... & ... & 2.172 & AGN2 & 44.80 \\
453 & XMDS J022802.3$-$042546 & 02:28:02.357 & $-$4:25:46.815 & 2.06 & 0.32	 & 1.6 & 23.27 & 20.71 & R & ... & ... & ... & ... & 0.568 & AGN2 & 43.48 \\
470 & XMDS J022804.5$-$041818 & 02:28:04.571 & $-$4:18:18.119 & 2.48 & $-$0.51 & 0.1 & 26.05 & 25.00 & R & ... & ... & ... & ... & 1.886 & AGN2 & 44.79 \\
    &                       &              &		   &	  &	  & 1.3 & 25.53 & 24.53 & R & ... & ... & ... & ... & 0.418 & SFG & 43.19 \\
    &                       &              &		   &	  &	  & 3.6 & 24.38 & 22.95 & R & ... & ... & ... & ... & 1.082 & AGN2 & 44.20 \\
487 & XMDS J022643.6$-$043317 & 02:26:43.649 & $-$4:33:18.291 & 2.74 & 0.42	& 0.4 & 19.70 & 17.80 & R & 17.98 & 495 & 1401 & 0.308 & 0.489 & SFG & 43.00 \\
498 & XMDS J022629.2$-$043057 & 02:26:29.282 & $-$4:30:57.554 & 5.29 & $-$0.47 & 0.7 & 20.10 & 19.75 & B & 19.59 & 75 & 941 & 2.031 & 1.903 & AGN1 & 45.20 \\
503 & XMDS J022649.3$-$042920 & 02:26:49.366 & $-$4:29:21.129 & 2.05 & 0.03	& 0.5 & 23.25 & 20.20 & R & 20.42 & 121 & 295 & 0.6335 & 0.723 & AGN2 & 43.58 \\
505 & XMDS J022649.0$-$042745 & 02:26:49.004 & $-$4:27:46.432 & 2.03 & $-$0.53 & 0.3 & 20.04 & 18.92 & R & 19.11 & 203 & 1834 & 0.327 & 0.084 & AGN2 & 42.86 \\
521 & XMDS J022658.8$-$042321 & 02:26:58.864 & $-$4:23:21.563 & 5.19 & $-$0.11 & 1.2 & 24.73 & 21.93 & R & 22.32 & 127 & 566 & 1.3253 & 1.754 & AGN2 & 44.74 \\
523 & XMDS J022622.1$-$042221 & 02:26:22.126 & $-$4:22:21.538 & 4.41 & $-$0.63 & 0.6 & 19.28 & 18.50 & B & 18.77 & 155 & 1326 & 2.0060 & 1.586 & AGN1 & 45.11 \\
551 & XMDS J022342.0$-$043533 & 02:23:42.086 & $-$4:35:33.683 & 2.64 & $-$0.40 & 1.2 & 21.41 & 20.70 & B & 20.84 & 82 & 721 & ... & 1.128 & AGN1 & 44.28 \\
561 & XMDS J022424.1$-$043228 & 02:24:24.168 & $-$4:32:28.884 & 2.81 & $-$0.69 & 0.8 & 18.97 & 18.65 & B & 18.81 & 146 & 1087 & ... & 1.678 & AGN1 & 44.72 \\
564 & XMDS J022350.7$-$043157 & 02:23:50.768 & $-$4:31:57.899 & 2.17 & $-$0.50 & 0.2 & 19.99 & 19.23 & B & 19.17 & 179 & 503 & ... & 0.224 & AGN2 & 42.51 \\
565 & XMDS J022356.8$-$043115 & 02:23:56.802 & $-$4:31:15.137 & 1.92 & $-$0.48 & 1.5 & 24.50 & 22.79 & R & 23.33 & 24 & $< 241$ & ... & 1.051 & SFG & 44.06 \\
567 & XMDS J022432.4$-$043036 & 02:24:32.468 & $-$4:30:36.864 & 2.71 & $-$0.62 & 0.1 & 20.80 & 19.91 & B & 20.49 & 66 & 433 & ... & 0.588 & AGN1 & 43.59 \\
571 & XMDS J022330.2$-$043004 & 02:23:30.265 & $-$4:30:04.246 & 2.76 & $-$0.43 & 1.7 & 20.63 & 19.94 & B & 20.18 & 89 & 1402 & 2.666 & 2.404 & AGN1 & 45.21 \\
577 & XMDS J022438.9$-$042705 & 02:24:38.940 & $-$4:27:05.814 & 12.66 & $-$0.67 & 0.9	& 18.87 & 17.40 & R & 17.54 & 637 & 6630 & 0.252 & 0.188 & AGN2 & 43.39 \\
578 & XMDS J022350.7$-$042703 & 02:23:50.729 & $-$4:27:03.790 & 1.34 & $-$0.70 & 0.3 & 21.74 & 21.40 & B & 21.59 & 42 & $< 241$ & ... & 1.033 & AGN1 & 43.88 \\
602 & XMDS J022351.2$-$042054 & 02:23:51.280 & $-$4:20:54.416 & 2.57 & 0.23	& 1.2 & 20.35 & 19.11 & R & 19.21 & 73 & 3752 & 0.181 & 0.097 & SFG & 42.42 \\
626 & XMDS J022326.0$-$043534 & 02:23:25.969 & $-$4:35:35.294 & 3.78 & $-$0.45 & 1.0 & 25.00 & 22.33 & R & 22.30 & 64 & 409 & ... & 1.149 & SFG & 44.45 \\
708 & XMDS J022605.3$-$045803 & 02:26:05.398 & $-$4:58:03.890 & 2.30 & $-$0.58 & 0.5 & 21.33 & 21.20 & B & 21.36 & 69 & $< 241$ & ... & 1.476 & AGN2 & 44.50 \\
709 & XMDS J022606.7$-$045722 & 02:26:06.723 & $-$4:57:23.365 & 2.11 & $-$0.67 & 0.3 & 20.93 & 20.14 & B & 20.32 & 109 & 626 & ... & 0.661 & AGN2 & 43.60 \\
710 & XMDS J022627.4$-$045710 & 02:26:27.430 & $-$4:57:11.053 & 5.40 & $-$0.46 & 0.0 & 21.85 & 21.14 & B & 21.51 & 45 & 1559 & ... & 2.320 & AGN1 & 45.35 \\
718 & XMDS J022615.1$-$045355 & 02:26:15.130 & $-$4:53:55.676 & 2.14 & $-$0.31 & 0.2 & 23.06 & 21.13 & R & 21.48 & 63 & $< 241$ & ... & 0.896 & SFG & 43.94 \\
720 & XMDS J022628.9$-$045252 & 02:26:28.949 & $-$4:52:53.487 & 2.52 & $-$0.58 & 2.5 & 21.86 & 19.82 & R & 20.99 & ... & ... & ... & 2.027 & AGN1 & 44.88 \\
731 & XMDS J022554.1$-$044921 & 02:25:54.200 & $-$4:49:21.907 & 1.25 & $-$0.17 & 0.7 & 23.80 & 21.73 & R & 22.05 & 48 & $< 241$ & ... & 0.534 & AGN2 & 43.17 \\
738 & XMDS J022556.1$-$044724 & 02:25:56.111 & $-$4:47:24.727 & 3.33 & $-$0.56 & 0.2 & 20.92 & 20.16 & B & 20.75 & 107 & 589 & 1.010 & 0.898 & AGN2 & 44.25 \\
739 & XMDS J022617.1$-$044724 & 02:26:17.199 & $-$4:47:24.925 & 2.33 & 0.38	& 0.9 & 18.90 & 17.38 & R & 17.52 & 432 & 3233 & 0.140 & 0.184 & SFG & 42.14 \\
742 & XMDS J022514.3$-$044659 & 02:25:14.304 & $-$4:47:00.049 & 4.13 & $-$0.57 & 1.0 & 18.63 & 18.17 & B & 18.12 & 243 & 2697 & 1.924 & 1.615 & AGN1 & 45.04 \\
743 & XMDS J022625.2$-$044647 & 02:26:25.287 & $-$4:46:47.949 & 2.28 & $-$0.43 & 0.0 & 24.51 & 24.40 & B & 24.01 & 25 & 247 & ... & 1.556 & AGN2 & 44.56 \\
746 & XMDS J022512.6$-$044633 & 02:25:12.644 & $-$4:46:33.803 & 7.48 & 0.49	 & 1.0 & 22.20 & 19.41 & R & 19.59 & 107 & $< 241$ & ... & 0.219 & SFG & 43.10 \\
747 & XMDS J022640.4$-$044606 & 02:26:40.488 & $-$4:46:07.281 & 3.15 & 0.44	 & 1.0 & 24.76 & 24.07 & B & 24.51 & 31 & 460 & ... & 1.243 & AGN2 & 44.53 \\
    &                       &              &		   &	  &	 & 2.0 & 25.24 & 24.08 & R & 24.53 & ... & ... & & 1.221 & AGN2 & 44.51 \\
748 & XMDS J022610.9$-$044550 & 02:26:10.984 & $-$4:45:51.052 & 2.05 & $-$0.33 & 0.7 & 23.13 & 22.15 & B & 22.75 & 20 & $< 241$ & ... & 2.886 & AGN1 & 45.17 \\
755 & XMDS J022600.1$-$044412 & 02:26:00.170 & $-$4:44:13.349 & 2.80 & $-$0.55 & 2.6 & 24.76 & 22.03 & R & 22.41 & 46 & 576 & ... & 0.730 & AGN2 & 43.83 \\
760 & XMDS J022531.4$-$044210 & 02:25:31.431 & $-$4:42:10.616 & 2.71 & 1.00	& 2.7 & 23.86 & 22.06 & R & 22.28 & 34 & $< 241$ & 1.2274 & 0.833 & AGN2 & 45.96 \\
779 & XMDS J022321.8$-$045740 & 02:23:22.045 & $-$4:57:38.385 & 4.72 & 0.66	& 0.9 & 22.04 & 19.46 & R & 19.64 & 340 & 5583 & ... & 0.637 & AGN2 & 44.02 \\
780 & XMDS J022332.0$-$045740 & 02:23:32.191 & $-$4:57:38.712 & 2.45 & $-$0.56 & 0.6 & 20.32 & 19.62 & B & 19.66 & 262 & 1250 & ... & 0.963 & AGN2 & 44.07 \\
782 & XMDS J022326.3$-$045708 & 02:23:26.549 & $-$4:57:05.982 & 2.91 & $-$0.31 & 1.5 & 20.98 & 20.48 & B & 20.32 & 132 & 579 & 0.826 & 0.839 & AGN2 & 43.98 \\
787 & XMDS J022317.9$-$045527 & 02:23:18.081 & $-$4:55:25.250 & 1.42 & 0.18	& 0.5 & 24.22 & 22.65 & R & 23.23 & 14 & $< 241$ & ... & 0.981 & AGN2 & 43.90 \\
788 & XMDS J022353.7$-$045510 & 02:23:53.880 & $-$4:55:08.174 & 4.86 & $-$0.52 & 0.4 & 22.29 & 21.43 & B & 21.52 & 51 & $< 241$ & ... & 0.958 & AGN1 & 44.36 \\
789 & XMDS J022329.1$-$045452 & 02:23:29.341 & $-$4:54:50.778 & 2.40 & $-$0.64 & 1.2 & 20.44 & 19.92 & B & 20.01 & 140 & $< 241$ & 0.604 & 0.646 & AGN2 & 43.56 \\
800 & XMDS J022403.8$-$045120 & 02:24:04.068 & $-$4:51:18.278 & 2.03 & $-$0.58 & 0.2 & 22.70 & 20.95 & R & 21.07 & 130 & 874 & ... & 0.874 & AGN2 &  43.88 \\
801 & XMDS J022344.4$-$045120 & 02:23:44.634 & $-$4:51:18.223 & 1.74 & $-$0.42 & 1.0 & 22.36 & 21.50 & B & 21.65 & 29 & $< 241$ & ... & 0.959 & AGN2 & 43.92 \\
807 & XMDS J022333.0$-$044924 & 02:23:33.185 & $-$4:49:22.513 & 1.78 & $-$0.49 & 0.9 & 20.66 & 20.26 & B & 20.22 & 37 & 290 & 2.302 & 2.039 & AGN1 & 44.87 \\
817 & XMDS J022354.5$-$044815 & 02:23:54.772 & $-$4:48:13.852 & 1.69 & $-$0.66 & 1.2 & 18.25 & 18.01 & B & 18.15 & 251 & 3418 & 2.458 & 2.428 & AGN1 & 44.91 \\
820 & XMDS J022319.4$-$044732 & 02:23:19.582 & $-$4:47:30.407 & 3.87 & $-$0.38 & 0.7 & 21.18 & 18.48 & R & 18.68 & 531 & 3756 & 0.293 & 0.640 & AGN2 & 43.03 \\
825 & XMDS J022330.6$-$044633 & 02:23:30.803 & $-$4:46:31.754 & 3.35 & $-$0.50 & 0.7 & 24.57 & 23.37 & R & ... & 18 & $< 241$ & ... & 1.735 & AGN2 & 44.84 \\
828 & XMDS J022318.8$-$044616 & 02:23:19.040 & $-$4:46:13.919 & 1.73 & $-$0.51 & 0.1 & 21.23 & 20.76 & B & 21.15 & 45 & 356 & ... & 0.673 & AGN1 & 43.53 \\
840 & XMDS J022330.9$-$044235 & 02:23:31.149 & $-$4:42:33.036 & 2.07 & $-$0.21 & ...	& ... & ... & ... & ... & 9 & $< 241$ & ... & 2.297 & AGN2 & 44.94 \\
842 & XMDS J022402.4$-$044140 & 02:24:02.650 & $-$4:41:38.339 & 3.63 & 0.16 & 3.8 & 14.87 & 13.54 & R & ... & 13528 & 40594 & 0.043 & 0.010 & SFG & 41.24 \\
844 & XMDS J022343.2$-$044105 & 02:23:43.453 & $-$4:41:03.352 & 1.95 & 0.75 & 2.8 & 26.04 & 24.47 & R & 24.18 & 26 & 363 & ... & 1.493 & AGN2 & 44.57 \\
    &                       &              &              &      &      & 4.5 & 23.70 & 23.24 &    B & 23.25 & ... & ... & ... & 1.970 & AGN2 & 44.87 \\
846 & XMDS J022317.2$-$044035 & 02:23:17.415 & $-$4:40:33.523 & 3.72 & $-$0.38 & 1.1 & 18.81 & 18.23 & B & 18.74 & 580 & 4587 & 0.842 & 0.765 & AGN1 & 44.11 \\
1197 & XMDS J022720.2$-$045738 & 02:27:20.271 & $-$4:57:38.992 & 1.90 & 0.32 & 1.5 & 23.05 & 21.70 & R & 22.48 & 39 & 306 & ... & 1.116 & AGN2 & 44.18 \\
1199 & XMDS J022651.6$-$045714 & 02:26:51.736 & $-$4:57:14.797 & 6.39 & $-$0.44 & 0.8	& 21.93 & 19.78 & R & 20.11 & 104 & 830 & 0.331 & 0.290 & AGN2 & 43.37 \\
1201 & XMDS J022723.4$-$045608 & 02:27:23.526 & $-$4:56:08.628 & 1.65 & $-$0.43 & 1.5	& 22.18 & 21.34 & B & 21.48 & 54 & 313 & ... & 0.168 & AGN2 & 42.12 \\
1219 & XMDS J022701.6$-$045158 & 02:27:01.690 & $-$4:51:58.996 & 1.15 & 0.07 & ...	& ... & ... & ... & ... & 23 & $< 241$ & ... & 2.112 & SFG & 44.62 \\
1226 & XMDS J022711.7$-$045038 & 02:27:11.761 & $-$4:50:38.630 & 8.55 & $-$0.51 & 1.1	& 21.17 & 20.42 & B & 21.10 & 97 & 734 & ... & 0.946 & AGN1 & 44.59 \\
1227 & XMDS J022736.8$-$045033 & 02:27:36.895 & $-$4:50:34.200 & 2.41 & 0.00 & 0.9 & 21.86 & 19.44 & R & 19.60 & 169 & 864 & ... & 0.445 & AGN2 & 43.26 \\
1231 & XMDS J022731.9$-$044957 & 02:27:31.946 & $-$4:49:57.627 & 0.98 & $-$0.54 & 0.8	& 22.63 & 20.95 & R & 21.22 & 73 & 416 & ... & 0.741 & AGN2 & 43.39 \\
1236 & XMDS J022729.0$-$044857 & 02:27:29.058 & $-$4:48:57.908 & 1.18 & 0.09 & 2.5 & 27.35 & 24.03 & R & 24.71 & 30 & 621 & ... & 1.513 & AGN2 & 44.28 \\
1246 & XMDS J022712.8$-$044636 & 02:27:12.916 & $-$4:46:37.102 & 2.34 & $-$0.59 & 1.5	& 18.03 & 17.69 & B & 17.75 & 451 & 2830 & ... & 1.446 & AGN1 & 44.48 \\
1247 & XMDS J022633.1$-$044637 & 02:26:33.213 & $-$4:46:38.477 & 3.17 & $-$0.32 & 2.0	& 25.01 & 21.29 & R & 21.66 & 72 & $< 241$ & ... & 1.197 & AGN2 & 44.42 \\
1248 & XMDS J022725.4$-$044619 & 02:27:25.522 & $-$4:46:19.686 & 4.62 & $-$0.54 & 0.1	& 17.02 & 15.29 & R & 15.36 & 655 & 1974 & ... & 0.034 & SFG & 41.09 \\
1252 & XMDS J022716.0$-$044539 & 02:27:16.045 & $-$4:45:39.562 & 4.62 & $-$0.60 & 1.6	& 20.42 & 20.02 & B & 20.12 & 143 & 1125 & ... & 0.590 & AGN1 & 43.82 \\
1264 & XMDS J022751.3$-$044251 & 02:27:51.403 & $-$4:42:51.783 & 3.19 & $-$0.36 & 1.7	& 20.19 & 19.82 & B & 19.77 & 63 & 1035 & ... & 1.694 & AGN1 & 44.79 \\
1265 & XMDS J022712.6$-$044221 & 02:27:12.698 & $-$4:42:21.727 & 6.28 & 0.10 & 1.8 & 19.79 & 17.90 & R & 18.03 & 435 & 4214 & 0.205 & 0.232 & SFG & 42.93 \\
\end{supertabular}		      
}				      
\end{center} 
\end{landscape}

\section{Results of spectral fits} \label{appspectra}
We report here the spectral fit results from spectra of individual sources (see
section~\ref{singlespec}). In Table~\ref{z0tab} we list the column density, photon index,
reduced $\chi^2$ and number of degrees of freedom obtained from analysis of all sources for which we could leave
free both $\Gamma$ and N$_H$, together with the hardness ratio values. In
Table~\ref{alltab} we report spectral fit results obtained by fixing the photon index to
$\Gamma = 2.0$ (columns 2 and 3) and $\Gamma = 1.7$ (columns 4 and 5), or inserting photometric (or spectroscopic, when available) redshift in the
model (columns from 6 to 9). In column 10 we report the EPIC cameras used to extract the spectrum.

For three sources 
(XMDS 161, 282 and 1199, 
see Table~\ref{alltab}) the spectral model with $\Gamma = 2.0$ gives a poor fit ($\chi^2_\nu >
2$). 
XMDS 1199 
shows a moderate X-ray absorption (N$_H \sim 10^{21}$
cm$^{-2}$) and $\Gamma$ frozen to 1.7 gives a better fit ($\chi^2_\nu =
1.35$). The other two sources have instead very steep spectra, in fact the spectral fit
obtained with free photon index (see Table~\ref{z0tab}) gives $\Gamma > 2$ for
both of them, but, while for XMDS 282 the fit with $\Gamma$ free is good
($\chi^2_\nu = 1.07$), we were not able to find an acceptable fit with a simple
power law model for
XMDS 161, whose spectrum exhibits a significant soft excess.

On the other hand, when $\Gamma = 1.7$, there are 10 sources having
$\chi^2_\nu > 2$: all of them are bright, soft sources, for which the spectral
fit with free photon index gives $\Gamma > 2$.

For sources XMDS 124 and 779 no stable solutions were found fixing the photon
index to 2.0. 

When the redshift was
introduced in the spectral model, the photon index was left free when the
$\chi^2$ statistics could be used, otherwise it was fixed to 2.0. 
In the two cases cited above (XMDS 124 and 779) and for 
XMDS 739
we however had to fix the photon index to $\Gamma = 1.7$, because for $\Gamma = 2.0$
no stable solution was found. For 
XMDS 453 
no stable solution was found with either value of $\Gamma$.

\begin{table*}
\begin{center}
\begin{tabular}{llllrr}
\hline \hline
XMDS ID	& logN$_H$ (cm$^{-2})$		& $\Gamma$ 		 & $\chi^2_\nu$ (d.o.f.)	& HR$_{cb}$	& HR$_{dc}$\\
\hline
112     & $20.44^{+0.51}_{-0.03}$	& $1.65^{+0.32}_{-0.16}$ & 0.90 (19)	& -0.60	& -0.65 \\
120     & $21.08^{+0.18}_{-0.28}$	& $1.98^{+0.22}_{-0.19}$ & 1.02 (32) & -0.61	& -0.48 \\
140     & $20.41^{+0.39}_{-0.00}$   	& $1.60^{+0.20}_{-0.16}$ & 0.84 (10)	& -0.72	& -0.33 \\
142     & $20.41^{+0.50}_{-0.00}$       & $2.28^{+0.42}_{-0.24}$ & 1.15 (10)	& -0.74	& -0.37 \\
144     & $20.98^{+0.30}_{-0.30}$	& $2.63^{+0.56}_{-0.26}$ & 1.00	(12) & -0.60	& -0.51 \\
161     & $20.41^{+0.12}_{-0.00}$	& $2.77^{+0.08}_{-0.05}$ & 1.40	(74) & -0.81	& -0.60 \\
179     & $20.41^{+0.40}_{-0.00}$	& $2.34^{+0.35}_{-0.25}$ & 1.30	(11) & -0.65	& -0.52 \\
281     & $20.97^{+0.29}_{-0.35}$	& $2.37^{+0.25}_{-0.47}$ & 0.69	(12) & -0.72	& -0.33 \\
282     & $20.41^{+0.35}_{-0.00}$   	& $2.45^{+0.25}_{-0.16}$ & 1.07	(19) & -0.72	& -0.62 \\
416     & $20.41^{+0.48}_{-0.00}$	& $2.03^{+0.36}_{-0.23}$ & 1.24	(13) & -0.60	& -0.52 \\
440     & $20.68^{+0.45}_{-0.26}$	& $2.02^{+0.50}_{-0.29}$ & 0.34	(10) & -0.58	& -0.52 \\
561     & $20.42^{+0.57}_{-0.00}$	& $2.46^{+0.51}_{-0.21}$ & 0.74	(13) & -0.76	& -0.55 \\
577     & $20.41^{+0.38}_{-0.00}$	& $2.35^{+0.19}_{-0.13}$ & 0.60	(25) & -0.74	& -0.52 \\
738     & $20.41^{+0.58}_{-0.00}$    	& $2.36^{+0.28}_{-0.26}$ & 0.43	(8) & -0.69	& -0.39 \\
742     & $20.41^{+0.38}_{-0.00}$	& $2.13^{+0.38}_{-0.32}$ & 1.26	(7) & -0.66	& -0.50 \\
788     & $20.41^{+0.44}_{-0.00}$	& $1.81^{+0.27}_{-0.20}$ & 0.62	(11) & -0.56	& -0.80 \\
789     & $20.42^{+0.54}_{-0.00}$ 	& $2.32^{+0.49}_{-0.22}$ & 0.58 (11) & -0.74	& -0.47 \\
820     & $20.41^{+0.61}_{-0.00}$	& $1.48^{+0.51}_{-0.22}$ & 1.21	(8) & -0.52	& -0.41 \\
1199    & $21.22^{+0.23}_{-0.30}$	& $2.13^{+0.32}_{-0.32}$ & 1.06	(9) & -0.58	& -0.32 \\
1226    & $20.54^{+0.36}_{-0.13}$	& $1.83^{+0.21}_{-0.15}$ & 0.80	(22) & -0.62	& -0.49 \\
1248    & $20.41^{+0.47}_{-0.00}$	& $1.98^{+0.23}_{-0.15}$ & 1.01	(20) & -0.67	& -0.33 \\
1252    & $20.41^{+0.21}_{-0.00}$	& $2.20^{+0.20}_{-0.19}$ & 1.24 (15) & -0.66	& -0.69 \\
\hline
\end{tabular}
\caption{Spectral parameters obtained using simple power law model with both $\Gamma$ and N$_H$ free parameters for sources for
which $\chi^2$ statistics can be used. The quoted errors correspond to the 90 per
cent confidence level for one interesting parameter. The formal 0.00 errors on logN$_H$ are
the result of having fixed a minimum column density (the galactic value).}
\label{z0tab}
\end{center}
\end{table*}

\begin{landscape}
\begin{table*}
\begin{center}
\begin{tabular}{llllllllll}
\hline \hline
XMDS ID	& \multicolumn{4}{c}{Results if $z = 0$} & \multicolumn{4}{c}{Results if $z = z_{ph}$} & EPIC\\
	& \multicolumn{2}{c}{$\Gamma = 2.0$}	&  \multicolumn{2}{c}{$\Gamma = 1.7$}	& & & & & camera\\	
	& logN$_H$ (cm$^{-2})$	& $\chi^2_\nu$ (d.o.f.)	&  logN$_H$ (cm$^{-2})$	& $\chi^2_\nu$ (d.o.f.)	& logN$_H$ (cm$^{-2})$		& $\Gamma$ 		 & $\chi^2_\nu$ (d.o.f.)& $z$ & \\
\hline
4	& $21.64^{+0.13}_{-0.14}$	& --		& $21.52^{+0.15}_{-0.17}$	& --		& $21.81^{+0.15}_{-0.16}$       & 2.0	& --	& 0.265$^a$ & pn \\
18      & $21.26^{+0.15}_{-0.19}$	& --		& $21.07^{+0.20}_{-0.30}$	& --		& $22.12^{+0.19}_{-0.24}$       & 2.0	& --	& 1.612 & pn \\
36      & $21.48^{+0.19}_{-0.22}$	& --		& $21.31^{+0.23}_{-0.32}$	& --		& $22.13^{+0.25}_{-0.29}$       & 2.0	& --	& 1.043 & M1 M2 \\
40      & $22.22^{+0.22}_{-0.23}$	& --		& $22.14^{+0.24}_{-0.25}$	& --		& $23.42^{+0.23}_{-0.25}$       & 2.0	& --	& 1.955 & pn \\
55      & $21.86^{+0.27}_{-0.33}$	& --		& $21.73^{+0.29}_{-0.38}$	& --		& $22.61^{+0.28}_{-0.36}$	& 2.0	& --	& 1.029$^a$ & pn \\
60      & $20.42^{+0.42}_{-0.01}$	& --		& $20.41^{+0.21}_{-0.00}$	& --		& $20.41^{+0.53}_{-0.00}$       & 2.0	& --	& 0.478$^a$ & pn \\
71      & $21.94^{+0.19}_{-0.23}$	& --		& $21.81^{+0.22}_{-0.27}$	& --		& $22.91^{+0.19}_{-0.25}$       & 2.0	& --	& 1.401 & M1 M2 \\
91      & $20.41^{+0.23}_{-0.00}$	& --		& $20.41^{+0.16}_{-0.00}$	& --		&      				& 	&	& 	& pn \\
94	& $21.77^{+0.14}_{-0.15}$	& --		& $21.67^{+0.15}_{-0.17}$	& --		& $22.03^{+0.16}_{-0.17}$	& 2.0	& --	& 0.349 & pn \\
111    	& $21.50^{+0.22}_{-0.25}$	& --		& $21.35^{+0.25}_{-0.33}$	& --		&				&	&	&	& pn \\
112	& $21.00^{+0.18}_{-0.24}$	& 1.01 (20)	& $20.58^{+0.33}_{-0.17}$	& 0.86 (20)	& $20.61^{+0.85}_{-0.20}$	& $1.68^{+0.21}_{-0.19}$ & 0.89 (19) & 0.878$^a$ &M1 \\
120	& $21.09^{+0.11}_{-0.13}$	& 0.99 (33)	& $20.75^{+0.18}_{-0.25}$	& 1.16 (33)	&	&	&	&	& M1 M2 \\
124	&				&		& $20.90^{+0.77}_{-0.48}$	& --		& $20.73^{+0.99}_{-0.32}$       & 1.7	& --	& 0.371 & pn \\
133     & $21.50^{+0.15}_{-0.17}$	& --		& $21.37^{+0.18}_{-0.23}$	& --		& $21.98^{+0.18}_{-0.21}$       & 2.0	& --	& 0.723 & pn \\
139     & $20.42^{+0.31}_{-0.00}$	& --		& $20.41^{+0.15}_{-0.00}$	& --		& $20.41^{+0.47}_{-0.00}$   	& 2.0	& --	& 0.729 & m1 M2 \\
140     & $20.93^{+0.22}_{-0.34}$	& 1.45 (11)	& $20.48^{+0.39}_{-0.07}$	& 0.85 (11)	& $20.41^{+0.36}_{-0.00}$       & $1.60^{+0.21}_{-0.16}$ & 0.84 (10) & 0.22$^a$ &pn \\
142     & $20.41^{+0.20}_{-0.00}$	& 1.38 (11)	& $20.42^{+0.00}_{-0.00}$	& 2.75 (11)	& $20.41^{+0.66}_{-0.00}$       & $2.28^{+0.37}_{-0.24}$ & 1.15 (10) & 0.573 & pn \\
144     & $20.42^{+0.27}_{-0.00}$	& 1.25 (13)	& $20.41^{+0.00}_{-0.00}$	& 2.26 (13)	& $21.36^{+0.36}_{-0.53}$       & $2.61^{+0.50}_{-0.26}$ & 0.99 (12) & 0.864 & pn \\
149     & $20.93^{+0.26}_{-0.47}$	& --		& $20.51^{+0.46}_{-0.10}$	& --		& $21.08^{+0.34}_{-0.67}$	& 2.0	& --	& 0.493 & pn \\
161     & $20.42^{+0.00}_{-0.00}$	& 13.24 (75)	& $20.41^{+0.00}_{-0.00}$	& 22.11 (75)	& $20.41^{+0.00}_{-0.00}$	& $2.77^{+0.07}_{-0.05}$ & 1.40 (74) & 0.053$^a$ & pn \\
179     & $20.41^{+0.14}_{-0.00}$   	& 1.63 (12)	& $20.42^{+0.00}_{-0.00}$   	& 2.96 (12)	& $20.41^{+0.45}_{-0.00}$       & $2.34^{+0.32}_{-0.25}$ & 1.30 (11) & 0.495$^a$ & pn \\
\hline
\multicolumn{10}{l}{$^a$: spectroscopic redshift}\\
\end{tabular}
\caption{Spectral parameters obtained using a simple power law model with $z = 0$ and $\Gamma
= 2.0$ in columns 2 and 3, $\Gamma = 1.7$ in columns 4 and 5 and using 
photometric (or spectroscopic when available) redshift in colums 6, 7 and 8. 
In this case, the photon index is a free parameter when $\chi^2$ statistics was used.
The quoted errors correspond to the 90 per
cent confidence level for one interesting parameter. 
The formal 0.00 errors on logN$_H$ are
the result of having fixed a minimum column density (the galactic value).}
\label{alltab}
\end{center}
\end{table*}

\setcounter{table}{1}%
\begin{table*}
\begin{center}
\begin{tabular}{llllllllll}
\hline \hline
XMDS ID	& \multicolumn{4}{c}{Results if $z = 0$} & \multicolumn{4}{c}{Results if $z = z_{ph}$} & EPIC\\
	& \multicolumn{2}{c}{$\Gamma = 2.0$}	&  \multicolumn{2}{c}{$\Gamma = 1.7$}	& & & & & camera\\	
	& logN$_H$ (cm$^{-2})$	& $\chi^2_\nu$ (d.o.f.)	&  logN$_H$ (cm$^{-2})$	& $\chi^2_\nu$ (d.o.f.)	& logN$_H$ (cm$^{-2})$		& $\Gamma$ 		 & $\chi^2_\nu$	(d.o.f.)& $z_{ph}$ & \\
\hline
272     & $20.41^{+0.13}_{-0.00}$	& --		& $20.41^{+0.09}_{-0.00}$	& --		& $20.41^{+0.04}_{-0.00}$       & 2.0	& --	& 0.797 & pn \\
281     & $20.57^{+0.28}_{-0.15}$	& 0.75 (13)	& $20.41^{+0.18}_{-0.00}$	& 1.34 (13)	& $21.35^{+0.36}_{-0.66}$       & $2.32^{+0.19}_{-0.22}$ & 0.67 (12) & 0.93 & pn \\
282     & $20.41^{+0.00}_{-0.00}$ 	& 2.18 (20)	& $20.42^{+0.00}_{-0.00}$ 	& 4.71 (20)	& $20.41^{+0.32}_{-0.00}$       & $2.45^{+0.26}_{-0.16}$ & 1.07 (19) & 0.189 & pn \\
406     & $22.29^{+0.21}_{-0.24}$	& --		& $22.20^{+0.22}_{-0.25}$	& --		& $22.88^{+0.21}_{-0.24}$	& 2.0	& --	& 0.713 & pn \\
416     & $20.42^{+0.34}_{-0.00}$	& 1.16 (14)	& $20.41^{+0.19}_{-0.00}$	& 1.55 (14)	& $20.41^{+1.17}_{-0.00}$       & $2.03^{+0.33}_{-0.23}$ & 1.24 (13) & 1.668 & pn \\
440     & $20.65^{+0.31}_{-0.23}$	& 0.31 (11)	& $20.42^{+0.25}_{-0.00}$	& 0.62 (11)	& $21.50^{+0.73}_{-1.09}$       & $2.00^{+0.36}_{-0.27}$ & 0.34 910) & 2.574 & pn \\
453	& $22.24^{+0.39}_{-0.38}$	& --		& $22.10^{+0.41}_{-0.66}$	& --		&				&	&	&	& M2 pn \\
498     & $20.42^{+0.44}_{-0.00}$	& --		& $20.41^{+0.23}_{-0.00}$	& --		& $20.41^{+1.29}_{-0.00}$       & 2.0	& --	& 2.031$^a$ & pn \\
521     & $21.49^{+0.22}_{-0.30}$	& --		& $21.35^{+0.26}_{-0.46}$	& --		& $22.29^{+0.17}_{-0.40}$       & 2.0	& --	& 1.325$^a$ & M1 M2 \\
561     & $20.41^{+0.12}_{-0.00}$	& 1.61 (14)	& $20.41^{+0.00}_{-0.00}$	& 3.26 (14)	& $20.42^{+1.32}_{-0.01}$       & $2.48^{+0.48}_{-0.23}$ & 0.74 (13) & 1.678 & pn \\
564     & $20.42^{+0.59}_{-0.01}$	& --		& $20.41^{+0.41}_{-0.00}$	& --		& $20.41^{+0.63}_{-0.00}$	& 2.0	& --	& 0.224 & M1 M2 \\
577	& $20.41^{+0.09}_{-0.00}$	& 1.31 (26)	& $20.41^{+0.00}_{-0.00}$	& 3.08 (26)	& $20.41^{+0.29}_{-0.00}$       & $2.35^{+0.18}_{-0.13}$ & 0.60	(25) & 0.252$^a$ &M1 M2 \\
710     & $20.83^{+0.25}_{-0.41}$	& --		& $20.42^{+0.26}_{-0.00}$	& --		& $21.74^{+0.24}_{-1.33}$       & 2.0	& --	& 2.320 & pn \\
718     & $21.29^{+0.27}_{-0.37}$    	& --		& $21.09^{+0.34}_{-0.66}$    	& --		& $21.81^{+0.34}_{-0.48}$       & 2.0	& --	& 0.896 & pn \\
738	& $20.42^{+0.18}_{-0.00}$	& 0.97 (9)	& $20.41^{+0.00}_{-0.00}$	& 2.40 (9)	& $20.41^{+0.99}_{-0.00}$       & $2.36^{+0.44}_{-0.26}$ & 0.43 (8) & 1.010$^a$ & pn \\
739     & $22.31^{+0.37}_{-0.56}$	& --		& $21.94^{+0.64}_{-0.52}$	& --		& $22.05^{+0.67}_{-0.59}$       & 1.7	& --	& 0.140$^a$ & pn \\
742     & $20.41^{+0.28}_{-0.00}$	& 1.16 (8)	& $20.41^{+0.21}_{-0.00}$	& 1.75 (8)	& $20.41^{+1.06}_{-0.00}$       & $2.13^{+0.38}_{-0.32}$ & 1.26 (7) & 1.924$^a$ & pn \\
746	& $22.50^{+0.14}_{-0.15}$	& --		& $22.41^{+0.15}_{-0.16}$	& --		& $22.71^{+0.15}_{-0.15}$       & 2.0	& --	& 0.219 & M1 M2 \\
755     & $20.41^{+0.26}_{-0.00}$   	& --		& $20.41^{+0.15}_{-0.00}$   	& --		& $20.41^{+0.41}_{-0.00}$       & 2.0	& --	& 0.730 & pn \\
779	&				&		& $21.58^{+0.39}_{-0.30}$	& --		& $22.10^{+0.50}_{-0.46}$       & 1.7 	& --	& 0.637 & pn \\
782    	& $20.42^{+0.50}_{-0.01}$	& --		& $20.42^{+0.32}_{-0.00}$	& --		& $20.41^{+0.79}_{-0.00}$       & 2.0	& --	& 0.826$^a$ & pn \\
\hline
\multicolumn{10}{l}{$^a$: spectroscopic redshift}\\									    
\end{tabular}							    
\caption{continued}
\end{center}
\end{table*}

\setcounter{table}{1}%
\begin{table*}
\begin{center}
\begin{tabular}{llllllllll}
\hline \hline
XMDS ID	& \multicolumn{4}{c}{Results if $z = 0$} & \multicolumn{4}{c}{Results if $z = z_{ph}$} & EPIC\\
	& \multicolumn{2}{c}{$\Gamma = 2.0$}	&  \multicolumn{2}{c}{$\Gamma = 1.7$}	& & & & & camera\\	
	& logN$_H$ (cm$^{-2})$	& $\chi^2_\nu$ (d.o.f.)	&  logN$_H$ (cm$^{-2})$	& $\chi^2_\nu$ (d.o.f.)	& logN$_H$ (cm$^{-2})$		& $\Gamma$ 		 & $\chi^2_\nu$	(d.o.f.)& $z_{ph}$ & \\
\hline
788     & $20.61^{+0.35}_{-0.20}$	& 0.71 (12)	& $20.41^{+0.27}_{-0.00}$	& 0.63 (12)	& $20.41^{+0.73}_{-0.00}$       & $1.81^{+0.24}_{-0.20}$ & 0.62 (11) & 0.958 & pn \\
789     & $20.41^{+0.16}_{-0.00}$	& 1.00 (12)	& $20.42^{+0.00}_{-0.00}$	& 2.45 (12)	& $20.41^{+0.76}_{-0.00}$       & $2.33^{+0.41}_{-0.23}$ & 0.58 (11) & 0.604$^a$ & pn \\
820     & $21.08^{+0.22}_{-0.32}$	& 1.41 (9)	& $20.79^{+0.32}_{-0.37}$	& 1.14 (9)	& $20.41^{+0.74}_{-0.00}$       & $1.50^{+0.40}_{-0.23}$ & 1.21 (8) & 0.293$^a$ & pn \\
825	& $21.41^{+0.12}_{-0.14}$	& --		& $21.22^{+0.16}_{-0.21}$	& --		& $22.36^{+0.16}_{-0.18}$       & 2.0	& --	& 1.735 & M1 M2 \\
828     & $20.42^{+0.41}_{-0.00}$	& --		& $20.41^{+0.26}_{-0.00}$	& --		& $20.41^{+0.64}_{-0.00}$	& 2.0	& --	& 0.673 & pn \\
846     & $20.68^{+0.38}_{-0.26}$   	& --		& $20.41^{+0.36}_{-0.00}$   	& --		& $20.85^{+0.56}_{-0.44}$       & 2.0	& --	& 0.842$^a$ & pn \\
1199    & $21.16^{+0.18}_{-0.25}$   	& 2.02 (10)	& $20.95^{+0.25}_{-0.40}$   	& 1.35 (10)	& $21.37^{+0.26}_{-0.40}$       & $2.11^{+0.52}_{-0.34}$ & 1.06 (9) & 0.331$^a$ & pn \\
1226    & $20.81^{+0.17}_{-0.25}$	& 0.86 (23)	& $20.42^{+0.19}_{-0.00}$	& 0.83 (23)	& $20.81^{+0.55}_{-0.040}$	& $1.86^{+0.19}_{-0.20}$ & 0.78 (22) & 0.946 & pn \\
1227    & $21.67^{+0.27}_{-0.26}$	& --		& $21.55^{+0.28}_{-0.31}$	& --		& $21.97^{+0.32}_{-0.31}$       & 2.0	& --	& 0.445 & pn \\
1248    & $20.42^{+0.38}_{-0.00}$	& 0.97 (21)	& $20.42^{+0.16}_{-0.00}$	& 1.40 (21)	& $20.41^{+0.33}_{-0.00}$       & $1.98^{+0.23}_{-0.15}$ & 1.01 (20) & 0.034 &M1 M2 \\
1252	& $20.41^{+0.12}_{-0.00}$	& 1.34 (16)	& $20.41^{+0.00}_{-0.00}$	& 2.39 (16)	& $20.41^{+0.19}_{-0.00}$       & $2.20^{+0.20}_{-0.19}$ & 1.24 (16) & 0.590 &pn \\
1264	& $20.42^{+0.28}_{-0.00}$	& --		& $20.42^{+0.20}_{-0.00}$	& --		& $20.41^{+0.81}_{-0.00}$ 	& 2.0	& --	& 1.694 & pn \\
1265	& $22.20^{+0.15}_{-0.16}$	& --		& $22.11^{+0.16}_{-0.18}$	& --		& $22.40^{+0.16}_{-0.17}$	& 2.0	& --	& 0.205$^a$ & pn \\
\hline
\multicolumn{10}{l}{$^a$: spectroscopic redshift}\\
\end{tabular}							    
\caption{continued}
\end{center}
\end{table*}

\end{landscape}

\end{document}